\documentclass{aa}  
\usepackage{graphicx}
\usepackage{color}
\usepackage[english]{babel}
\usepackage{graphicx}
\usepackage{hyperref}
\usepackage{lscape}
\usepackage{multirow}
\usepackage{natbib}
\usepackage[varg]{txfonts}
\usepackage{url}
\usepackage{breakurl}
\usepackage{xspace}
\bibpunct{(}{)}{;}{a}{}{,}
\newcounter{Rco}
\newcommand{\ionw}[3]{\mbox{\ion{#1}{#2}~$\lambda\,#3\,\mathrm{\AA}$}\xspace}

\newcommand{\Ionst}[1]{\setcounter{Rco}{#1}\Roman{Rco}}

\newcommand{\Ionw}[3]{\mbox{#1\,{\scriptsize\Ionst{#2}}~$\lambda\,#3$\,\AA}\xspace}

\newcommand{\Ionww}[3]{\mbox{#1\,{\scriptsize\Ionst{#2}}~$\lambda\lambda\,#3$\,\AA}\xspace}

\newcommand{\logg}{\mbox{$\log g$}\xspace}
\newcommand{\loggw}[1]{\mbox{$\log g\hspace{-0.5mm} =\hspace{-0.5mm}  #1$}}

\newcommand{\Teff}{\mbox{$T_\mathrm{eff}$}\xspace}
\newcommand{\Teffw}[1]{\mbox{$\Teff\hspace{-0.5mm} =\hspace{-0.5mm} #1 \,\mathrm{K}$}}
\newcommand{\ebv}{\mbox{$E_{B-V}$}}
\newcommand{\ebvw}[1]{\mbox{$\ebv\hspace{-0.5mm} =\hspace{-0.5mm} #1$}}

\newcommand{\mmspr}{\hbox{}\hspace{+0.5cm}}
\newcommand{\mmspa}{\hspace{+0.5cm}\hbox{}}

\newcommand{\tmspr}{\hbox{}\hspace{+1.8mm}}

\newcommand{\vmspl}{\hbox{}\hspace{+2.5mm}}
\newcommand{\noten}[1]{\multicolumn{2}{c}{#1\textcolor{white}{\hspace{0.2mm}$\times$\hspace{0.2mm}$10^{+0}$}}}
\newcommand{\two}{\multicolumn{2}{c}{}}
\newcommand{\rel}{\object{RX\,J0503.9$-$2854}\xspace}
\newcommand{\re}{\object{RE\,0503$-$289}\xspace}
\begin{document}
\title{Complete spectral energy distribution \\ of the hot, helium-rich white dwarf \rel
           \thanks
           {Based on observations with the NASA/ESA Hubble Space Telescope, obtained at the Space Telescope Science 
            Institute, which is operated by the Association of Universities for Research in Astronomy, Inc., under 
            NASA contract NAS5-26666.
           }\fnmsep
           \thanks
           {Based on observations made with the NASA-CNES-CSA Far Ultraviolet Spectroscopic Explorer.
           }\fnmsep
           \thanks
           {Based on observations made with ESO Telescopes at the La Silla Paranal Observatory under program
            IDs 072.D-0362, 165.H-0588, and 167.D-0407.}
      }

\titlerunning{The complete spectral energy distribution of the helium-rich white dwarf \rel}

\author{D\@. Hoyer\inst{1}
        \and
        T\@. Rauch\inst{1}
        \and
        K\@. Werner\inst{1}
        \and
        J\@. W\@. Kruk\inst{2}
        \and
        P\@. Quinet\inst{3,4}
        }

\institute{Institute for Astronomy and Astrophysics,
           Kepler Center for Astro and Particle Physics,
           Eberhard Karls University,
           Sand 1,
           72076 T\"ubingen,
           Germany \\
           \email{rauch@astro.uni-tuebingen.de}
           \and
           NASA Goddard Space Flight Center, Greenbelt, MD\,20771, USA
           \and
           Physique Atomique et Astrophysique, Universit\'e de Mons -- UMONS, 7000 Mons, Belgium
           \and
           IPNAS, Universit\'e de Li\`ege, Sart Tilman, 4000 Li\`ege, Belgium}

\date{Received 10 October 2016 / Accepted 17 October 2016}

\abstract {In the line-of-sight toward the DO-type white dwarf \rel, the density of the
           interstellar medium (ISM) is very low, and thus the contamination of the stellar
           spectrum almost negligible. This allows to identify many metal lines in a wide
           wavelength range from the extreme ultraviolet to the near infrared.  
          }
          {In previous spectral analyses, many metal lines in the ultraviolet spectrum of
           \rel have been identified. A complete line list of observed and identified lines
           is presented here. 
          }
          {We compared synthetic spectra that had been calculated from model atmospheres in
           non-local thermodynamical equilibrium, with observations.
          }
          {In total, we identified 1272 lines (279 of them were newly assigned) in the
           wavelength range from the extreme ultraviolet to the near infrared.
           287 lines remain unidentified.
           A close inspection of the EUV shows that still no good fit to
           the observed shape of the stellar continuum flux can be achieved although
           He, C, N, O, Al, Si, P, S, Ca, Sc, Ti, V, Cr, Mn, Fe, Cr, Ni
           Zn, Ga, Ge, As, Kr, Zr, Mo, Sn, Xe, and Ba are included in the stellar
           atmosphere models.           
          }
          {There are two possible reasons for the deviation between observed and synthetic flux in the EUV may have two
           reasons. Opacities from hitherto unconsidered elements in
           the model-atmosphere calculationmay be missing, and/or the effective temperature is
           slightly lower than previously determined. 
          }
         

\maketitle

\section{Introduction}
\label{sect:intro}

The white dwarf (WD) \rel \citep[henceforth \re, \object{WD\,0501$-$289}][]{mccooksion1999,mccooksion1999cat} 
was discovered in the
ROSAT (ROentgen SATellite)
wide field camera all-sky survey of extreme-ultraviolet (EUV) sources \citep{poundsetal1993}.
\citet{barstowetal1993} reported its discovery by the Extreme Ultraviolet Explorer (EUVE),
and identified it with a peculiar He-rich DO-type WD, namely \object{MCT\,0501$-$2858} 
in the Montreal-Cambridge-Tololo survey of southern hemisphere blue stars \citep{demersetal1986}. 
They found that \re is located in a direction with very low density of the interstellar medium (ISM).
In the line of sight (LOS) toward \re, \citet{vennesetal1994} measured a column density of 
$\log (N_\mathrm{\ion{H}{i}}\,/\,\mathrm{cm^{-2}}) =  17.75 - 18.00$ using EUVE photometry data. 
\citet{rauchetal2016kr} resolved at least two ISM components in the LOS toward \re based on
high-resolution and high signal-to-noise ultraviolet (UV) spectroscopy performed by
Far Ultraviolet Spectroscopic Explorer (FUSE)
and
HST/STIS (Hubble Space Telescope / Space Telescope Imaging Spectrograph) and 
measured a very low (\ebvw{0.015 \pm 0.002}) interstellar reddening.

The almost negligible contamination by ISM line absorption allows us to identify even weak lines of
many species from so far He up to trans-iron elements as heavy as Ba (Table\,\ref{tab:idhistory}). 
For reliable abundance analyses of these elements, a precise \Teff and  \logg determination is a prerequisite
to keep error propagation as small as possible. 
An initial constraint of \Teffw{60\,000 - 70\,000} was given by \citet{vennesetal1994} from EUV photometry.
The first spectral analysis by means of non-local thermodynamic equilibrium (NLTE) stellar atmosphere models
considering opacities of H, He, and C was published by \citet{barstowetal1994}. They found
\Teffw{60\,000 - 80\,000} and $\log (g \mathrm{/ cm/s^2}) = 7.5 - 8.0$. \citet{dreizlerwerner1996}
used ultraviolet (UV) spectra in addition and NLTE model
atmospheres and determined \Teffw{70\,000 \pm 5\,000} and \loggw{7.5 \pm 0.5}.
Recently, \citet{rauchetal2016kr} analyzed optical and ultraviolet (FUSE and HST/STIS) spectra and
significantly reduced the error limits to $\pm 2\,000\,\mathrm{K}$ and $\pm 0.1$, respectively.
Table\,\ref{tab:previous} summarizes previous analyses.

\begin{table*}
\caption{History of \Teff and \logg determinations \citep[cf\@.,][]{ringatPhD2013}. PM denotes photometry.}
\label{tab:previous}
\setlength{\tabcolsep}{.3em}
\centering
\begin{tabular}{r@{\,$-$\,}llllp{30mm}l}
\hline
\hline
\noalign{\smallskip}
\multicolumn{2}{c}{\Teff\,/\,kK}               & \logg                 & Model atmosphere       & Method              & Comment                                            & Reference                  \\
\hline                                                          
\noalign{\smallskip}                                            
60&90                                          &                       &                        & EUV, PM             & very low $N_{\mathrm{H}\,\textsc{i}}$                 & \citet{barstowetal1993}     \\
                                                                
60&80                                          &                       &                        & EUV, OPT            & very low $N_{\mathrm{H}\,\textsc{i}}$                 & \citet{barstowetal1993}    \\
                                                                
60&80                                          & 7.5$-$8.0             & He, HHeC               & NLTE, OPT, UV, EUV  & EUV problem\tablefootmark{a}                       & \citet{barstowetal1994}    \\
                                                                
60&70                                          &                       &                        & EUV, PM             & very low $N_{\mathrm{H}\,\textsc{i}}$                 & \citet{vennesetal1994}     \\
                                                                
\multicolumn{2}{l}{\vmspl 70\tablefootmark{b}} & 7.0                   & HeCNOSiFeNi            & \vmspl LTE, EUV, UV &                                                    & \citet{polomskietal1995}   \\
                                                                
\multicolumn{2}{l}{\vmspl 65\tablefootmark{c}} & 7.5\tablefootmark{d}  & HHeC                   & NLTE, OPT           &  no H detectable, upper limit 5\,\% (mass fraction) & \citet{werner1996}         \\
                                                                
\multicolumn{2}{l}{\vmspl 70}                  & 7.5                   & HHeCNOSi               & NLTE, OPT, UV       & $M = 0.49\,M_\odot$                                 & \citet{dreizlerwerner1996} \\
                                                                
66.6&70.4                                      & 7.13-7.27             & HHe                    & \vmspl LTE, UV      & $M = 0.40\,M_\odot$                                 & \citet{vennesetal1998}     \\
                                                                
\multicolumn{2}{l}{\vmspl 70\tablefootmark{e}} & 7.5\tablefootmark{e}  &                        & NLTE, diffusion     & no good fit achieved                                        & \citet{dreizler1999}       \\
                                                                
69&75                                          & 7.26$-$7.63           & HHeC                   & NLTE, OPT, UV, EUV  &  EUV problem\tablefootmark{a}                      & \citet{barstowetal2000}    \\
\multicolumn{2}{l}{}                           &                       & HHeCNOSiFeNi           &                     &                                                    &                            \\
                                                                
                                                                
65&70                                          & 7.5\tablefootmark{e}  & HeCNi, HeONi           & NLTE, EUV           & EUV problem\tablefootmark{a}                       & \citet{werneretal2001}     \\

\multicolumn{2}{l}{\vmspl 70\tablefootmark{e}} & 7.5\tablefootmark{e}  & HHeCNOSiFeNi+          & NLTE, UV            & EUV problem\tablefootmark{a}                       & \citet{barstowetal2007}    \\
     \multicolumn{2}{l}{}                      &                       & PS                     & \vmspl LTE          &                                                    &                            \\
68&72                                          & 7.4$-$7.6             & HeCNOAlSiPS+           & NLTE, OPT, UV       & $M = 0.514^{+0.15}_{-0.05}\,M_\odot$                 & \citet{rauchetal2016kr}    \\
\multicolumn{2}{l}{}                           &                       & CaScTiVCrMnFeCrNi+     &                     &                                                    &                            \\
\multicolumn{2}{l}{}                           &                       & ZnGaGeAsKrZrMoSnXeBa   &                     &                                                    &                            \\
\hline

\end{tabular}
\tablefoot{
\tablefoottext{a}{Sect\@.\,\ref{sect:euvproblem},}
\tablefoottext{b}{adopted upper limit of \citet{vennesetal1994},}
\tablefoottext{c}{adopted value close to lower limit of \citet{barstowetal1994},}
\tablefoottext{d}{adopted from \citet{barstowetal1994},}
\tablefoottext{e}{adopted from \citet{dreizlerwerner1996}}
}
\end{table*}

\begin{table}\centering 
  \caption{Photospheric abundances (mass fraction) of \re. The reference for the 1$^\mathrm{st}$ line identifications
           is given in the final column.}
\label{tab:idhistory}
\setlength{\tabcolsep}{.4em}
\begin{tabular}{lr@{.}lp{52mm}}
\hline
\hline
\noalign{\smallskip}                                                                                          
Element      & \multicolumn{2}{c}{Abundance} & 1$^{st}$ Line identifications\\
\hline                   
\noalign{\smallskip}                                                                                   
\mmspr He~~~ & $ 9$&$73\times 10^{-1}$ & \citet{barstowetal1994} \\
\mmspr C     & $ 2$&$22\times 10^{-2}$ & \citet{barstowetal1994} \\
\mmspr N     & $ 5$&$49\times 10^{-5}$ & \citet{dreizlerwerner1996} \\
\mmspr O     & $ 2$&$94\times 10^{-3}$ & \citet{polomskietal1995},\newline \citet{dreizlerwerner1996} \\
\mmspr Al    & $ 5$&$01\times 10^{-5}$ & \citet{rauchetal2016zr} \\
\mmspr Si    & $ 1$&$60\times 10^{-4}$ & \citet{polomskietal1995},\newline \citet{dreizlerwerner1996} \\
\mmspr P     & $ 1$&$06\times 10^{-6}$ & \citet{vennesetal1998,barstowetal2007} \\
\mmspr S     & $ 3$&$96\times 10^{-5}$ & \citet{barstowetal2007} \\
\mmspr Ni    & $ 7$&$25\times 10^{-5}$ & \citet{barstowetal2000} \\
\mmspr Zn    & $ 1$&$13\times 10^{-4}$ & \citet{rauchetal2014zn} \\
\mmspr Ga    & $ 3$&$44\times 10^{-5}$ & \citet{werneretal2012},\newline \citet{rauchetal2015ga} \\
\mmspr Ge    & $ 1$&$58\times 10^{-4}$ & \citet{werneretal2012},\newline \citet{rauchetal2012ge} \\
\mmspr As    & $ 1$&$60\times 10^{-5}$ & \citet{werneretal2012} \\
\mmspr Se    & \multicolumn{2}{c}{}   & \citet{werneretal2012} \\
\mmspr Kr    & $ 5$&$04\times 10^{-4}$ & \citet{werneretal2012},\newline \citet{rauchetal2016kr} \\
\mmspr Zr    & $ 3$&$00\times 10^{-4}$ & \citet{rauchetal2016zr} \\
\mmspr Mo    & $ 1$&$88\times 10^{-4}$ & \citet{rauchetal2016mo} \\
\mmspr Sn    & $ 2$&$06\times 10^{-4}$ & \citet{werneretal2012} \\
\mmspr Te    & \multicolumn{2}{c}{}   & \citet{werneretal2012} \\
\mmspr I     & \multicolumn{2}{c}{}   & \citet{werneretal2012} \\
\mmspr Xe    & $ 1$&$26\times 10^{-4}$ & \citet{werneretal2012},\newline \citet{rauchetal2015xe},\newline \citet{rauchetal2016zr}  \\
\mmspr Ba    & $ 3$&$57\times 10^{-4}$ & \citet{rauchetal2014ba} \\
\hline
\end{tabular}
\end{table}

\section{Observations}
\label{sect:observations}

In this paper, we used the observed spectra that are briefly described in the following. 
If they are compared to synthetic spectra, the latter are convolved with Gaussians to model 
the respective instrument's resolution.
\vspace{-4mm} 

\paragraph{Extreme ultraviolet} observations by the EUVE observatory were performed using
the  short-wavelength  ($70\,\mathrm{\AA} < \lambda < 190\,\mathrm{\AA}$),
the medium-wavelength ($140\,\mathrm{\AA} < \lambda < 380\,\mathrm{\AA}$), and
the   long-wavelength ($280\,\mathrm{\AA} < \lambda < 760\,\mathrm{\AA}$) 
spectrometers with a resolving power of $R \approx 300$. 
Details of the data reduction are given by \citet{dupuisetal1995}.
\vspace{-4mm} 

\paragraph{Far ultraviolet} spectra 
($910\,\mathrm{\AA} < \lambda < 1190\,\mathrm{\AA}$, $R \approx 20\,000$)
were obtained with FUSE.
Their data IDs are
M1123601 (2000-12-04), 
M1124201 (2001-02-02), and
P2041601 (2000-12-05). The spectra were shifted to rest wavelengths and co-added. For details see \citet{werneretal2012}.
\vspace{-4mm} 

\paragraph{Ultraviolet} spectroscopy was performed with HST/STIS on 2014-08-14.
Two observations with grating E140M
($1144\,\mathrm{\AA} < \lambda < 1709\,\mathrm{\AA}$, $R \approx 45\,800$) and
two observations with grating E230M
($1690\,\mathrm{\AA} < \lambda < 2366\,\mathrm{\AA}$, 
 $2277\,\mathrm{\AA} < \lambda < 3073\,\mathrm{\AA}$, $R \approx 30\,000$)
were co-added.
These observations are retrievable from the Barbara A\@. Mikulski Archive for Space Telescopes (MAST).
\vspace{-4mm} 

\paragraph{Optical} spectra 
($3290\,\mathrm{\AA} < \lambda <  4524\,\mathrm{\AA}$,
 $4604\,\mathrm{\AA} < \lambda <  5609\,\mathrm{\AA}$,
 $5673\,\mathrm{\AA} < \lambda <  6641\,\mathrm{\AA}$) 
were obtained on 2000-09-09 and 2001-04-08 in the framework of the Supernova Ia Progenitor Survey project 
\citep[SPY,][]{napiwotzkietal2001, napiwotzkietal2003}.
The Ultraviolet and Visual Echelle Spectrograph (UVES) attached to the Very Large Telescope (VLT) located at the
European Southern Observatory (ESO) on Cerro Paranal in Chile was employed
to achieve a resolution of about 0.2\,\AA. 
In addition, we use a spectrum  taken with the 
Echelle Multi Mode Instrument (EMMI) attached to the
New Technology Telescope (NTT)  (1992-01, 
$4094\,\mathrm{\AA} < \lambda <  4994\,\mathrm{\AA}$, resolution of about 3.0\,\AA).
\vspace{-4mm} 

\paragraph{Near infrared} spectroscopy 
($9500\,\mathrm{\AA} < \lambda <  13420\,\mathrm{\AA}$,  $R \approx 950$) 
was performed on 2003-12-10 using the Son-of-Isaac (SofI) instrument at the NTT.
The spectrum used here was digitized with Dexter\footnote{\url{http://dc.zah.uni-heidelberg.de/sdexter}}
from Fig.\,1 in \citet{dobbieetal2005}.

\section{Model atmospheres and atomic data}
\label{sect:models}

The stellar model atmospheres used for this paper were calculated with our T\"ubingen NLTE Model Atmosphere Package
\citep[TMAP\footnote{\url{http://astro.uni-tuebingen.de/~TMAP}},][]{werneretal2003,tmap2012}.
They assume plane-parallel geometry, are chemically homogeneous, and in hydrostatic and radiative 
equilibrium. An adaptation is the New Generation Radiative Transport (NGRT) code 
\citep{dreizlerwolff1999, schuhetal2002} that can consider diffusion in addition to calculate
stratified stellar atmospheres.

The T\"ubingen Model Atom Database
(TMAD\footnote{\url{http://astro.uni-tuebingen.de/~TMAD}})
provides ready-to-use model atoms in TMAP format for many species up to Ba.
TMAD has been constructed as part of the T\"ubingen contribution to the German Astrophysical Virtual Observatory 
(GAVO\footnote{\url{http://www.g-vo.org}}).

\citet{werneretal2012} discovered lines of trans-iron elements, namely
Ga (atomic number $Z = 31$), Ge (32), As (33), Se (34), Kr (36), Mo (42), Sn (50), Te (52), I (53), and Xe (54),
in the FUSE spectrum of \re. For precise abundance determinations of these species,
reliable atomic data is mandatory. For example, reliable transition probabilities are required, not only for lines that are
identified in the observation but for the complete model atoms that are considered in the model-atmosphere calculations.
Due to the lack of such data, \citet{werneretal2012} were restricted to abundance determinations of Kr and Xe only.

We initiated the calculation of new transition probabilities that were then used to determine the abundance of the respective
element. Table\,\ref{tab:transprob} gives an overview of the so far calculated data. To provide easy access to
this data, the registered Tübingen Oscillator Strengths Service (TOSS) 
has been created within the GAVO project.

\begin{table}\centering 
  \caption{Newly calculated transition probabilities.}
\label{tab:transprob}
\begin{tabular}{rll}
\hline
\hline
Element & Ions & Reference \\
\hline
\noalign{\smallskip}
Zn & \sc{iv - v  } & \citet{rauchetal2014zn} \\
Ga & \sc{iv - vi } & \citet{rauchetal2015ga} \\
Ge & \sc{v  - vi } & \citet{rauchetal2012ge} \\
Kr & \sc{iv - vii} & \citet{rauchetal2016kr} \\
Zr & \sc{iv - vii} & \citet{rauchetal2016zr} \\
Tc & \sc{ii - vi } & \citet{werneretal2015tc} \\
Mo & \sc{iv - vii} & \citet{rauchetal2016mo} \\
Xe & \sc{iv - vii} & \citet{rauchetal2015xe,rauchetal2016zr} \\
Ba & \sc{v  - vii} & \citet{rauchetal2014ba} \\
\hline
\end{tabular}
\end{table}

To construct model atoms for the use within TMAP, the elements given in Table\,\ref{tab:transprob}
require the calculation of so-called super levels and super lines with our Iron Opacity and Interface
\citep[IrOnIc,][]{rauchdeetjen2003} due to the
very high number of atomic levels and lines. We transferred the TOSS data into Kurucz's data 
format\footnote{\url{http://kurucz.harvard.edu/atoms.html}} that can be read by IrOnIc.

\section{Radial velocity and gravitational redshift}
\label{sect:lines}

\begin{figure}
   \resizebox{\hsize}{!}{\includegraphics{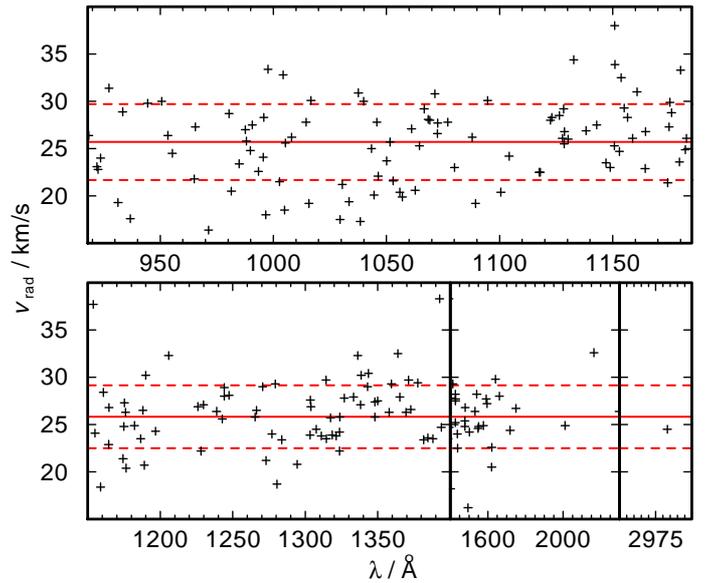}}
    \caption{Determination of $v_\mathrm{rad}$ from individual lines in the FUSE (top panel) and
             HST/STIS observations (bottom).
             The full horizontal lines indicate the average $v_\mathrm{rad}$ for FUSE and
             HST/STIS, respectively. The dashed lines show the 1\,$\sigma$ error.
            }
   \label{fig:vrad}
\end{figure}

To shift the observation to rest wavelength, we determined the radial velocity $v_\mathrm{rad}$ of \re from 
FUSE and HST/STIS spectra. To measure the wavelengths of the line centers, we used 
IRAF\footnote{IRAF is distributed by the National Optical Astronomy 
Observatory, which is operated by the Associated Universities for 
Research in Astronomy, Inc., under cooperative  agreement with the 
National Science Foundation.} to fit Gaussians to the line profiles.
In total, we evaluated 
100 lines in the FUSE wavelength range and 
103 lines in the STIS wavelength range (Fig.\,\ref{fig:vrad}). The averages are 
$v_\mathrm{rad}^\mathrm{FUSE} = 25.7 \pm 4.2\,\mathrm{km/s}$ and
$v_\mathrm{rad}^\mathrm{STIS} = 25.8 \pm 3.7\,\mathrm{km/s}$. We adopted the mean value of 
$v_\mathrm{rad} = 25.7^{+3.6}_{-4.0}\,\mathrm{km/s}$. From this value, the gravitational redshift $z$
has to be subtracted. To calculate $z$ and the respective radial velocity, we created the GAVO tool
T\"ubingen Gravitational REDshift calculator (TGRED, Fig.\,\ref{fig:TGRED}).
For \re, we derive $v^\mathrm{gred}_\mathrm{rad} = 15.5^{+6.7}_{-4.6}\,\mathrm{km/s}$. The true radial velocity
is then $v^\mathrm{RE\,0503-289}_\mathrm{rad} = 10.2^{+8.2}_{-8.6}\,\mathrm{km/s}$.

\section{Line identification}
\label{sect:lines}

To unambiguously identify lines in our observed spectra (Sect.\,\ref{sect:observations}),
we used the best synthetic model of \citet{rauchetal2016zr} and calculated additional spectra
with oscillator strengths set to zero for individual elements. This allows to find weak lines,
even if they are blended by stronger lines. The detection limit is an equivalent width of 
$W_\lambda = 2\,\mathrm{m\AA}$. Table\,\ref{tab:linestat} shows the total numbers of lines 
identified in the four wavelength ranges and the numbers of lines that were suited to determine
$W_\lambda$ and $v_\mathrm{rad}$. The current line lists are presented in 
Tables\,\ref{tab:lineids_EUVE} - \ref{tab:lineids_sofi}, a regularly updated version is available
at \url{http://astro.uni-tuebingen.de/~hoyer/objects/RE0503-289}.

\begin{table}\centering 
\caption{Statistics of the identified (in brackets: newly identified in this paper) and unidentified lines in the observed spectra. 
         The last two columns give the numbers of lines that were used to measure their equivalent widths $W_\lambda$ and 
         $v_\mathrm{rad}$
         (Fig.\,\ref{fig:vrad}), respectively.}
\label{tab:linestat}
\setlength{\tabcolsep}{.6em}
\begin{tabular}{rrr@{}lrrr}
\hline
\hline
\noalign{\smallskip}
   Wavelength    & \multicolumn{4}{c}{Numbers of lines}                         &             &                  \\
\cline{2-5}                                                                     
                 & \multicolumn{4}{c}{}                                         &             &                  \vspace{-5.5mm}\\
                 &       & \multicolumn{3}{c}{}                                 &  $W_\lambda$ &  $v_\mathrm{rad}$ \vspace{-2mm}\\
        Range    & Total & \multicolumn{2}{c}{Identified}  & Unidentified       &             &                  \\
\hline                   
\noalign{\smallskip}
EUV              &    74 &         74&(\tmspr 35)          &            0\mmspa &           0 &                0 \\
FUV              &   616 &        536&(\tmspr 55)          &           76\mmspa &         148 &              100 \\
NUV              &   790 &        579&(120)                &          211\mmspa &         252 &              103 \\
optical          &    83 &         83&(\tmspr 69)          &            0\mmspa &           0 &                0 \\
NIR              &     2 &          2&(\tmspr \tmspr 0)    &            0\mmspa &           0 &                0 \\
\hline
\end{tabular}
\end{table}

\section{Visualization and online line list}
\label{sect:tvis}

In the framework of the Tübingen (GAVO project, we have developed the registered 
T\"ubingen VISisualization tool (TVIS) 
that allows the user to plot any data in an easy way on the WWW. The plotter itself is written in HTML5 and Javascript. 
To strongly increase the security of this web application, no Flash or Java is necessary to use it, meaning that 
TVIS will even work when Flash is dead and Java applets are blocked by the browsers. 

The comparison of our best model spectra with the available observation of \re in the EUV, FUV, NUV, and optical wavelength ranges
was realized with TVIS and is shown at 
{\small \url{http://astro.uni-tuebingen.de/~TVIS/objects/RE0503-289}}.
Figures \ref{fig:FUSE_complete} to \ref{fig:OPTI_complete} show the FUV to optical range.

\begin{figure}
   \resizebox{\hsize}{!}{\includegraphics{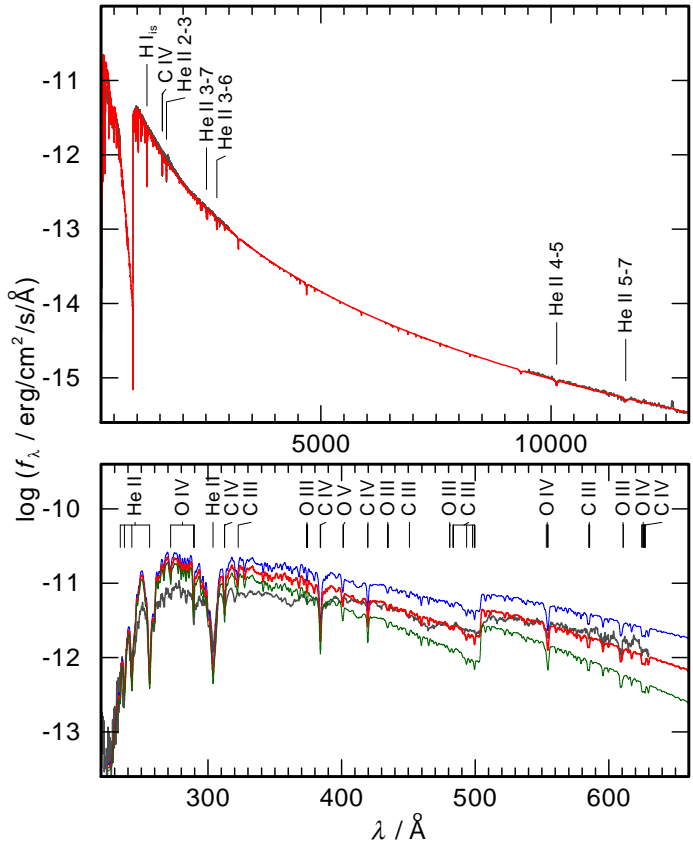}}
    \caption{Determination of \ebv. 
             Top: Reddening with \ebvw{0.00026} applied to our synthetic spectrum in the wavelength range
                  from the EUV to the NIR.
             Bottom: Same like top panel, for \ebvw{0.00016} (blue), 0.00026 (red), and 0.00036 (green) 
                     in the EUV wavelength range.
             Prominent lines are marked.
            }
   \label{fig:ebv}
\end{figure}

\begin{figure*}
   \resizebox{\hsize}{!}{\includegraphics{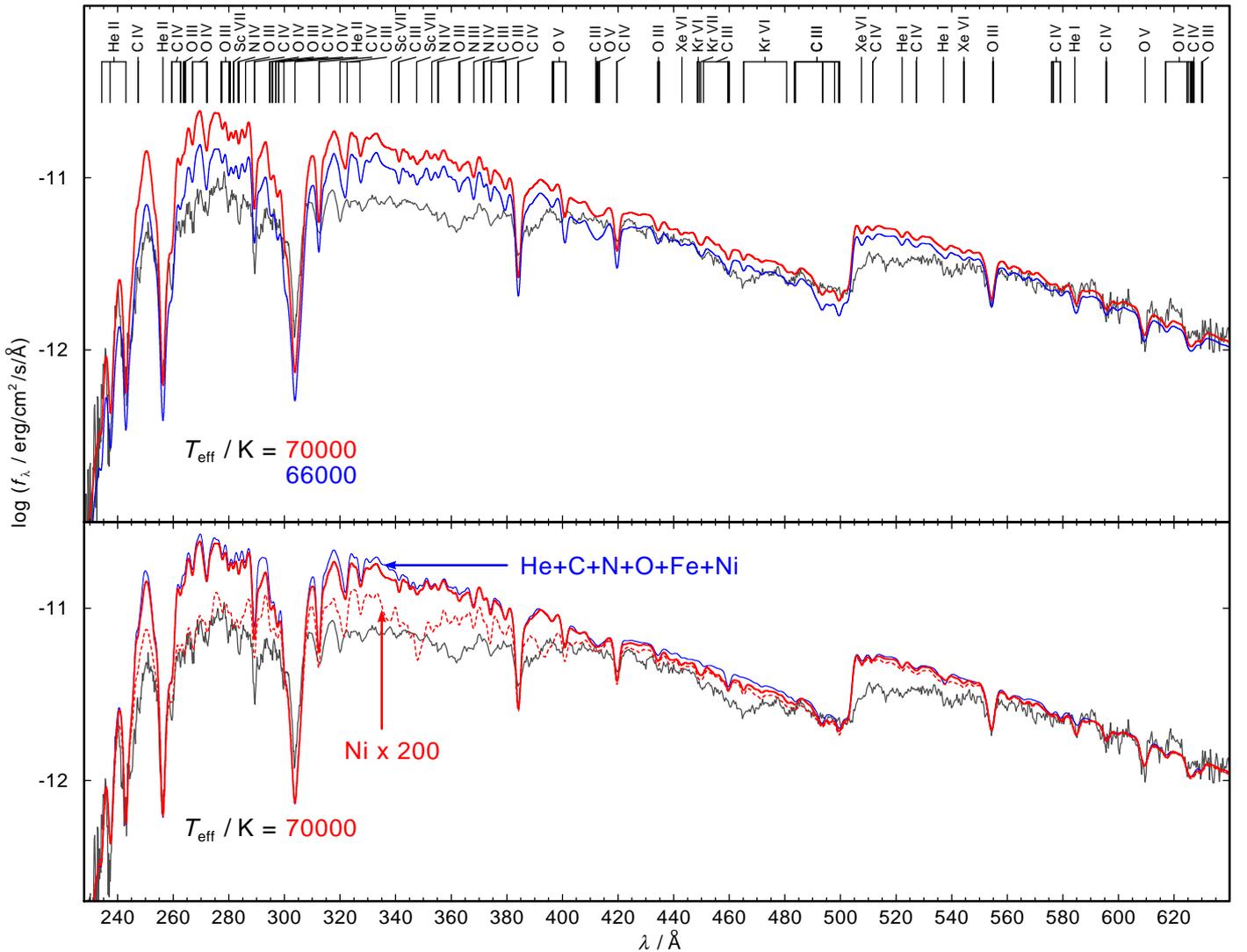}}
    \caption{Comparison of the EUVE observation 
             (gray line in both panels)
             with our models.
             Top panel: Two models with \Teffw{70\,000} (red) and \Teffw{66\,000} (blue).
                  Identified photospheric lines are marked at the top.
             Bottom panel: Three models with \Teffw{70\,000}.
                           Red, thick line: model from the top panel,
                           red, dashed line: model with 200 times increased Ni abundance, 
                           blue, thin line: model that considered only opacities of He, C, N, O, Fe, and Ni.
            }
   \label{fig:euvproblem}
\end{figure*}

\section{Is there still an EUV problem in RE\,0503$-$289?}
\label{sect:euvproblem}

To analyze the EUVE observation, \citet{barstowetal1995} used NLTE model atmospheres that were 
calculated with the code that is nowadays called TMAP. A synthetic spectrum (scaled to match the observed EUV flux)
that was calculated from a model with \Teffw{70\,000}, \loggw{7.0}, $C/He=1\,\%$, and $N/He=0.01\,\%$
(the latter being number ratios) reproduced well the observation. A major problem arose, however, from the fact 
that the model flux (reddened and interstellar neutral hydrogen absorption considered)
in the wavelength range $228\,\mathrm{\AA} < \lambda < 400\,\mathrm{\AA}$ 
was about an order of magnitude higher than observed. Only models with \Teff $< 65\,000\,\mathrm{K}$ produced an 
acceptable fit. \ionw{He}{i}{5875.62} (2p\,$^3$P$^\mathrm{o}$ - 3d\,$^3$D)  in the optical wavelength range 
\citep[e.g., in spectra taken with the TWIN spectrograph at the Calar Alto observatory in SPY spectra,][]{dreizlerwerner1996,rauchetal2016kr} 
establishes a stringent constraint of \Teffw{70\,000 \pm 2000}.

\citet{werneretal2001} calculated TMAP models that were composed of He, C, O, and the iron-group elements (Ca - Ni).
Interstellar \ion{He}{i} absorption was applied in addition to that of \ion{H}{i}.
The flux discrepancy was reduced (model flux three times higher than observed) but the basic problem, finding an agreement 
at \Teffw{70\,000}, was not solved.

\citet{ringatPhD2013} created the T\"ubingen EUV absorption tool (TEUV,
Fig.\ref{fig:TEUV}), that corrects synthetic stellar fluxes for ISM absorption for $\lambda < 911\,\mathrm{\AA}$. 
Presently, only radiative bound-free absorption 
of the lowest ionization states of H, He, C, N, and O is simulated.
Opacity Project data \citep{seatonetal1994} is used for the photoionization cross-sections. These
consider, for example, autoionization features.
For this paper, Si has been added to TEUV.
Two interstellar components with different radial and turbulent velocities, 
temperatures, and column densities can be considered. 
\citet{ringatPhD2013} calculated TMAP models (\Teffw{70\,000}, \loggw{7.5}) that included He, C, N, O, and the iron-group elements. 
Although Kurucz's line lists were strongly extended in 2009 \citep{kurucz2009,kurucz2011}, and about 
a factor of ten more iron-group lines were considered,the EUV model flux was about twice as high as that observed.
To match the observed EUV flux, \Teff had to be reduced to $\la 65\,000\,\mathrm{K}$.

\citet{rauchetal2016zr} determined \Teffw{70\,000\pm 2\,000} and \loggw{7.5\pm 0.1} in a detailed reanalysis of
optical and UV spectra. They included 27 elements, namely He, C, N, O, Al, Si, P, S, Ca, Sc, Ti, V, Cr, Mn, Fe, Cr, Ni,
Zn, Ga, Ge, As, Kr, Zr, Mo, Sn, Xe, and Ba, in their models. From these, we 
calculated the EUV spectrum ($228\,\mathrm{\AA} \le \lambda \le 910\,\mathrm{\AA}$) with
1601 atomic levels treated in NLTE, considering 2481 lines of the elements He - S and about
about 30 million lines of the elements with $Z\ge 20$. The frequency grid comprised
174\,873 points with $\Delta\lambda \le 0.005\,\mathrm{\AA}$.

Figure\,\ref{fig:ebv} demonstrates the determination of the interstellar reddening.
We apply the reddening data of 
\citet[][provided for $1.26\,\mathrm{\AA} \le \lambda \le 413\,\mathrm{\AA}$ and extrapolated toward the \ion{He}{i}
         ground-state threshold]{morrisonmccammon1983}
and
\citet[][$\lambda \ge 911\,\mathrm{\AA}$]{fitzpatrick1999}.
Between the \ion{He}{i} ground-state edge and the \ion{H}{i} Lyman edge, only absorption due to \ion{H}{i} is considered.
To determine \ebv, we normalized our models to the 2MASS H brightness \citep[$14.766\pm 0.063$,][]{cutrietal2003,cutrietal2003cat}.
To match the observed flux level between about 400\,\AA\ to 600\,\AA, \ebvw{0.00026\pm 0.00003} is necessary. This is
less than \ebvw{0.015 \pm 0.002} that was used by \citet{rauchetal2016kr} to reproduce the observed FUSE flux level. With the
Galactic reddening law of \citet[][$\log(N_\mathrm{\ion{H}{i}}/\ebv = 21.58\pm 0.1$]{groenewegenlamers1989} and
the total cloud column density of interstellar \ion{H}{i} of $1.5 \pm 0.2 \times 10^{18}\,\mathrm{cm^{-2}}$
\citep[measured from L\,$\beta$,][]{rauchetal2016kr}, we can calculate \ebvw{0.00039^{+0.00017}_{-0.00012}} which is within error 
limits well in agreement with our result.

A close look at the EUV wavelength range shows still a significant difference between model and observation (Fig.\,\ref{fig:euvproblem}, top panel),
most prominent between 250\,\AA\ and 400\,\AA\ and between 504\,\AA\ and 550\,\AA. Our present models reduced the deviation by about a factor of 
two compared the models of \citet{werneretal2001}. The EUV problem cannot be solved by using a cooler
model, even at \Teffw{66\,000}, which is already outside the error range of \Teffw{70\,000\pm 2000} given by \citet{rauchetal2016kr},
no sufficient improvement is achieved. 
The impact of metal opacities is demonstrated in Fig.\,\ref{fig:euvproblem} by a model that considered only opacities from 
He, C, N, O, Fe, and Ni with same abundance ratios like our best model.
To test the impact of additional opacity, we artificially increased the Ni abundance by 
factor of 200 to match the model's flux to the observed between 250\,\AA\ and 280\,\AA. This reduced the flux discrepancy between
300\,\AA\ and 400\,\AA\ as well while the wavelength region above the \ion{He}{i} ground-state threshold is unaffected.
However, we conclude that even in our advanced models opacity is missing from elements that are hitherto not considered.
To include, for example, other trans-iron elements requires detailed laboratory measurements of their spectra and the extensive 
calculation of transition probabilities.

\section{What is the nature of RE\,0503$-$289?}
\label{sect:evolution}

\begin{figure}
   \resizebox{\hsize}{!}{\includegraphics{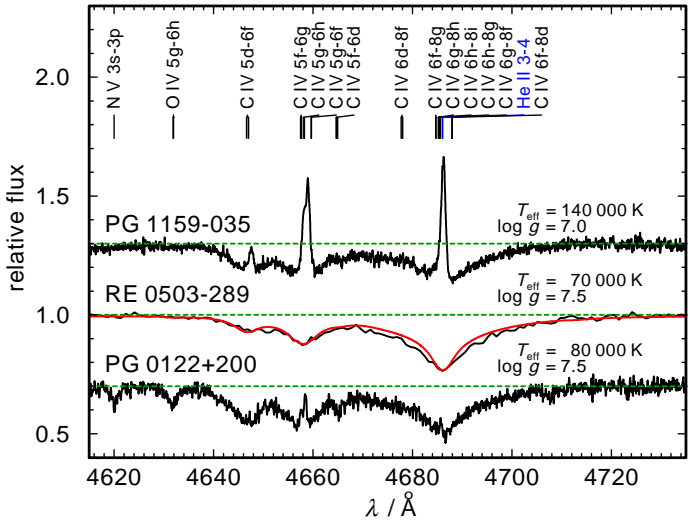}}
    \caption{Section of the optical spectra of 
             \object{PG\,1159$-$035} (from SPY, shifted by 0.3 in flux units), 
             \re (EMMI), and 
             \object{PG\,0122+200} (KECK, shifted by $-0.3$)
             (from top to bottom)
             around the PG\,1159 absorption trough.
             For \re, the synthetic spectrum of \citet{rauchetal2016zr} is overplotted (red line).
             The green, dashed lines indicate the continuum level.
            }
   \label{fig:trough}
\end{figure}

\re was first classified to be a DO-type WD \citep{barstowetal1993}.
Its optical spectrum exhibits an absorption trough around
\Ionww{C}{4}{4646.62 -  4687.95} and \Ionw{He}{2}{4685.80}.
This trough is the spectroscopic criterion for the H-deficient PG\,1159-type stars
\citep[e.g.,][]{wernerherwig2006}. Figure\,\ref{fig:trough} shows the comparison of
the wavelength region around this trough for the PG\,1159 prototype 
\object{PG\,1159$-$035} 
\citep[\object{V$\star$\,GW\,Vir}, \object{WD\,1159$-$035}, \Teffw{140\,000 \pm 5\,000}, \loggw{7.0 \pm 0.5},][]{jahnetal2007}
and the O(He)-type WD \object{KPD\,0005+5106} 
\citep[\object{WD\,0005+511}, \Teffw{195\,000 \pm 15\,000}, \loggw{6.7 \pm 0.2},][]{wernerrauch2015}. 
Both objects are at an earlier state of stellar evolution than \re.
The strengths of the PG\,1159 absorption troughs are almost the same for the much hotter
\object{PG\,1159$-$035} and \re, although their photospheric C abundances are significantly
different, $\approx 48\,\%$ by mass \citep{jahnetal2007} and $\approx 2\,\%$, respectively.
The cool PG\,1159-type star \object{PG\,0122+200} has about 22\,\% of C in its photosphere 
\citep{wernerrauch2014}.

\begin{figure}
   \resizebox{\hsize}{!}{\includegraphics{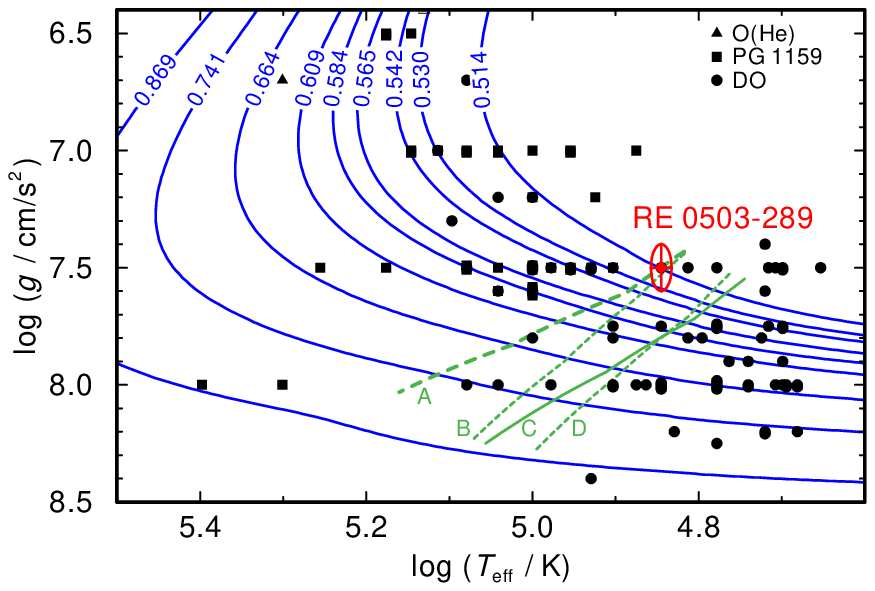}}
    \caption{Location of \re and related objects
             \citep{huegelmeyeretal2006,kepleretal2016,reindletal2014ohe,reindletal2014do,wernerherwig2006}
             in the log \Teff\ -- \logg plane 
             (cf., \url{http://www.star.le.ac.uk/~nr152/He.html} for stellar parameters).
             Evolutionary tracks for H-deficient WDs \citep{althausetal2009}
             labeled with their respective masses in $M_\odot$ are plotted for comparison. 
             Transition limits predicted by \citet{unglaubbues2000} are indicated (see text for details).
             }
   \label{fig:windlimit}
\end{figure}

In a log \Teff\ -- \logg diagram (Fig.\,\ref{fig:windlimit}), \re is located at the 
so-called PG\,1159 wind limit \citep[][their Fig.\,13, digitized with Dexter]{unglaubbues2000}
that was predicted for a ten-times-reduced mass-loss rate
\citep[line A, calculated with $\dot{M} = 1.29\times 10^{-15}L^{1.86}$ from][]{bloecker1995,pauldrachetal1988}.
This line approximately separates the regions that are populated by PG\,1159-type stars and
DO-type WDs. Lines B and C in Fig.\,\ref{fig:windlimit} show where the photospheric C content
is reduced by factors of 0.5 and 0.1, respectively, when using the mass-loss rate given above.
To the right of line D, no PG\,1159 star is located. 

\citet{werneretal2014} suggested a mass ratio $C/He = 0.02$ to conserve previously
assigned spectroscopic classes. However, PG\,1159 stars span a wide range of 
$C/He$ \citep[$0.03 - 0.33$,][]{werneretal2014}. 

\re is located close to line B of \citet{unglaubbues2000} 
in Fig.\,\ref{fig:windlimit}, that is, its C abundance should be already reduced by a factor of 0.5.
Thus, it is likely that \re had a $C/He \approx 0.05$ in its antecedent 
PG\,1159-star phase. Even now, its $C/He$ lies a bit higher than 0.02 and \re may be classified
as a PG\,1159 star as well. This is corroborated by the still high efficiency of
radiative levitation that is responsible for the extremely high overabundances of
trans-iron elements \citep{rauchetal2016mo}. However, the transition from a PG\,1159-type star to a 
DO-type star is smooth and \re is an ideal object to study this in detail.
Unfortunately, the strong radiative levitation of trans-iron elements wipes out all
information about their asymptotic giant branch (AGB) abundances
and \re is not suited to constrain AGB nucleosynthesis models.

\section{Results}
\label{sect:results}

\re fulfills criteria of PG\,1159 star and of DO-type WD classifications. 
The presence of the strong PG\,1159 absorption trough around
\Ionw{He}{2}{4685.80} (Fig.\,\ref{fig:trough}) shows that \re could be classified as
a PG\,1159 star, although its C abundance would then be the lowest of this group. 
It is located close to the so-called PG\,1159 wind limit (Fig.\,\ref{fig:windlimit}),
meaning that it is close to the regime in which gravitation will dominate and pull metals down, out 
of the atmosphere. The strongly increased abundances of trans-iron elements, however, 
indicate that radiative levitation is still efficiently counteracting this process. 
Thus, \re has not arrived in its final stage of evolution. Formally, due to its
\logg $> 7$, the DO-type WD classification is right.

In the observed spectra, we identified 1272 lines in the
wavelength range from the extreme ultraviolet to the near infrared. 
287 lines remain unidentified.
The best model of \citet{rauchetal2016zr} reproduces well most of the identified lines.

The EUV problem (Sect.\,\ref{sect:euvproblem}) -- the difference between observed and synthetic flux in the EUV 
is still present. Our advanced model atmospheres include opacities of 27 metals but their
flux in the EUV is still partly about a factor of approximately two too high compared with the observation. 
We expect that missing metal opacities are the reason for this discrepancy.

\begin{acknowledgements}
DH and TR are supported by the German Aerospace Center (DLR, grants 50\,OR\,1501 and 05\,OR\,1507, respectively).
The German Astrophysical Virtual Observatory (GAVO) project at T\"ubingen 
had been supported by the Federal Ministry of Education and Research (BMBF, 05\,AC\,6\,VTB, 05\,AC\,11\,VTB).
Financial support from the Belgian FRS-FNRS is also acknowledged. 
PQ is research director of this organization.
Some of the data presented in this paper were obtained from the
Mikulski Archive for Space Telescopes (MAST). STScI is operated by the
Association of Universities for Research in Astronomy, Inc., under NASA
contract NAS5-26555. Support for MAST for non-HST data is provided by
the NASA Office of Space Science via grant NNX09AF08G and by other
grants and contracts. 
We thank Ralf Napiwotzki for putting the reduced ESO/VLT spectra at our disposal.
The TEUV  tool    (\url{http://astro-uni-tuebingen.de/~TEUV}), 
the TGRED tool    (\url{http://astro-uni-tuebingen.de/~TGRED}),
the TIRO  service (\url{http://astro-uni-tuebingen.de/~TIRO}),
the TMAD  service (\url{http://astro-uni-tuebingen.de/~TMAD}),
the TOSS  service (\url{http://astro-uni-tuebingen.de/~TOSS}), and
the TVIS  tool    (\url{http://astro-uni-tuebingen.de/~TVIS}) used for this paper 
were constructed as part of the activities of the German Astrophysical Virtual Observatory.
This work used the profile-fitting procedure OWENS developed by M\@. Lemoine and the FUSE French Team.
This research has made use 
of NASA's Astrophysics Data System and
of the SIMBAD database operated at CDS, Strasbourg, France.
\end{acknowledgements}

\bibliographystyle{aa}
\bibliography{29869}

\onecolumn
\appendix

\section{Identified and unidentified lines in the spectrum of RE\,0503$-$289}
\label{app:lists}

\begin{landscape}
\begin{center}

\begin{longtable}{rlrlrlr@{\hspace{0.2mm}$\times$\hspace{0.2mm}}lr@{.}lr@{.}lr@{.}lcp{6cm}}
\caption{\label{tab:lineids_EUVE}Identified lines in the EUVE observations of \re. Lower and upper levels are the configuration or the energy (cm$^{-1}$, for elements with an atomic number $Z \geq 20$). 
         $f$ is the oscillator strength, 
         $W_\lambda$ the observed equivalent width, and 
         $v_\text{rad}$ the measured radial velocity. 
        ``unid.'' denotes observed but as yet unidentified lines. 
         The theoretical wavelengths correspond to those given 
         by NIST for He, C, N, O, Si, P, S, As, and Sn, 
         by \citet{kurucz1991,kurucz2009,kurucz2011} for Fe, and Ni, and
         by \citet{rauchetal2014zn,rauchetal2015ga,rauchetal2012ge,rauchetal2016kr,rauchetal2016zr,
                   rauchetal2016mo,rauchetal2015xe,rauchetal2014ba}
         for Zn, Ga, Ge, Kr, Zr, Mo, Xe, and Ba, respectively.}
\\ \hline \hline
\noalign{\smallskip}
\multicolumn{2}{c}{\multirow{2}{*}{Ion}} & \multicolumn{4}{c}{Levels}                             & \multicolumn{2}{c}{\multirow{2}{*}{$f$}} & \multicolumn{2}{c}{$W_\lambda$\,/} \hbox{}\hspace{1em} & \multicolumn{4}{c}{Wavelength\,/\,\AA}                         & \hbox{}\hspace{1em} $v_\text{rad}$\,/ & \multirow{2}{*}{Comment} \\
\cline{3-6}
\cline{11-14}
\noalign{\smallskip}
&                                        & \multicolumn{2}{c}{Lower}  & \multicolumn{2}{c}{Upper} & \two                                     & \multicolumn{2}{c}{m\AA}  \hbox{}\hspace{1em}          & \multicolumn{2}{c}{Theoretical} & \multicolumn{2}{c}{Observed} & \hbox{}\hspace{1em} km/s              & \\
\noalign{\smallskip}
\hline
\noalign{\smallskip}
\endfirsthead
\caption[]{Continued.}
\\ \hline \hline
\noalign{\smallskip}
\multicolumn{2}{c}{\multirow{2}{*}{Ion}} & \multicolumn{4}{c}{Levels}                             & \multicolumn{2}{c}{\multirow{2}{*}{$f$}} & \multicolumn{2}{c}{$W_\lambda$\,/}  \hbox{}\hspace{1em} & \multicolumn{4}{c}{Wavelength\,/\,\AA}                         & \hbox{}\hspace{1em} $v_\text{rad}$\,/ & \multirow{2}{*}{Comment} \\
\cline{3-6}
\cline{11-14}
\noalign{\smallskip}
&                                        & \multicolumn{2}{c}{Lower}  & \multicolumn{2}{c}{Upper} & \two                                     & \multicolumn{2}{c}{m\AA}  \hbox{}\hspace{1em}           & \multicolumn{2}{c}{Theoretical} & \multicolumn{2}{c}{Observed} & \hbox{}\hspace{1em} km/s              & \\
\noalign{\smallskip}
\hline
\noalign{\smallskip}
\endhead
\hline
\endfoot
\hline
\endlastfoot
He    & \textsc{ii}   & 1                                 &                                   & 6                                 &                                   & $7.80$&$10^{-3}$ & \two     & 234&347  & \two     &          & newly identified \\
He    & \textsc{ii}   & 1                                 &                                   & 5                                 &                                   & $1.39$&$10^{-2}$ & \two     & 237&331  & \two     &          & \citet{vennesetal1998} \\
He    & \textsc{ii}   & 1                                 &                                   & 4                                 &                                   & $2.90$&$10^{-2}$ & \two     & 243&026  & \two     &          & \citet{vennesetal1998} \\
C     & \textsc{iv}   & 2p                                & $^2$P$^\mathrm{ }_{1/2}$          & 6s                                & $^2$S$^\mathrm{ }_{1/2}$          & $1.32$&$10^{-3}$ & \two     & 247&341  & \two     &          & newly identified \\
C     & \textsc{iv}   & 2p                                & $^2$P$^\mathrm{ }_{3/2}$          & 6s                                & $^2$S$^\mathrm{ }_{1/2}$          & $1.32$&$10^{-3}$ & \two     & 247&407  & \two     &          & newly identified \\
He    & \textsc{ii}   & 1                                 &                                   & 3                                 &                                   & $7.91$&$10^{-2}$ & \two     & 256&316  & \two     &          & \citet{vennesetal1998} \\
C     & \textsc{iv}   & 2p                                & $^2$P$^\mathrm{o}_{1/2}$          & 5d                                & $^2$D$^\mathrm{ }_{3/2}$          & $4.57$&$10^{-2}$ & \two     & 259&468  & \two     &          &          \\
\end{longtable}

\begin{longtable}{rlrlrlr@{\hspace{0.2mm}$\times$\hspace{0.2mm}}lr@{.}lr@{.}lr@{.}lcp{6cm}}
\caption{\label{tab:lineids_FUSE}Like Table\,\ref{tab:lineids_EUVE}, for the FUSE observations.
        }
\\ \hline \hline
\noalign{\smallskip}
\multicolumn{2}{c}{\multirow{2}{*}{Ion}} & \multicolumn{4}{c}{Levels}                             & \multicolumn{2}{c}{\multirow{2}{*}{$f$}} & \multicolumn{2}{c}{$W_\lambda$\,/} \hbox{}\hspace{1em} & \multicolumn{4}{c}{Wavelength\,/\,\AA}                         & \hbox{}\hspace{1em} $v_\text{rad}$\,/ & \multirow{2}{*}{Comment} \\
\cline{3-6}
\cline{11-14}
\noalign{\smallskip}
&                                        & \multicolumn{2}{c}{Lower}  & \multicolumn{2}{c}{Upper} & \two                                     & \multicolumn{2}{c}{m\AA}  \hbox{}\hspace{1em}          & \multicolumn{2}{c}{Theoretical} & \multicolumn{2}{c}{Observed} & \hbox{}\hspace{1em} km/s              & \\
\noalign{\smallskip}
\hline
\noalign{\smallskip}
\endfirsthead
\caption[]{Continued.}
\\ \hline \hline
\noalign{\smallskip}
\multicolumn{2}{c}{\multirow{2}{*}{Ion}} & \multicolumn{4}{c}{Levels}                             & \multicolumn{2}{c}{\multirow{2}{*}{$f$}} & \multicolumn{2}{c}{$W_\lambda$\,/} \hbox{}\hspace{1em} & \multicolumn{4}{c}{Wavelength\,/\,\AA}                         & \hbox{}\hspace{1em} $v_\text{rad}$\,/ & \multirow{2}{*}{Comment} \\
\cline{3-6}
\cline{11-14}
\noalign{\smallskip}
&                                        & \multicolumn{2}{c}{Lower}  & \multicolumn{2}{c}{Upper} & \two                                     & \multicolumn{2}{c}{m\AA}  \hbox{}\hspace{1em}          & \multicolumn{2}{c}{Theoretical} & \multicolumn{2}{c}{Observed} & \hbox{}\hspace{1em} km/s              & \\
\noalign{\smallskip}
\hline
\noalign{\smallskip}
\endhead
\hline
\endfoot
\hline
\endlastfoot
H     & \textsc{i}    & 1                                 &                                   & 25                                &                                   & \two             & \two     & 913&215  & \two     &          & ISM multi-component \\
H     & \textsc{i}    & 1                                 &                                   & 24                                &                                   & \two             & \two     & 913&339  & \two     &          & ISM multi-component \\
H     & \textsc{i}    & 1                                 &                                   & 23                                &                                   & \two             & \two     & 913&480  & \two     &          & ISM multi-component \\
H     & \textsc{i}    & 1                                 &                                   & 22                                &                                   & \two             & \two     & 913&641  & \two     &          & ISM multi-component \\
H     & \textsc{i}    & 1                                 &                                   & 21                                &                                   & \two             & \two     & 913&826  & \two     &          & ISM multi-component \\
H     & \textsc{i}    & 1                                 &                                   & 20                                &                                   & \two             & \two     & 914&039  & \two     &          & ISM multi-component \\
H     & \textsc{i}    & 1                                 &                                   & 19                                &                                   & \two             & \two     & 914&286  & \two     &          & ISM multi-component \\
\end{longtable}

\begin{longtable}{rlrlrlr@{\hspace{0.2mm}$\times$\hspace{0.2mm}}lr@{.}lr@{.}lr@{.}lcp{6cm}}
\caption{\label{tab:lineids_STIS}Like Table\,\ref{tab:lineids_EUVE}, for the HST/STIS observations.
        }
\\ \hline \hline
\noalign{\smallskip}
\multicolumn{2}{c}{\multirow{2}{*}{Ion}} & \multicolumn{4}{c}{Levels}                             & \multicolumn{2}{c}{\multirow{2}{*}{$f$}} & \multicolumn{2}{c}{$W_\lambda$\,/} \hbox{}\hspace{1em} & \multicolumn{4}{c}{Wavelength\,/\,\AA}                         & \hbox{}\hspace{1em} $v_\text{rad}$\,/ & \multirow{2}{*}{Comment} \\
\cline{3-6}
\cline{11-14}
\noalign{\smallskip}
&                                        & \multicolumn{2}{c}{Lower}  & \multicolumn{2}{c}{Upper} & \two                                     & \multicolumn{2}{c}{m\AA}  \hbox{}\hspace{1em}          & \multicolumn{2}{c}{Theoretical} & \multicolumn{2}{c}{Observed} & \hbox{}\hspace{1em} km/s              & \\
\noalign{\smallskip}
\hline
\noalign{\smallskip}
\endfirsthead
\caption[]{Continued.}
\\ \hline \hline
\noalign{\smallskip}
\multicolumn{2}{c}{\multirow{2}{*}{Ion}} & \multicolumn{4}{c}{Levels}                             & \multicolumn{2}{c}{\multirow{2}{*}{$f$}} & \multicolumn{2}{c}{$W_\lambda$\,/} \hbox{}\hspace{1em} & \multicolumn{4}{c}{Wavelength\,/\,\AA}                         & \hbox{}\hspace{1em} $v_\text{rad}$\,/ & \multirow{2}{*}{Comment} \\
\cline{3-6}
\cline{11-14}
\noalign{\smallskip}
&                                        & \multicolumn{2}{c}{Lower}  & \multicolumn{2}{c}{Upper} & \two                                     & \multicolumn{2}{c}{m\AA}  \hbox{}\hspace{1em}          & \multicolumn{2}{c}{Theoretical} & \multicolumn{2}{c}{Observed} & \hbox{}\hspace{1em} km/s              & \\
\noalign{\smallskip}
\hline
\noalign{\smallskip}
\endhead
\hline
\endfoot
\hline
\endlastfoot
      &               &                                   &                                   &                                   &                                   & \two             & \two     & \two     & 1150&35  &          & unid.    \\
      &               &                                   &                                   &                                   &                                   & \two             & \two     & \two     & 1151&10  &          & unid.    \\
Zn    & \textsc{v}    & 201973                            &                                   & 288704                            &                                   & $2.26$&$10^{-1}$ & \two     & 1152&985 & \two     &          &          \\
O     & \textsc{iii}  & 2p$^3$                            & $^3$S$^\mathrm{ }_{1}$            & 2p$^4$                            & $^3$P$^\mathrm{ }_{2}$            & $1.41$&$10^{-1}$ & \two     & 1153&775 & 1153&92  & 37.7     &          \\
Zn    & \textsc{v}    & 208715                            &                                   & 295293                            &                                   & $7.17$&$10^{-2}$ & 29&8     & 1155&027 & 1155&12  & 24.1     &          \\
Zn    & \textsc{v}    & 285885                            &                                   & 372360                            &                                   & $3.17$&$10^{-1}$ & \two     & 1156&394 & \two     &          & newly identified \\
Ga    & \textsc{v}    & 246133                            &                                   & 332600                            &                                   & $3.44$&$10^{-1}$ & \two     & 1156&511 & \two     &          &          \\
Zn    & \textsc{v}    & 231997                            &                                   & 318436                            &                                   & $2.76$&$10^{-2}$ & \two     & 1156&885 & \two     &          &          \\
\end{longtable}

\begin{longtable}{rlrlrlr@{\hspace{0.2mm}$\times$\hspace{0.2mm}}lr@{.}lr@{.}lr@{.}lcp{6cm}}
\caption{\label{tab:lineids_optical}Like Table\,\ref{tab:lineids_EUVE}, for the optical observations.
        }
\\ \hline \hline
\noalign{\smallskip}
\multicolumn{2}{c}{\multirow{2}{*}{Ion}} & \multicolumn{4}{c}{Levels}                             & \multicolumn{2}{c}{\multirow{2}{*}{$f$}} & \multicolumn{2}{c}{$W_\lambda$\,/} \hbox{}\hspace{1em} & \multicolumn{4}{c}{Wavelength\,/\,\AA}                         & \hbox{}\hspace{1em} $v_\text{rad}$\,/ & \multirow{2}{*}{Comment} \\
\cline{3-6}
\cline{11-14}
\noalign{\smallskip}
&                                        & \multicolumn{2}{c}{Lower}  & \multicolumn{2}{c}{Upper} & \two                                     & \multicolumn{2}{c}{m\AA}  \hbox{}\hspace{1em}          & \multicolumn{2}{c}{Theoretical} & \multicolumn{2}{c}{Observed} & \hbox{}\hspace{1em} km/s              & \\
\noalign{\smallskip}
\hline
\noalign{\smallskip}
\endfirsthead
\caption[]{Continued.}
\\ \hline \hline
\noalign{\smallskip}
\multicolumn{2}{c}{\multirow{2}{*}{Ion}} & \multicolumn{4}{c}{Levels}                             & \multicolumn{2}{c}{\multirow{2}{*}{$f$}} & \multicolumn{2}{c}{$W_\lambda$\,/}  \hbox{}\hspace{1em} & \multicolumn{4}{c}{Wavelength\,/\,\AA}                         & \hbox{}\hspace{1em} $v_\text{rad}$\,/ & \multirow{2}{*}{Comment} \\
\cline{3-6}
\cline{11-14}
\noalign{\smallskip}
&                                        & \multicolumn{2}{c}{Lower}  & \multicolumn{2}{c}{Upper} & \two                                     & \multicolumn{2}{c}{m\AA}  \hbox{}\hspace{1em}           & \multicolumn{2}{c}{Theoretical} & \multicolumn{2}{c}{Observed} & \hbox{}\hspace{1em} km/s              & \\
\noalign{\smallskip}
\hline
\noalign{\smallskip}
\endhead
\hline
\endfoot
\hline
\endlastfoot
O     & \textsc{iv}   & 3s                                & $^2$P$^\mathrm{o}_{1/2}$          & 3p                                & $^2$D$^\mathrm{ }_{3/2}$          & $2.84$&$10^{-1}$ & \two     & 3348&055 & \two     &          & newly identified \\
O     & \textsc{iv}   & 3s                                & $^2$P$^\mathrm{o}_{3/2}$          & 3p                                & $^2$D$^\mathrm{ }_{5/2}$          & $2.55$&$10^{-1}$ & \two     & 3349&110 & \two     &          & newly identified \\
O     & \textsc{iv}   & 3s                                & $^4$P$^\mathrm{o}_{3/2}$          & 3p                                & $^4$D$^\mathrm{ }_{5/2}$          & $1.83$&$10^{-1}$ & \two     & 3381&212 & \two     &          & newly identified \\
O     & \textsc{iv}   & 3s                                & $^4$P$^\mathrm{o}_{1/2}$          & 3p                                & $^4$D$^\mathrm{ }_{3/2}$          & $1.45$&$10^{-1}$ & \two     & 3381&304 & \two     &          & newly identified \\
O     & \textsc{iv}   & 3s                                & $^4$P$^\mathrm{o}_{5/2}$          & 3p                                & $^4$D$^\mathrm{ }_{7/2}$          & $2.32$&$10^{-1}$ & \two     & 3385&518 & \two     &          & newly identified \\
O     & \textsc{iv}   & 3s                                & $^4$P$^\mathrm{o}_{1/2}$          & 3p                                & $^4$D$^\mathrm{ }_{1/2}$          & $1.45$&$10^{-1}$ & \two     & 3390&191 & \two     &          & newly identified \\
O     & \textsc{iv}   & 3s                                & $^4$P$^\mathrm{o}_{3/2}$          & 3p                                & $^4$D$^\mathrm{ }_{3/2}$          & $9.29$&$10^{-2}$ & \two     & 3396&803 & \two     &          & newly identified \\
\end{longtable}

\begin{longtable}{rlrlrlr@{\hspace{0.2mm}$\times$\hspace{0.2mm}}lr@{.}lr@{.}lr@{.}lcp{6cm}}
\caption{\label{tab:lineids_sofi}Like Table\,\ref{tab:lineids_EUVE}, for the SofI observations.
        }
\\ \hline \hline
\noalign{\smallskip}
\multicolumn{2}{c}{\multirow{2}{*}{Ion}} & \multicolumn{4}{c}{Levels}                             & \multicolumn{2}{c}{\multirow{2}{*}{$f$}} & \multicolumn{2}{c}{$W_\lambda$\,/} \hbox{}\hspace{1em} & \multicolumn{4}{c}{Wavelength\,/\,\AA}                         & \hbox{}\hspace{1em} $v_\text{rad}$\,/ & \multirow{2}{*}{Comment} \\
\cline{3-6}
\cline{11-14}
\noalign{\smallskip}
&                                        & \multicolumn{2}{c}{Lower}  & \multicolumn{2}{c}{Upper} & \two                                     & \multicolumn{2}{c}{m\AA}  \hbox{}\hspace{1em}          & \multicolumn{2}{c}{Theoretical} & \multicolumn{2}{c}{Observed} & \hbox{}\hspace{1em} km/s              & \\
\noalign{\smallskip}
\hline
\noalign{\smallskip}
\endfirsthead
\caption[]{Continued.}
\\ \hline \hline
\noalign{\smallskip}
\multicolumn{2}{c}{\multirow{2}{*}{Ion}} & \multicolumn{4}{c}{Levels}                             & \multicolumn{2}{c}{\multirow{2}{*}{$f$}} & \multicolumn{2}{c}{$W_\lambda$\,/}  \hbox{}\hspace{1em} & \multicolumn{4}{c}{Wavelength\,/\,\AA}                         & \hbox{}\hspace{1em} $v_\text{rad}$\,/ & \multirow{2}{*}{Comment} \\
\cline{3-6}
\cline{11-14}
\noalign{\smallskip}
&                                        & \multicolumn{2}{c}{Lower}  & \multicolumn{2}{c}{Upper} & \two                                     & \multicolumn{2}{c}{m\AA}  \hbox{}\hspace{1em}           & \multicolumn{2}{c}{Theoretical} & \multicolumn{2}{c}{Observed} & \hbox{}\hspace{1em} km/s              & \\
\noalign{\smallskip}
\hline
\noalign{\smallskip}
\endhead
\hline
\endfoot
\hline
\endlastfoot
He    & \textsc{ii}   & 4                                 &                                   & 5                                 &                                   & \noten{1.04}  & \two     & 10123&499 & \two     &          &                  \\
He    & \textsc{ii}   & 5                                 &                                   & 7                                 &                                   & $2.07$&$10^{-1}$ & \two     & 11626&406 & \two     &          &                  \\
\end{longtable}

\end{center}
\end{landscape}

\clearpage

\section{Observed spectra of RE\,0503$-$289 compared with our best model}
\label{app:plots}

\begin{figure*}
   \includegraphics[trim=0 -0 0 -8,height=23.0cm,angle=0]{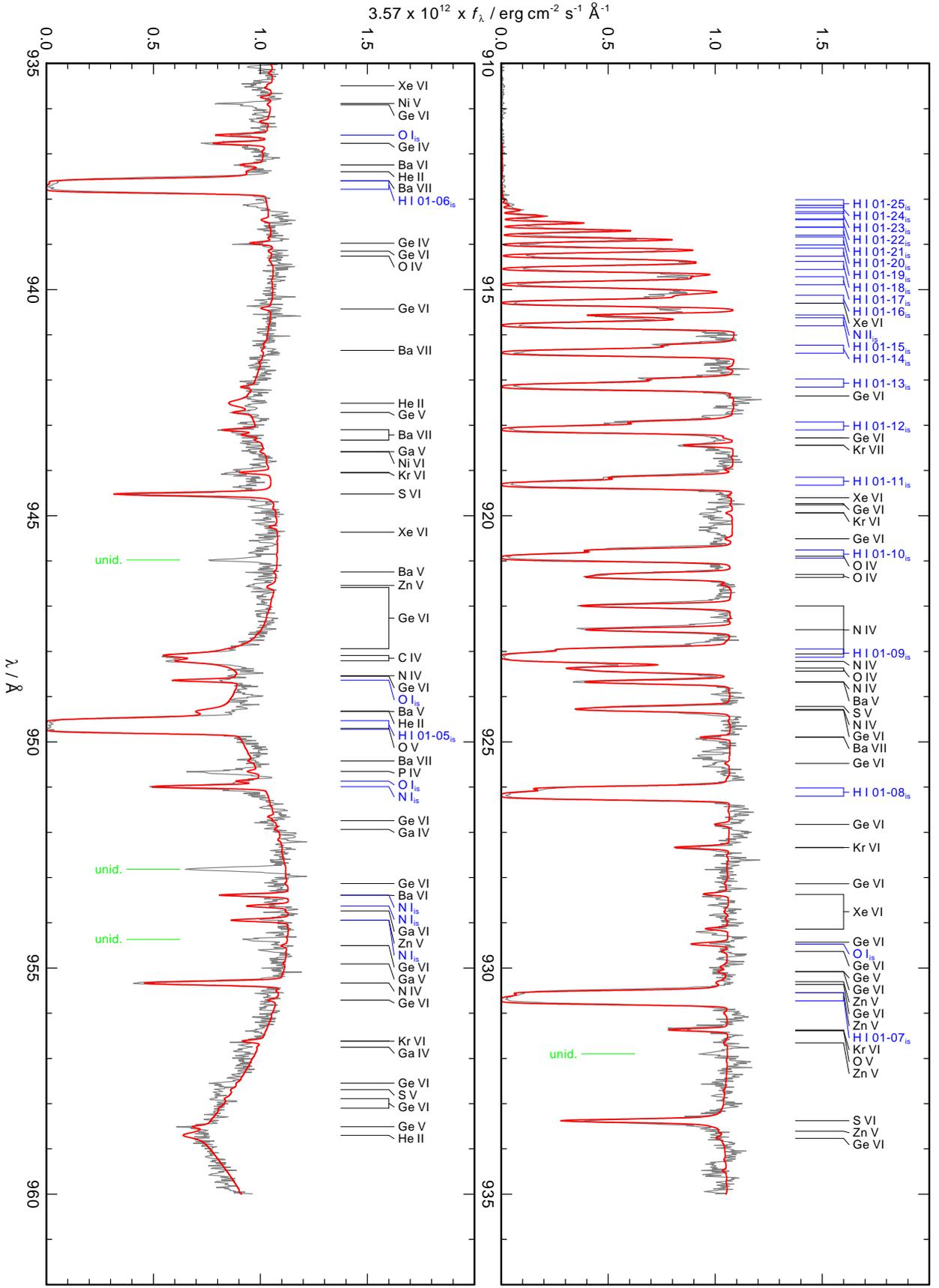}
  \caption{FUSE observation (gray) compared with the best model (red).
           Stellar lines are identified at top. ``unid.'' denotes unidentified lines.} 
  \label{fig:FUSE_complete}
\end{figure*}

In the following figures, we show the comparison of our
synthetic spectra with the
FUSE     (Fig.\ref{fig:FUSE_complete},
HST/STIS (Fig.\ref{fig:STIS_complete}, and
optical  (Fig.\ref{fig:OPTI_complete}
observations. 
A visualization via TVIS is available at \url{http://astro.uni-tuebingen.de/~TVIS/objects/RE0503-289}.
 
\addtolength{\textwidth}{6.3cm} 
\addtolength{\topmargin}{+4mm}
\addtolength{\evensidemargin}{+4mm}
\addtolength{\oddsidemargin}{+4mm}

\clearpage

\clearpage

\addtocounter{figure}{-1} 
\begin{figure*}
   \includegraphics[trim=0 -0 0 -8,height=23.0cm,angle=0]{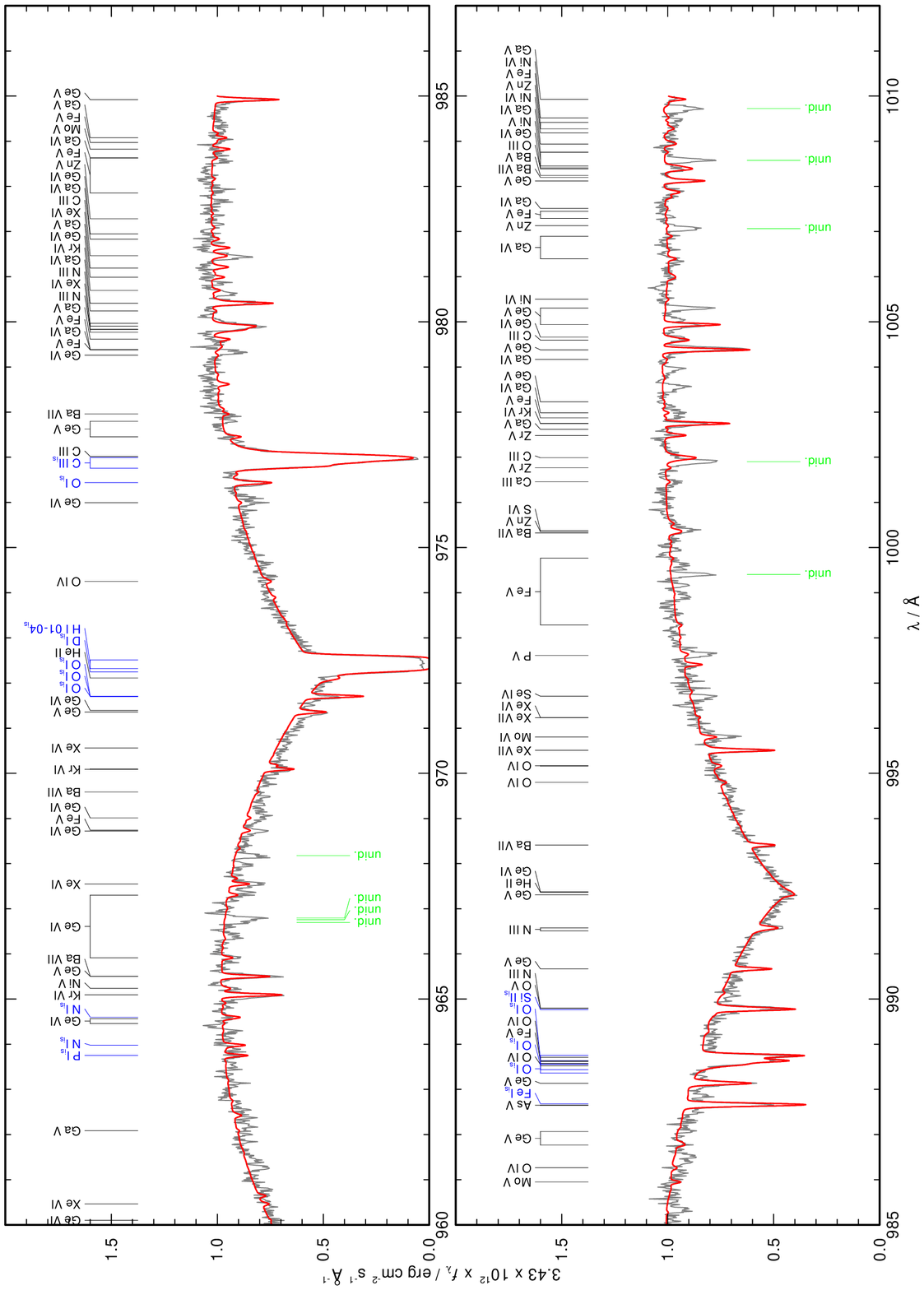}
  \caption{Figure\,\ref{fig:FUSE_complete} continued.} 
\end{figure*}

\clearpage

\addtocounter{figure}{-1} 
\begin{figure*}
   \includegraphics[trim=0 -0 0 -8,height=23.0cm,angle=0]{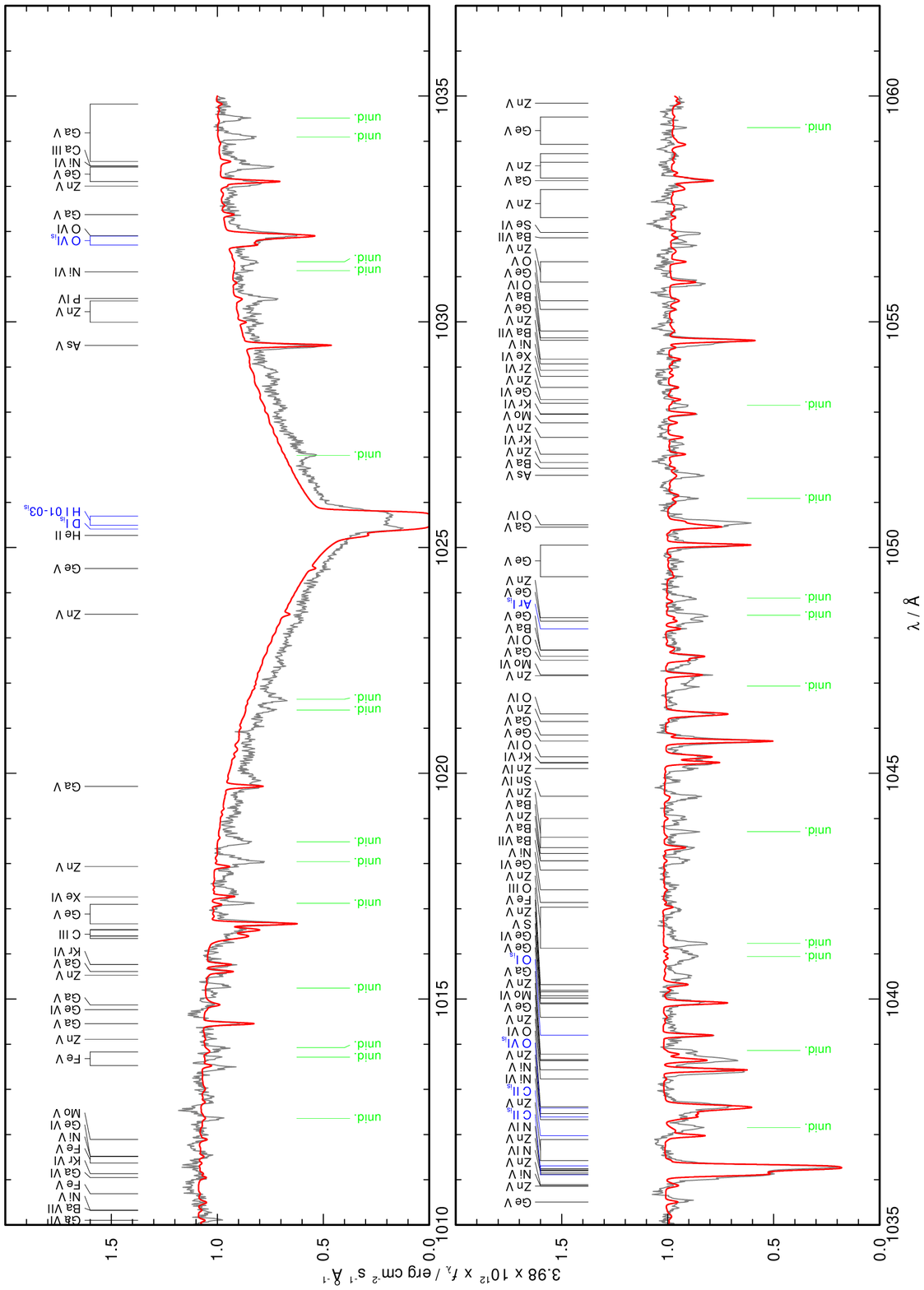}
  \caption{Figure\,\ref{fig:FUSE_complete} continued.} 
\end{figure*}

\clearpage

\addtocounter{figure}{-1} 
\begin{figure*}
   \includegraphics[trim=0 -0 0 -8,height=23.0cm,angle=0]{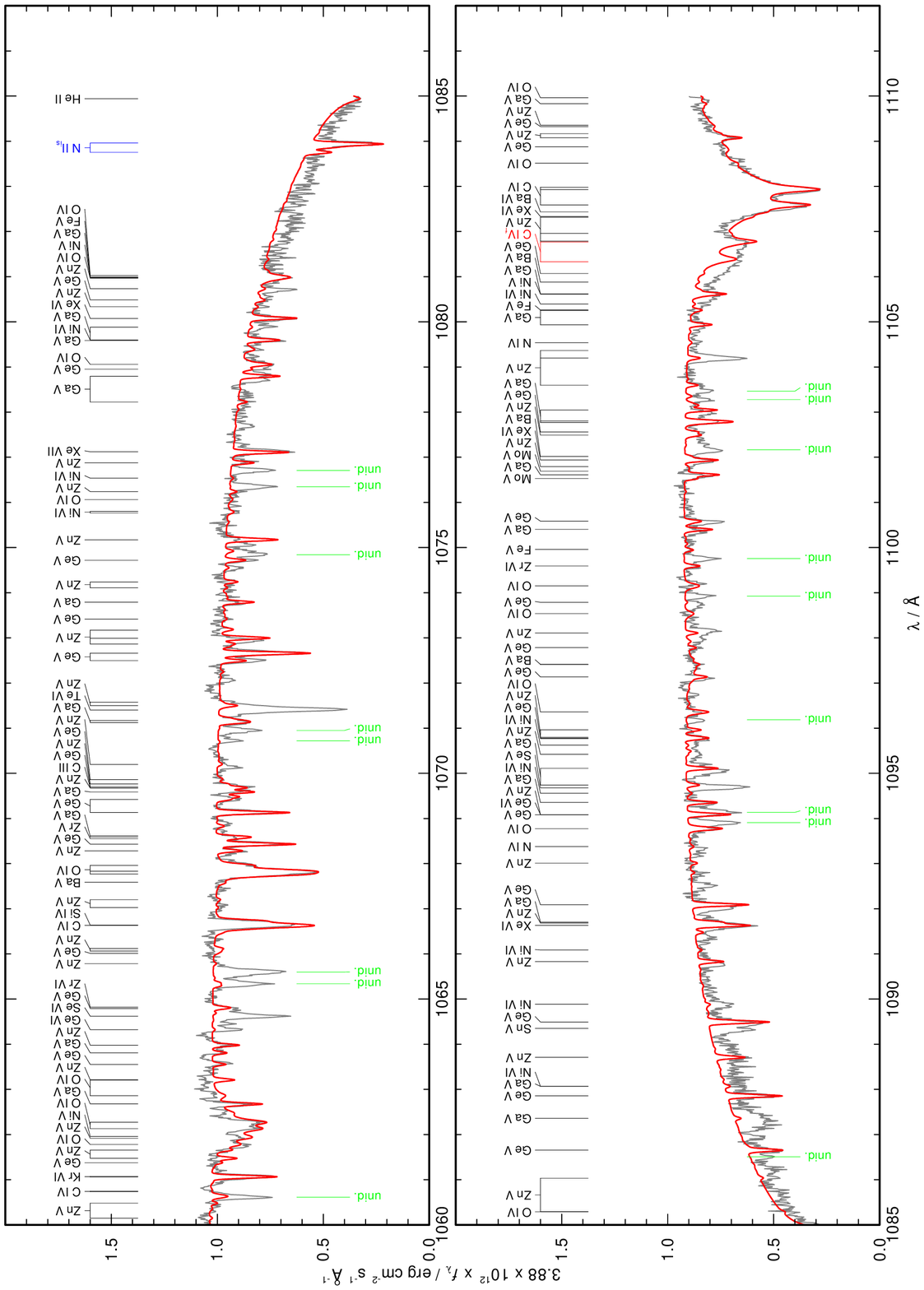}
  \caption{Figure\,\ref{fig:FUSE_complete} continued.} 
\end{figure*}

\clearpage

\addtocounter{figure}{-1} 
\begin{figure*}
   \includegraphics[trim=0 -0 0 -8,height=23.0cm,angle=0]{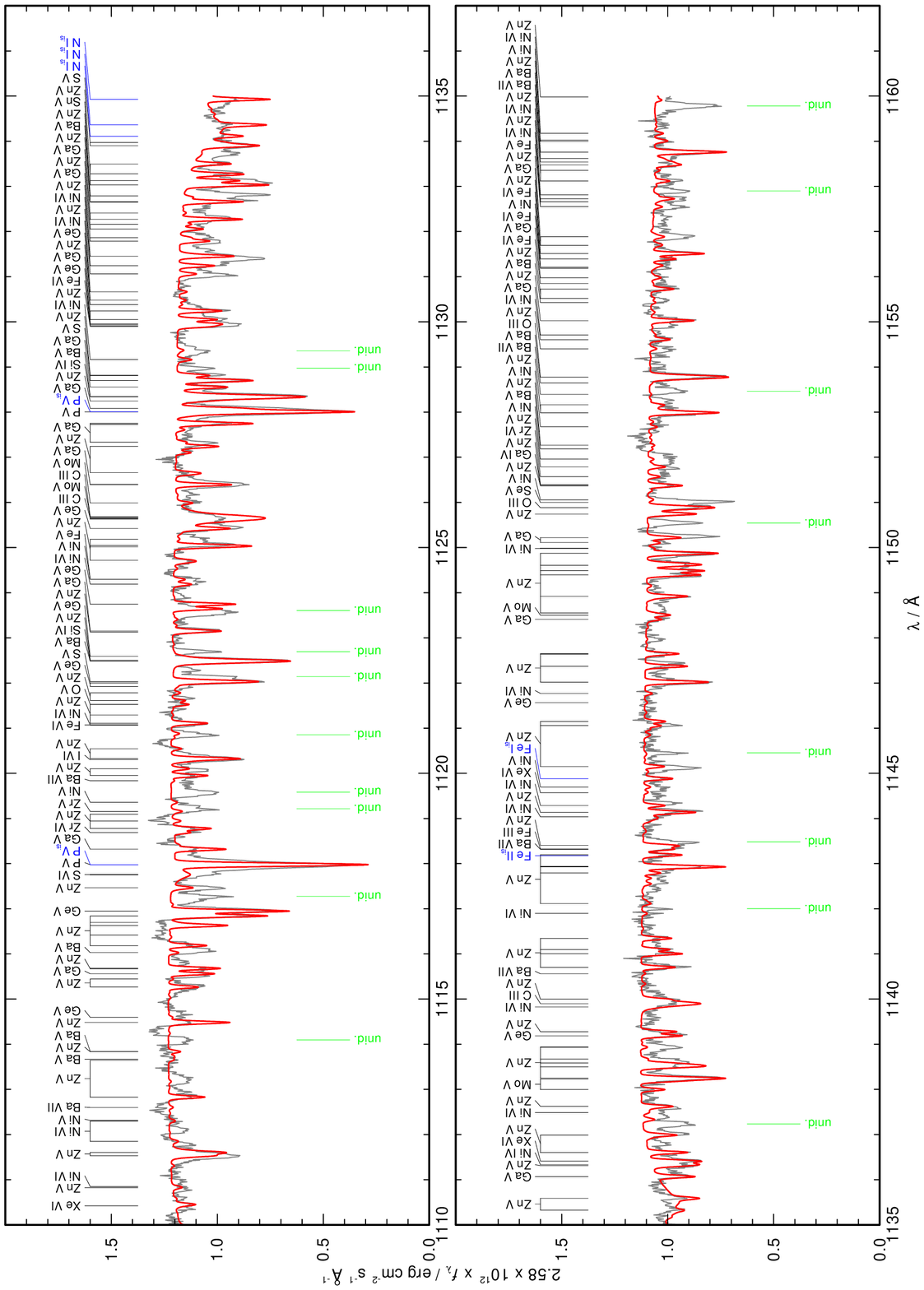}
  \caption{Figure\,\ref{fig:FUSE_complete} continued.} 
\end{figure*}

\clearpage

\addtocounter{figure}{-1} 
\begin{figure*}
   \includegraphics[trim=0 -0 0 -8,height=23.0cm,angle=0]{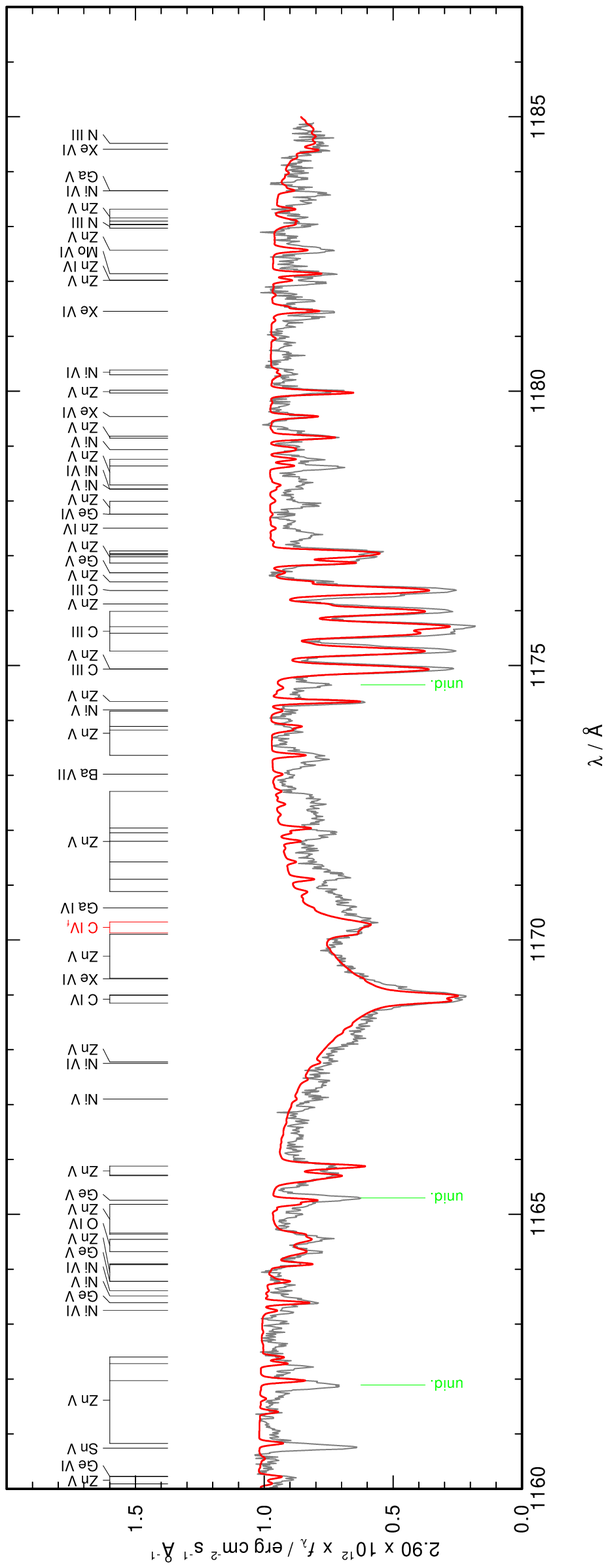}
  \caption{Figure\,\ref{fig:FUSE_complete} continued.} 
\end{figure*}

\clearpage

\begin{figure*}
   \includegraphics[trim=0 -0 0 -8,height=23.0cm,angle=0]{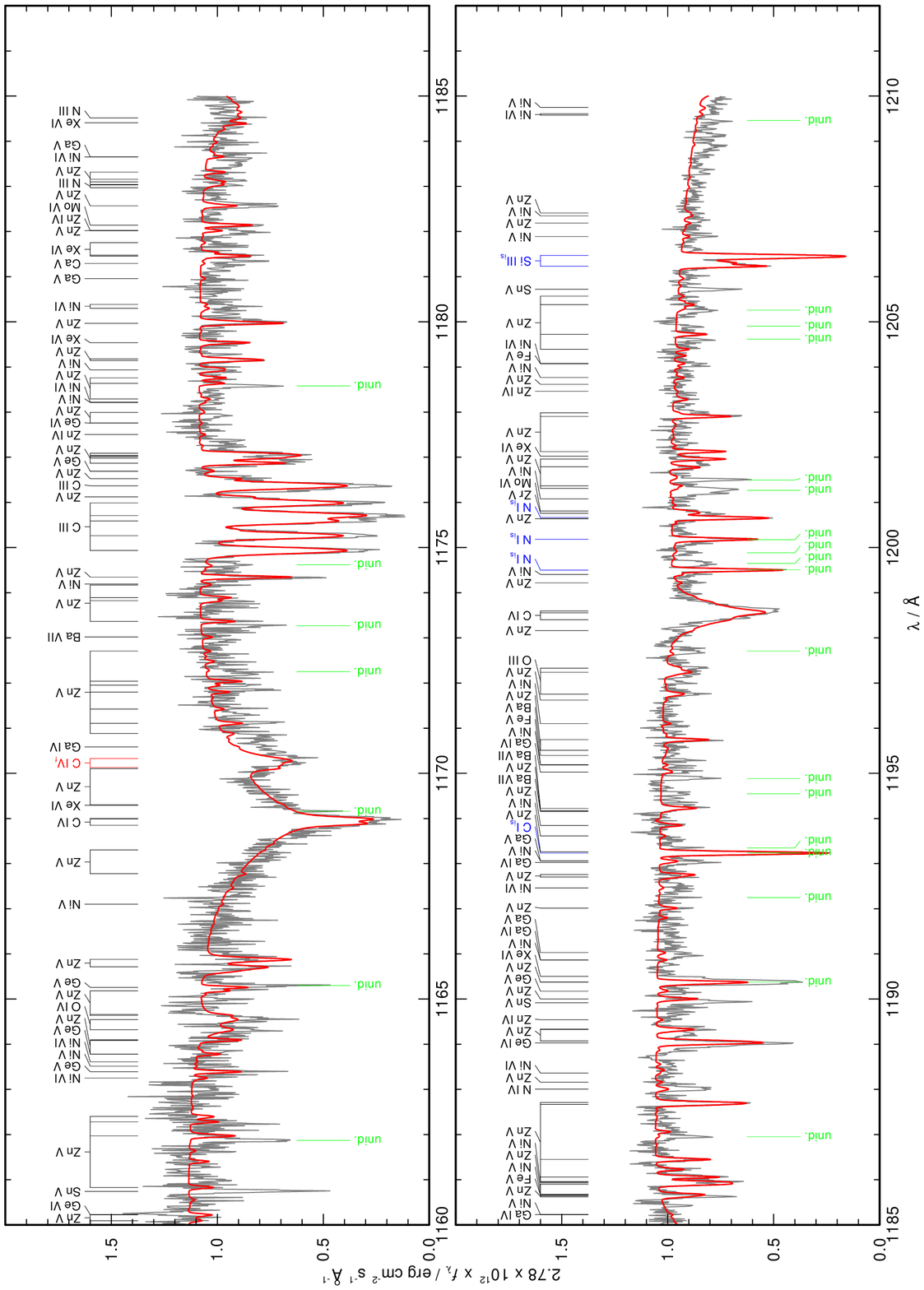}
  \caption{HST/STIS observation (gray) compared with the best model (red).
           Stellar lines are identified at top. ``unid.'' denotes unidentified lines.} 
  \label{fig:STIS_complete}
\end{figure*}

\clearpage

\addtocounter{figure}{-1} 
\begin{figure*}
   \includegraphics[trim=0 -0 0 -8,height=23.0cm,angle=0]{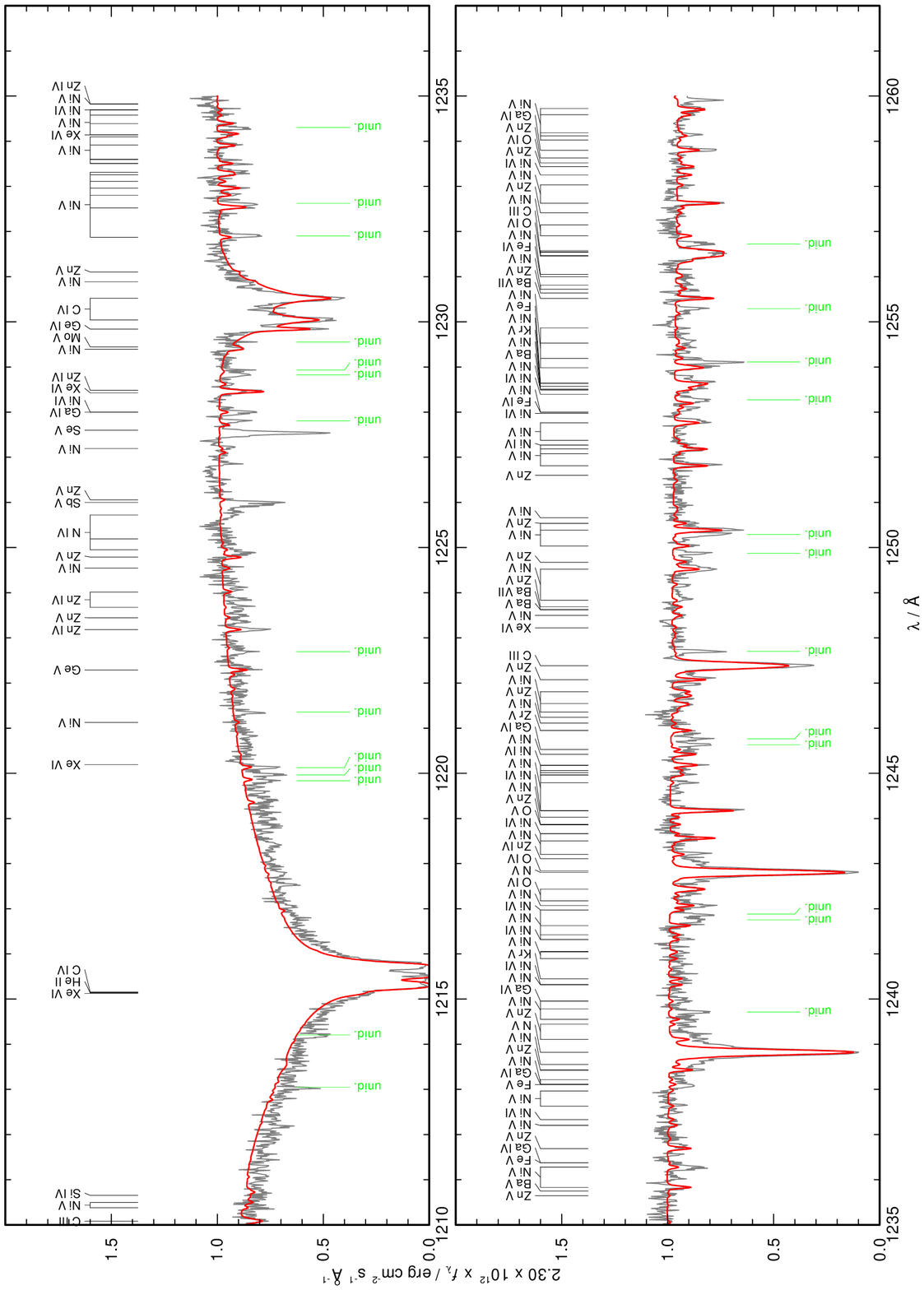}
  \caption{Figure\,\ref{fig:STIS_complete} continued.} 
\end{figure*}

\clearpage

\addtocounter{figure}{-1} 
\begin{figure*}
   \includegraphics[trim=0 -0 0 -8,height=23.0cm,angle=0]{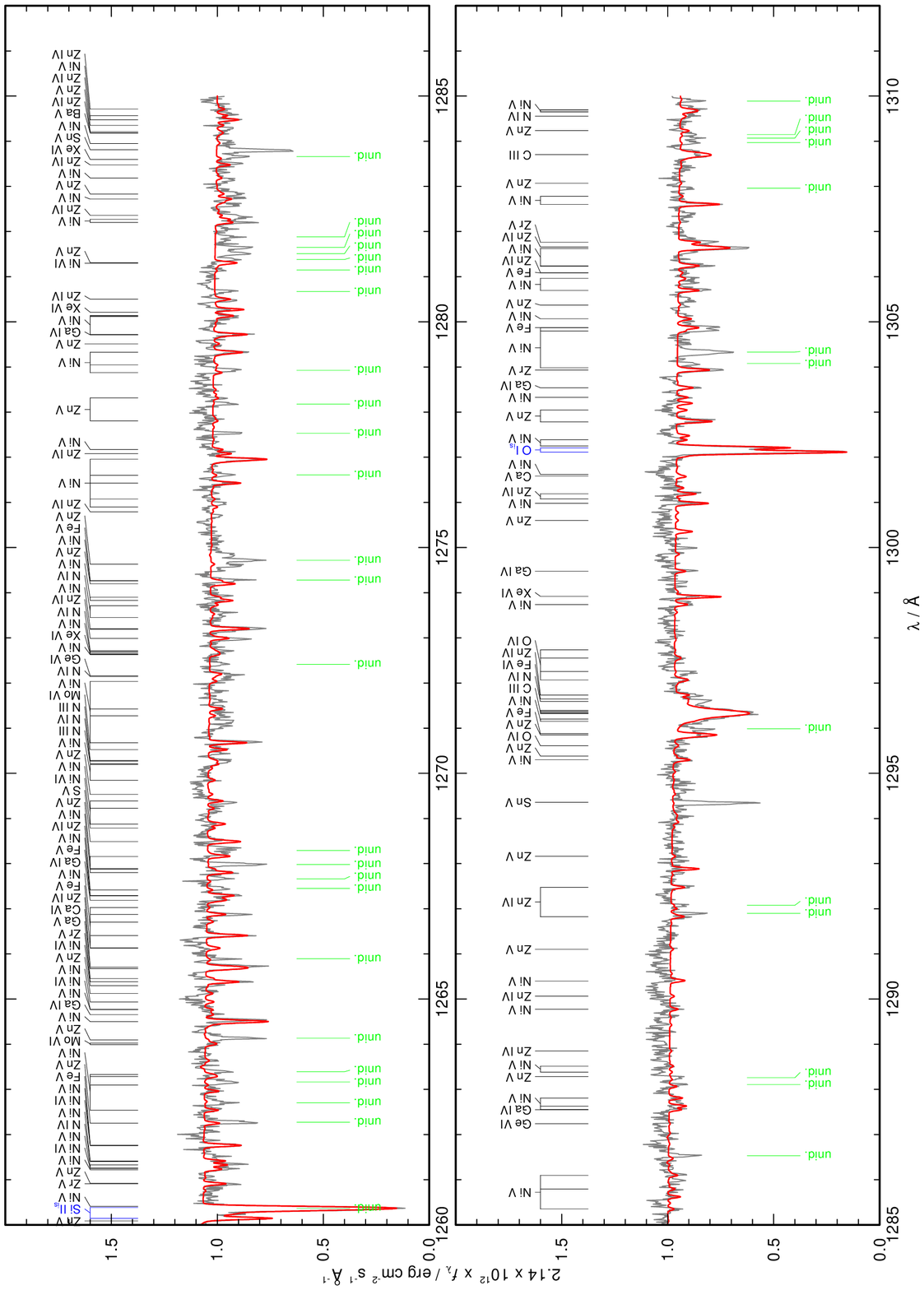}
  \caption{Figure\,\ref{fig:STIS_complete} continued.} 
\end{figure*}

\clearpage

\addtocounter{figure}{-1} 
\begin{figure*}
   \includegraphics[trim=0 -0 0 -8,height=23.0cm,angle=0]{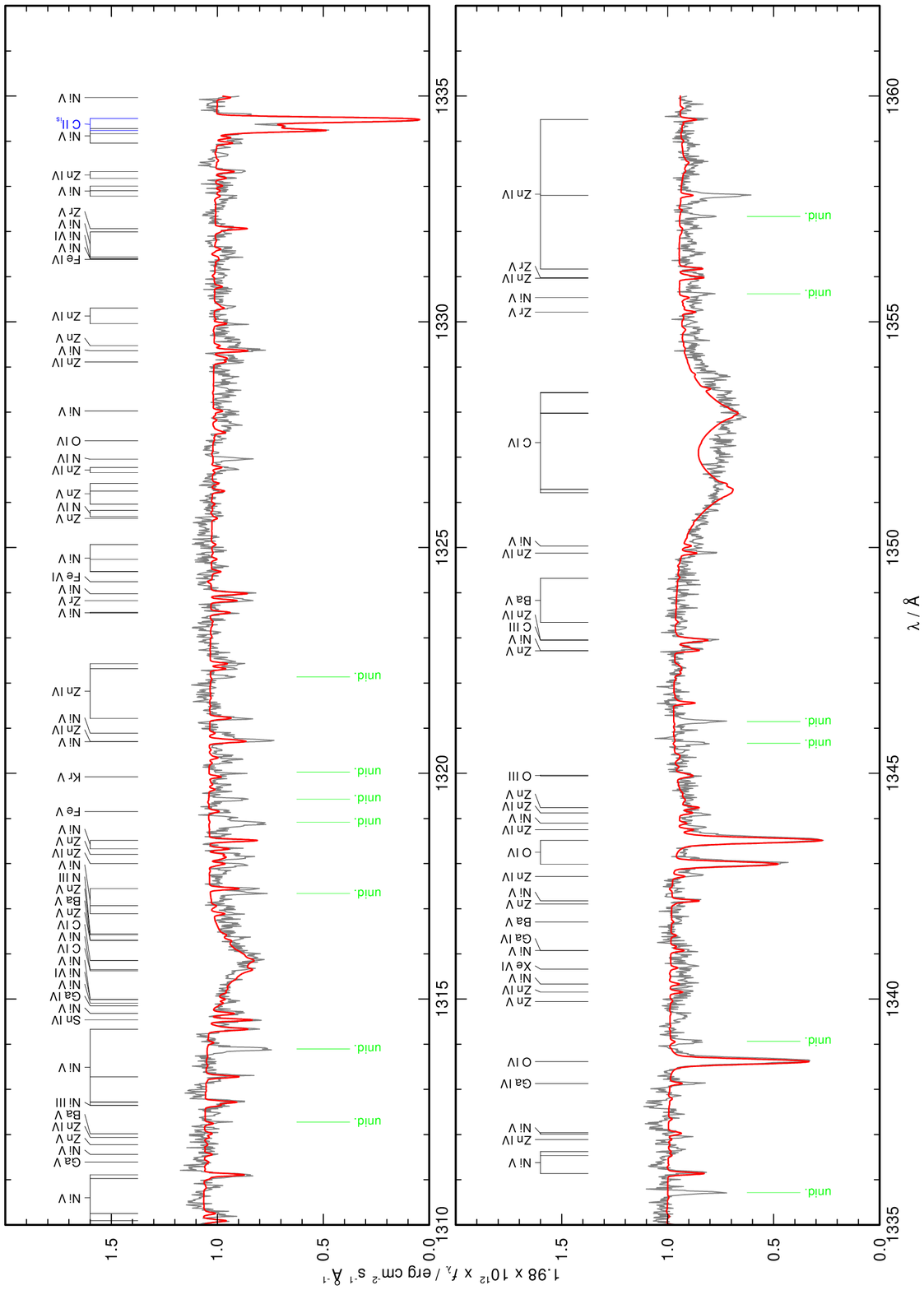}
  \caption{Figure\,\ref{fig:STIS_complete} continued.} 
\end{figure*}

\clearpage

\addtocounter{figure}{-1} 
\begin{figure*}
   \includegraphics[trim=0 -0 0 -8,height=23.0cm,angle=0]{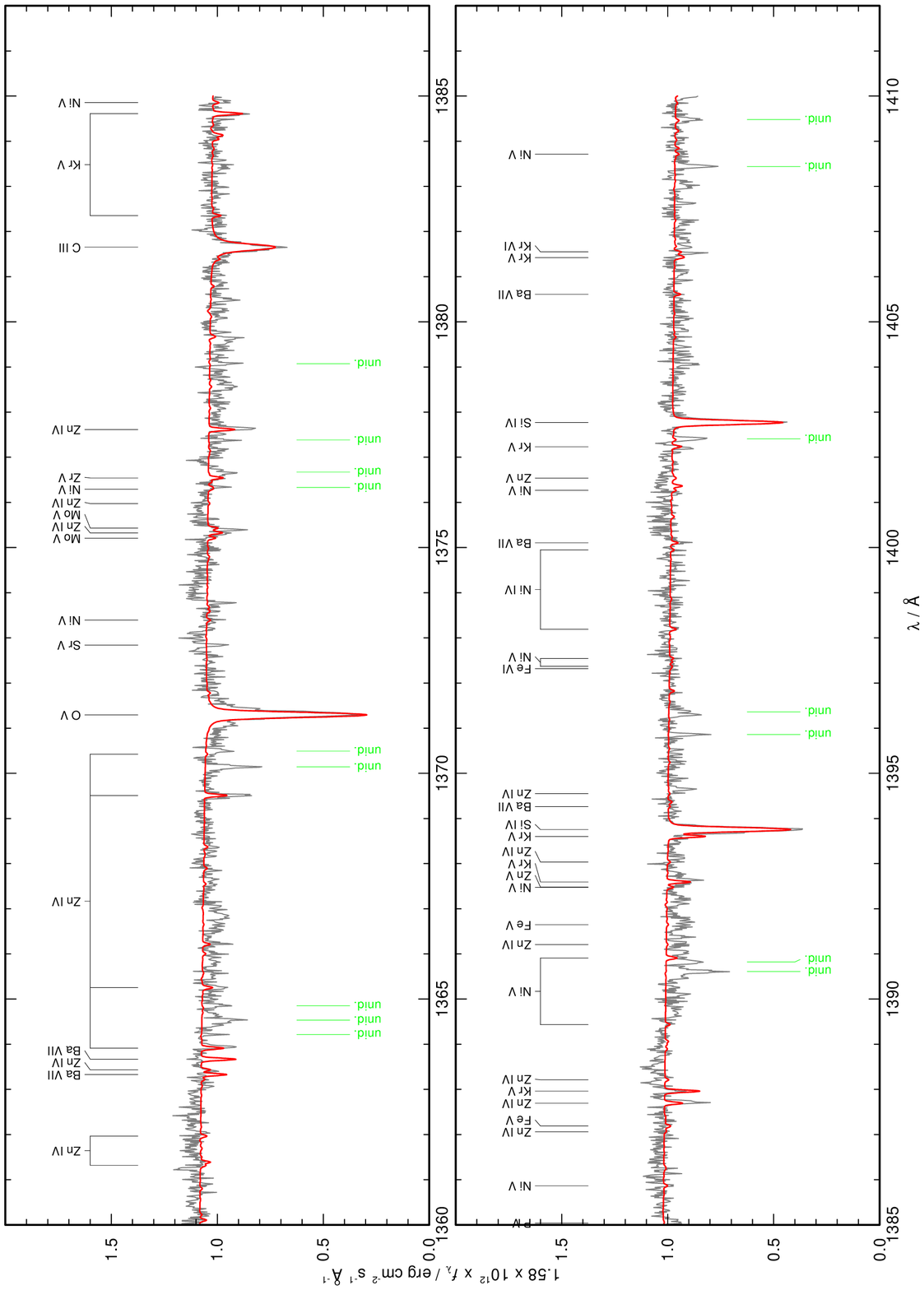}
  \caption{Figure\,\ref{fig:STIS_complete} continued.} 
\end{figure*}

\clearpage

\addtocounter{figure}{-1} 
\begin{figure*}
   \includegraphics[trim=0 -0 0 -8,height=23.0cm,angle=0]{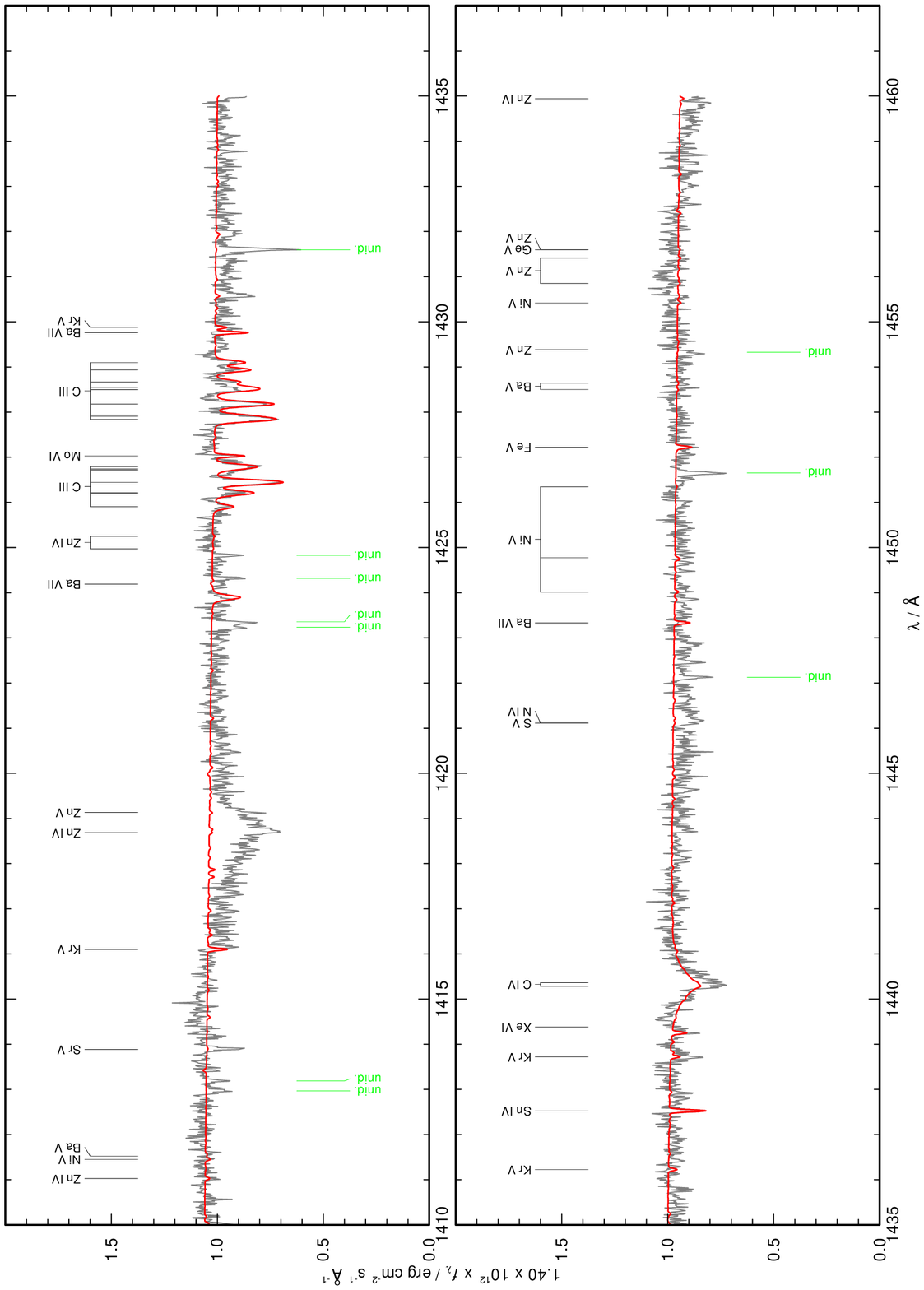}
  \caption{Figure\,\ref{fig:STIS_complete} continued.} 
\end{figure*}

\clearpage

\addtocounter{figure}{-1} 
\begin{figure*}
   \includegraphics[trim=0 -0 0 -8,height=23.0cm,angle=0]{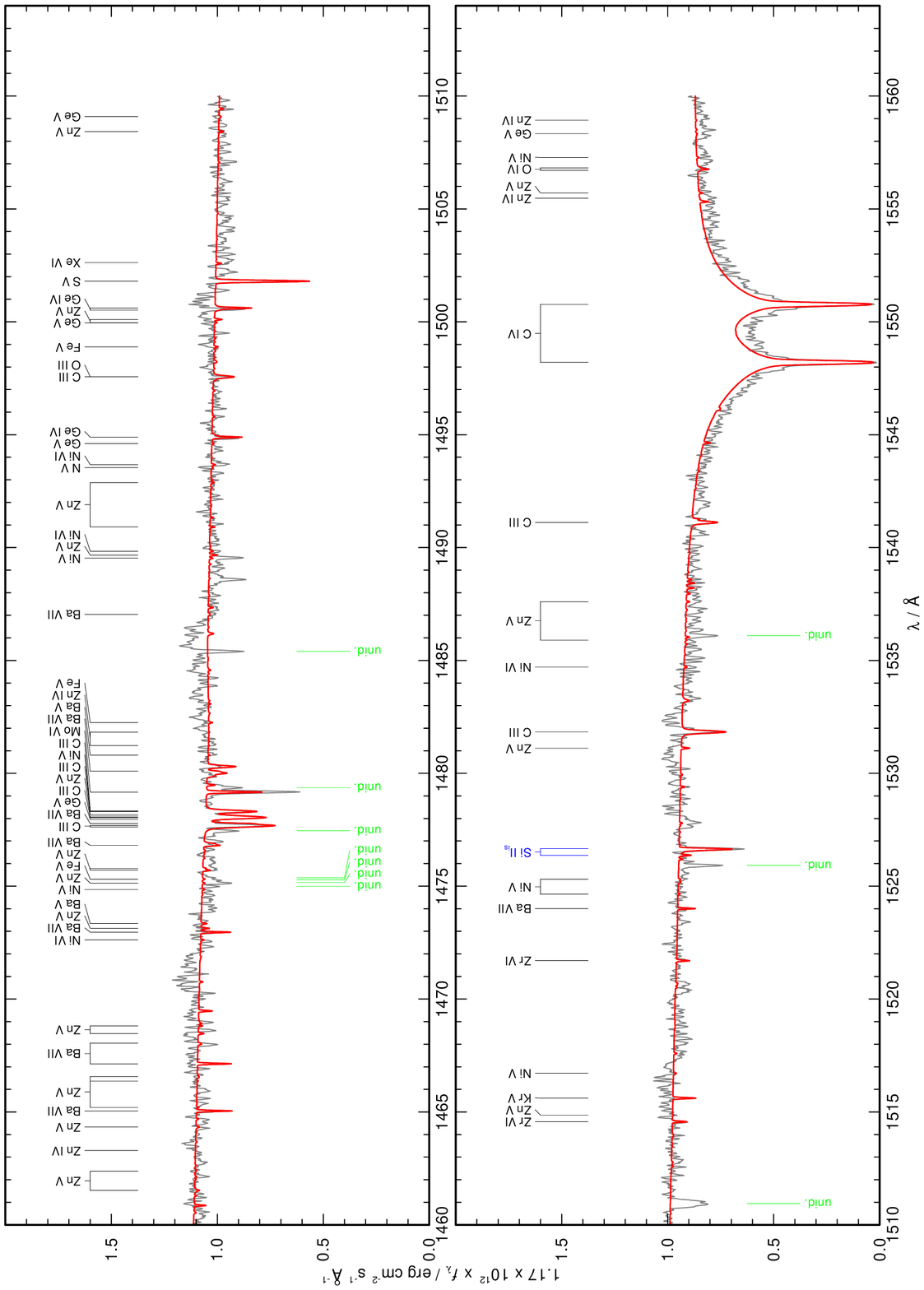}
  \caption{Figure\,\ref{fig:STIS_complete} continued.} 
\end{figure*}

\clearpage

\addtocounter{figure}{-1} 
\begin{figure*}
   \includegraphics[trim=0 -0 0 -8,height=23.0cm,angle=0]{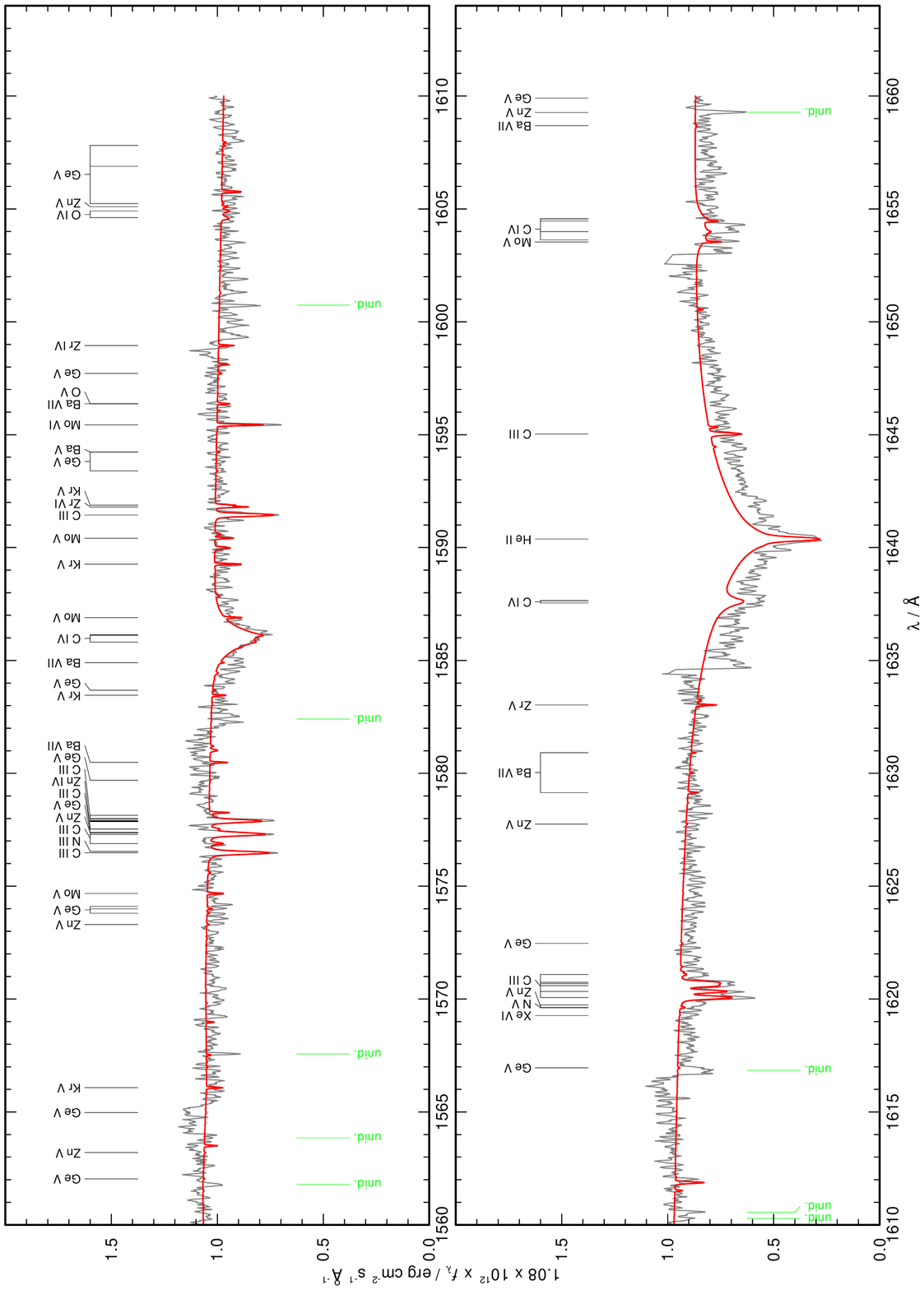}
  \caption{Figure\,\ref{fig:STIS_complete} continued.} 
\end{figure*}

\clearpage

\addtocounter{figure}{-1} 
\begin{figure*}
   \includegraphics[trim=0 -0 0 -8,height=23.0cm,angle=0]{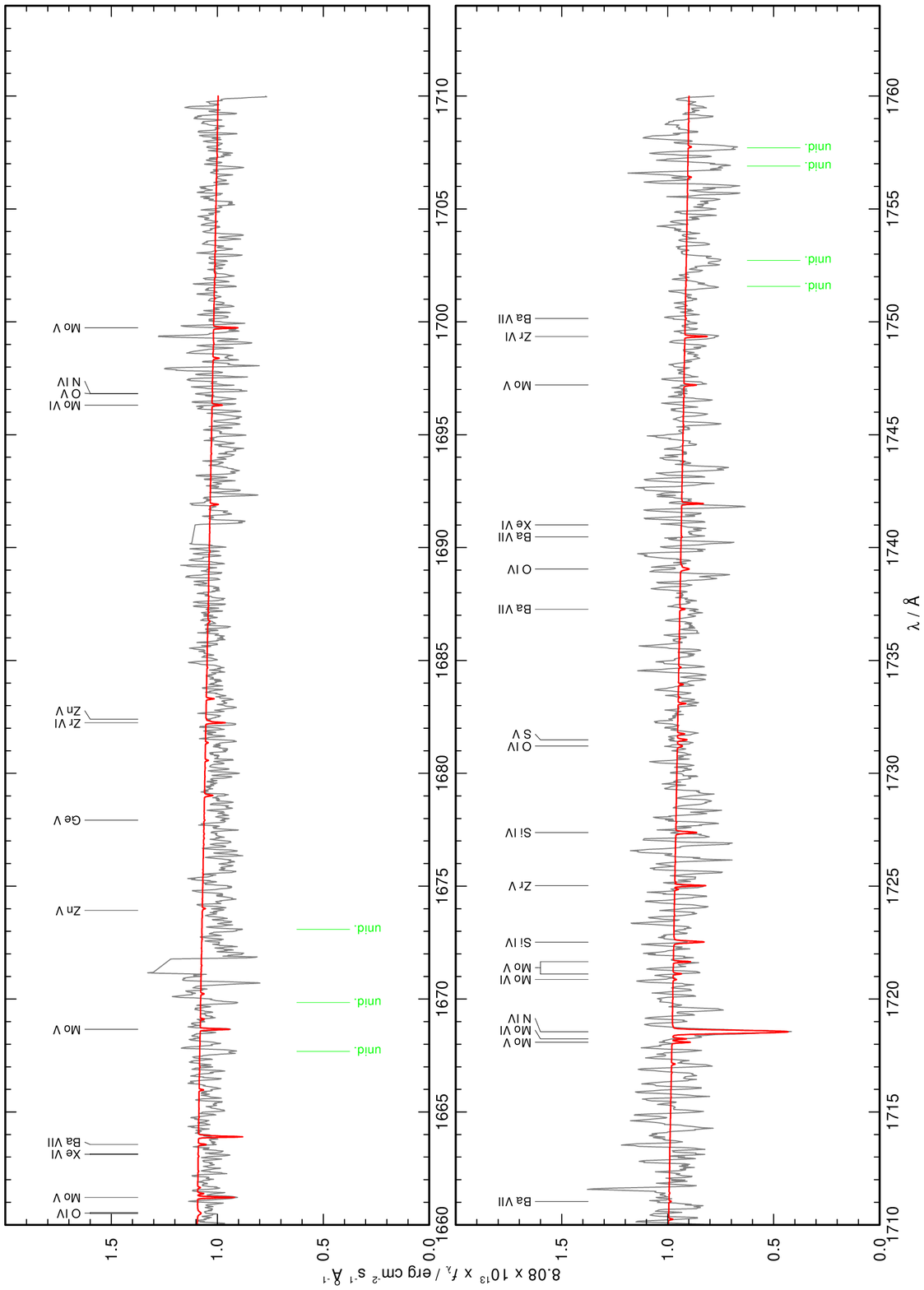}
  \caption{Figure\,\ref{fig:STIS_complete} continued.} 
\end{figure*}

\clearpage

\addtocounter{figure}{-1} 
\begin{figure*}
   \includegraphics[trim=0 -0 0 -8,height=23.0cm,angle=0]{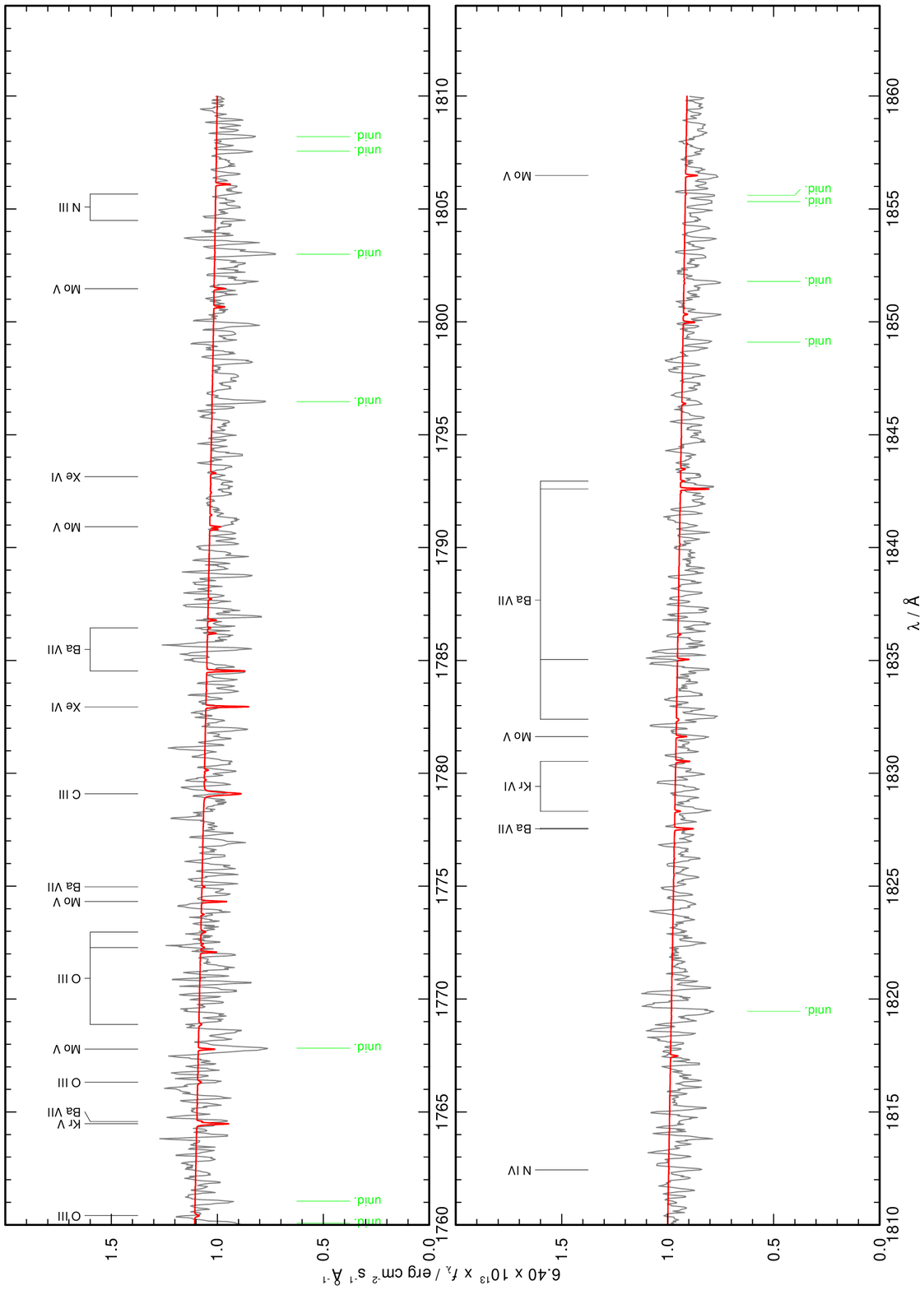}
  \caption{Figure\,\ref{fig:STIS_complete} continued.} 
\end{figure*}

\clearpage

\addtocounter{figure}{-1} 
\begin{figure*}
   \includegraphics[trim=0 -0 0 -8,height=23.0cm,angle=0]{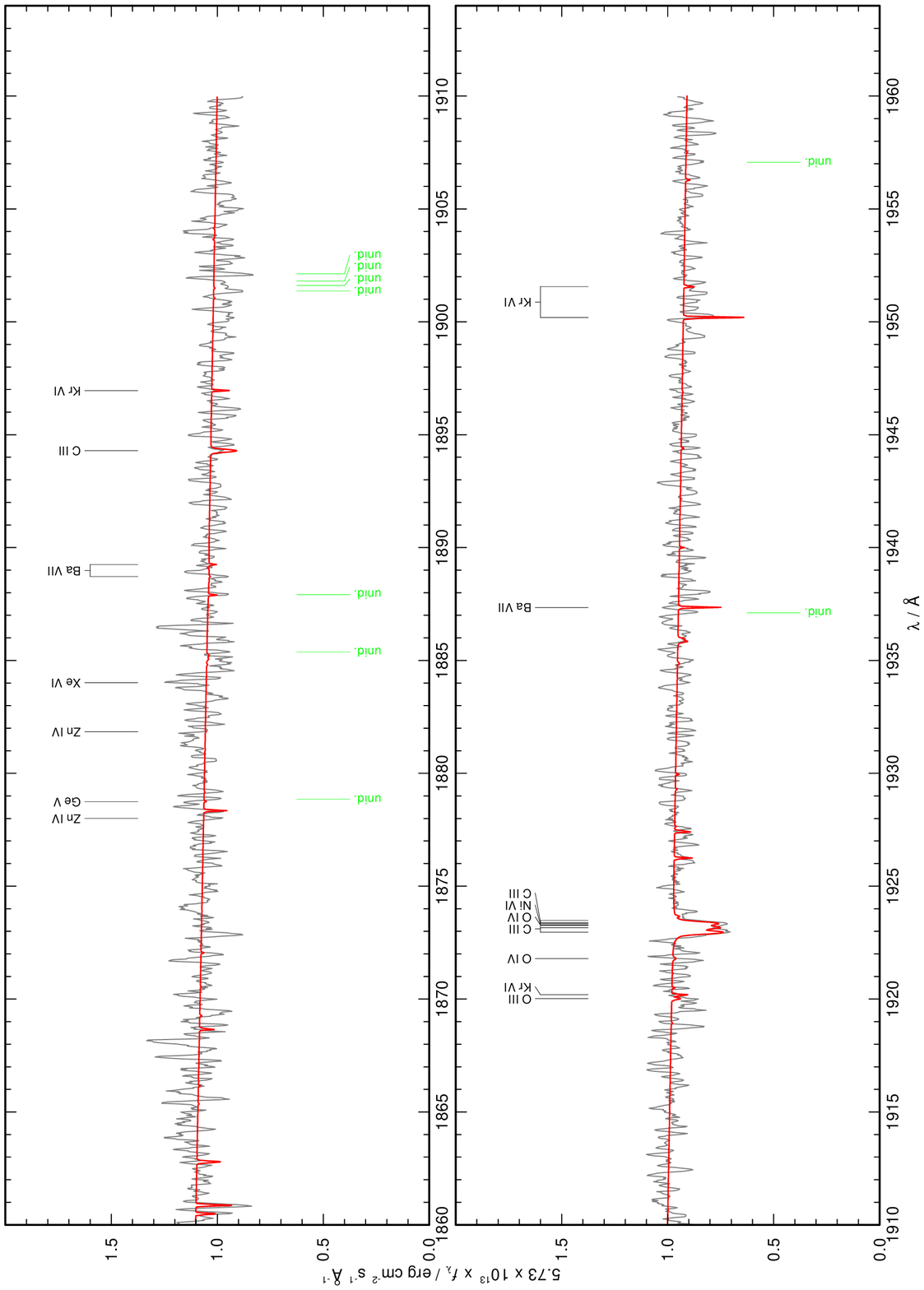}
  \caption{Figure\,\ref{fig:STIS_complete} continued.} 
\end{figure*}

\clearpage

\addtocounter{figure}{-1} 
\begin{figure*}
   \includegraphics[trim=0 -0 0 -8,height=23.0cm,angle=0]{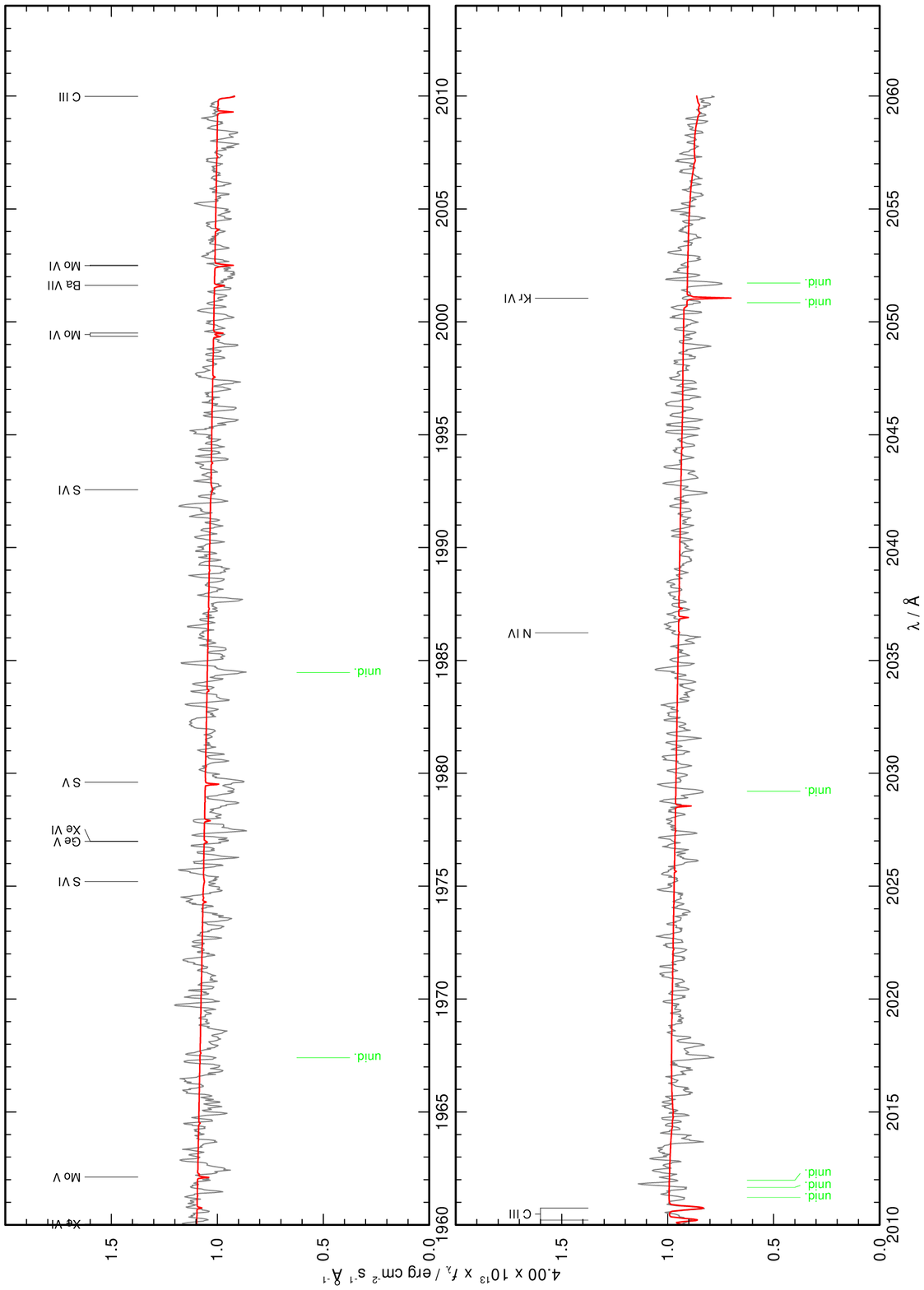}
  \caption{Figure\,\ref{fig:STIS_complete} continued.} 
\end{figure*}

\clearpage

\addtocounter{figure}{-1} 
\begin{figure*}
   \includegraphics[trim=0 -0 0 -8,height=23.0cm,angle=0]{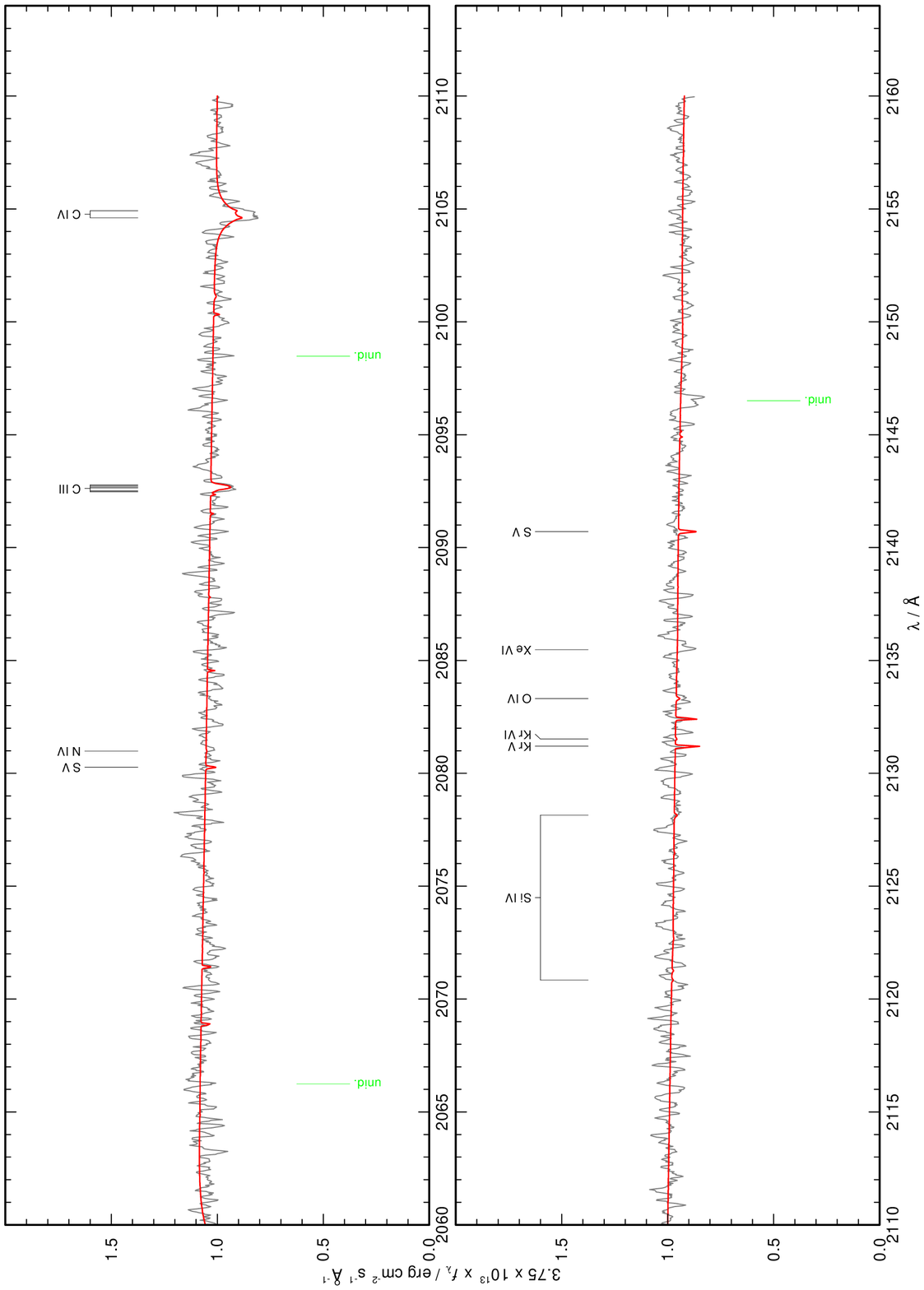}
  \caption{Figure\,\ref{fig:STIS_complete} continued.} 
\end{figure*}

\clearpage

\addtocounter{figure}{-1} 
\begin{figure*}
   \includegraphics[trim=0 -0 0 -8,height=23.0cm,angle=0]{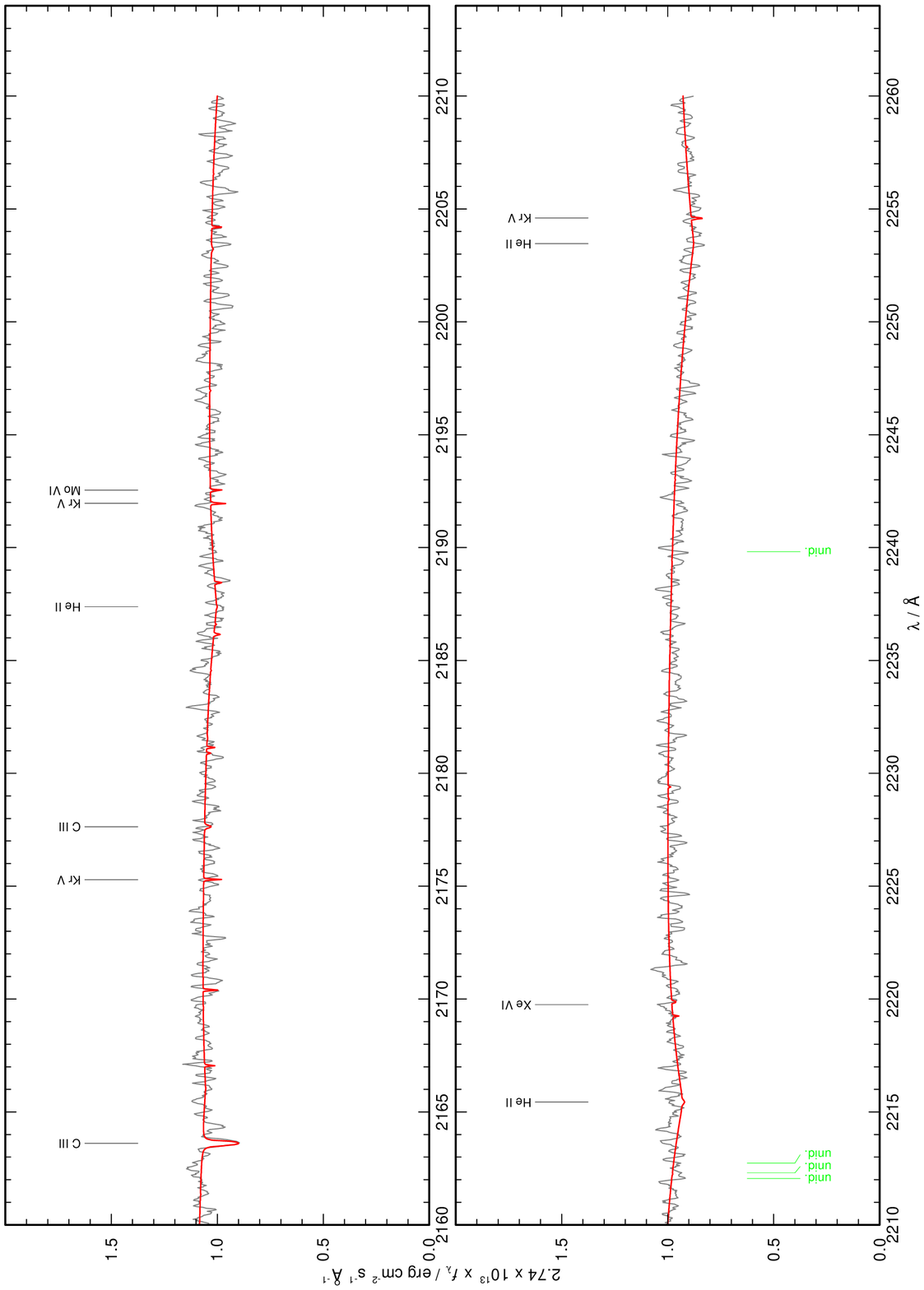}
  \caption{Figure\,\ref{fig:STIS_complete} continued.} 
\end{figure*}

\clearpage

\addtocounter{figure}{-1} 
\begin{figure*}
   \includegraphics[trim=0 -0 0 -8,height=23.0cm,angle=0]{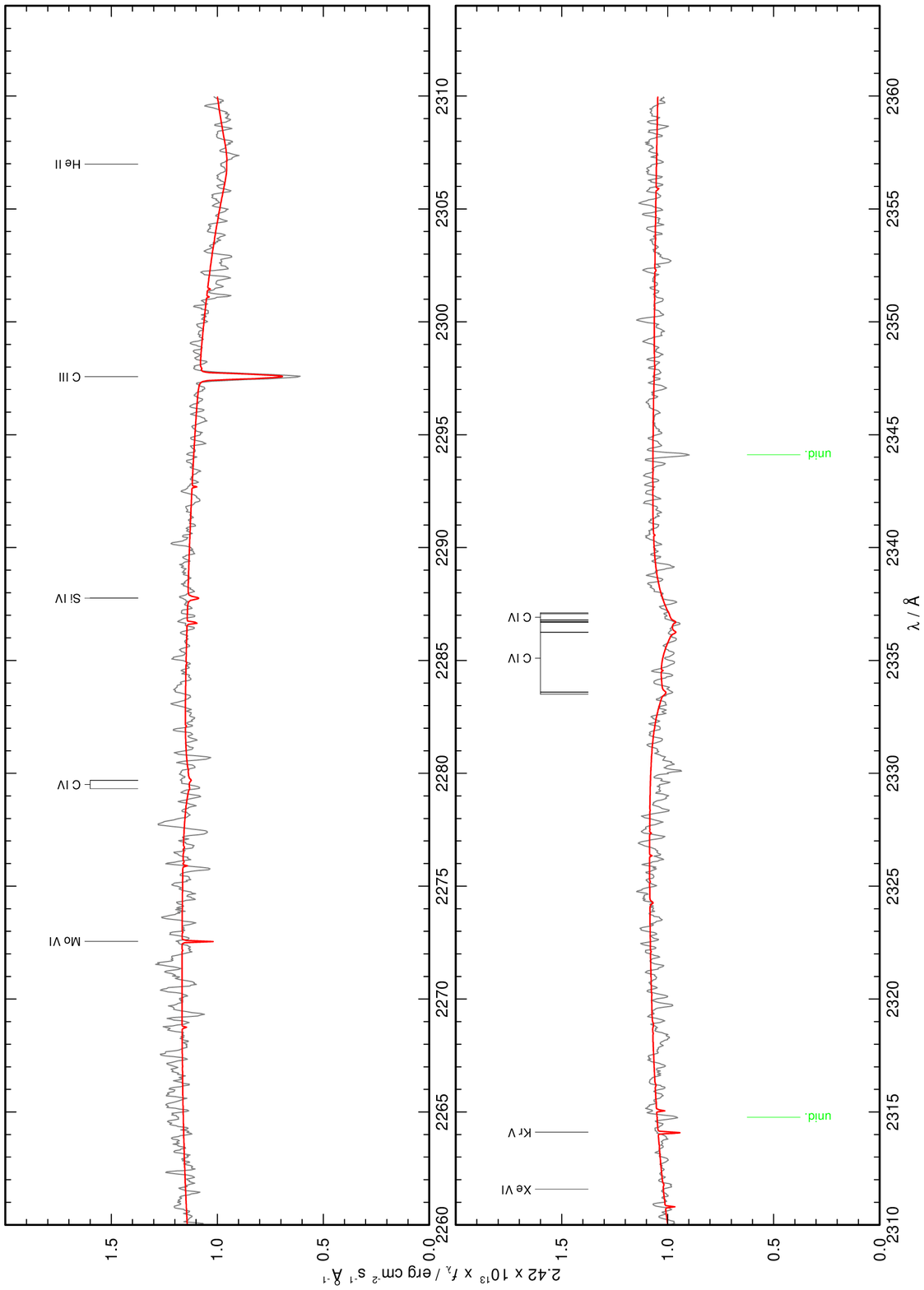}
  \caption{Figure\,\ref{fig:STIS_complete} continued.} 
\end{figure*}

\clearpage

\addtocounter{figure}{-1} 
\begin{figure*}
   \includegraphics[trim=0 -0 0 -8,height=23.0cm,angle=0]{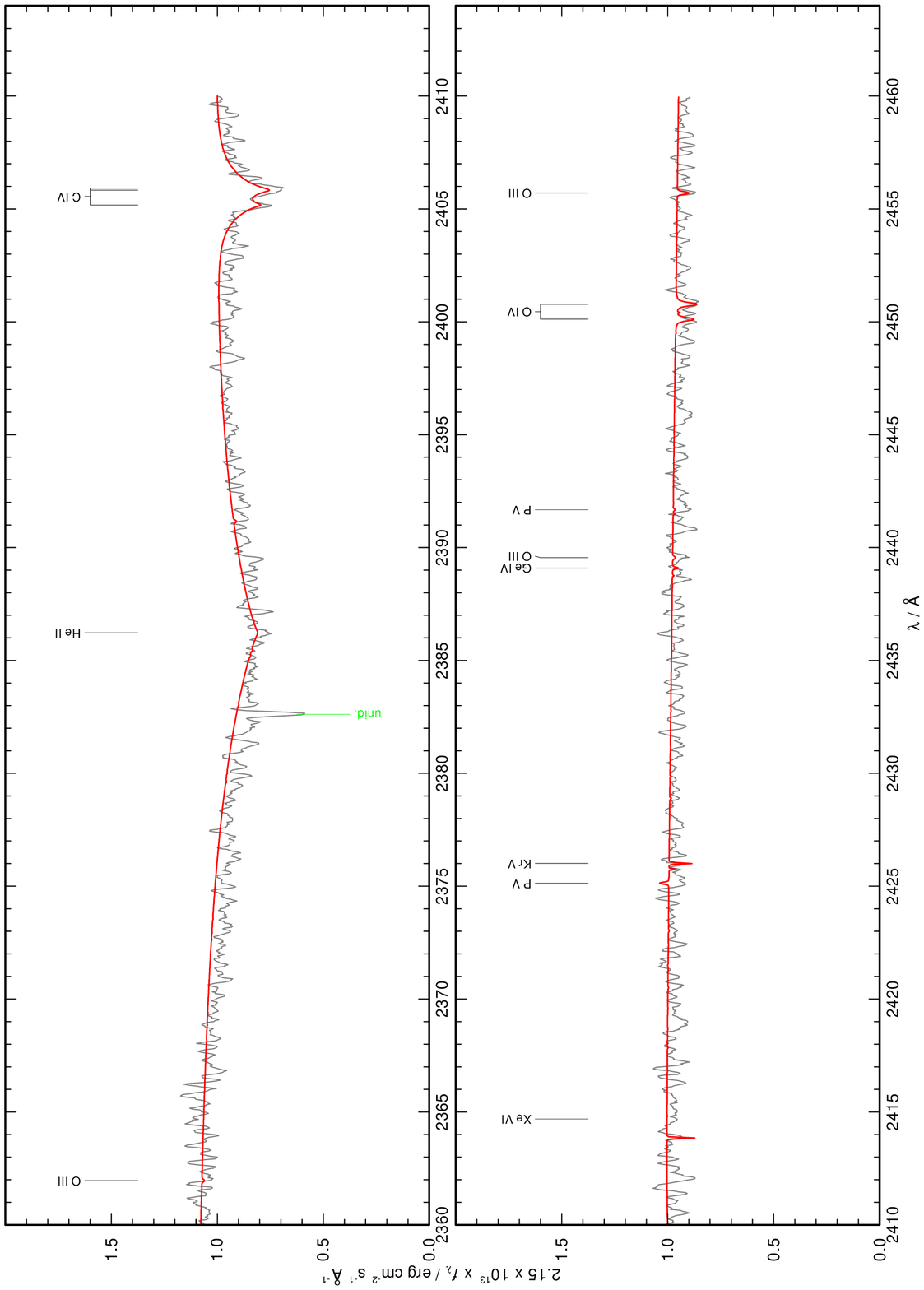}
  \caption{Figure\,\ref{fig:STIS_complete} continued.} 
\end{figure*}

\clearpage

\addtocounter{figure}{-1} 
\begin{figure*}
   \includegraphics[trim=0 -0 0 -8,height=23.0cm,angle=0]{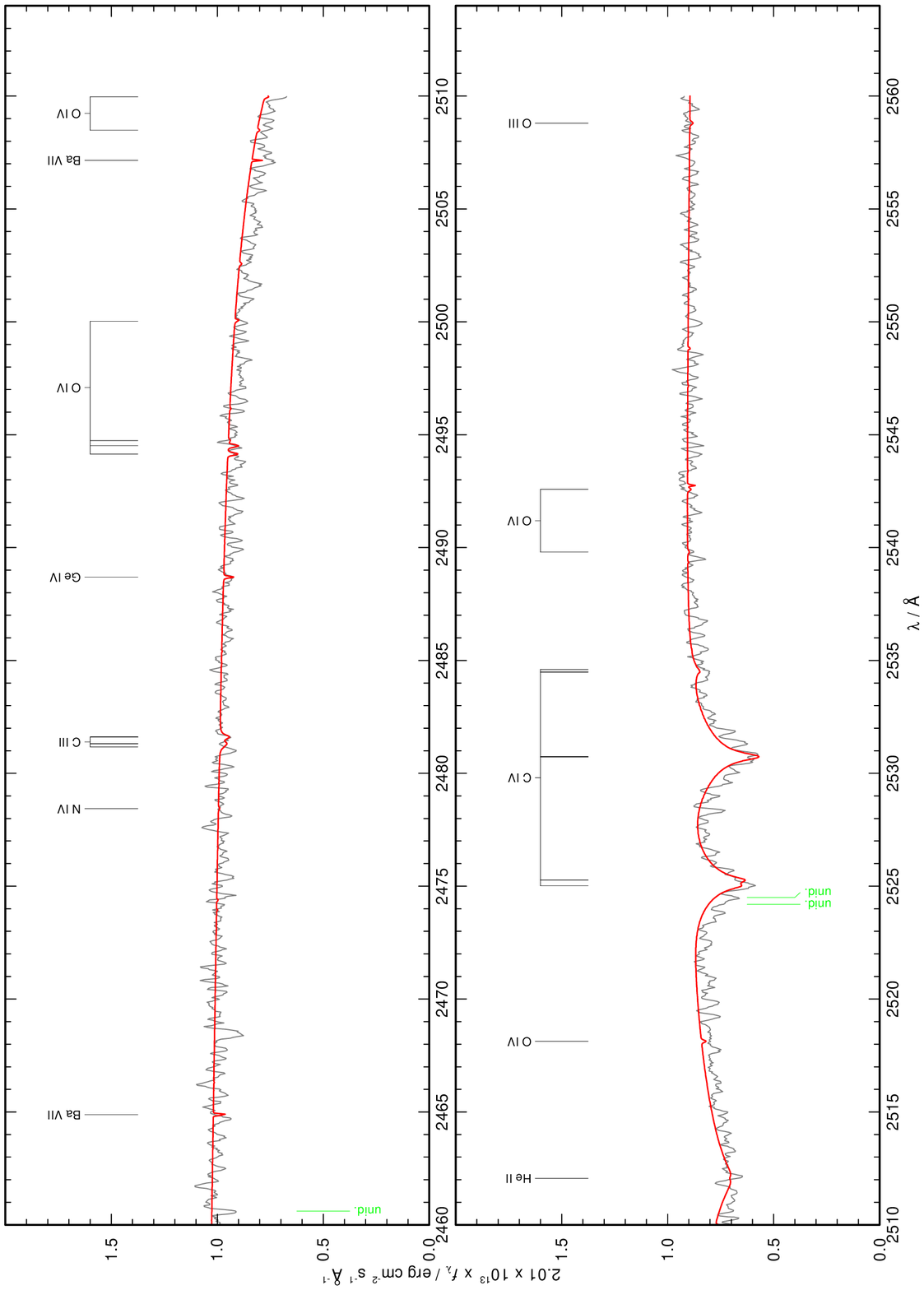}
  \caption{Figure\,\ref{fig:STIS_complete} continued.} 
\end{figure*}

\clearpage

\addtocounter{figure}{-1} 
\begin{figure*}
   \includegraphics[trim=0 -0 0 -8,height=23.0cm,angle=0]{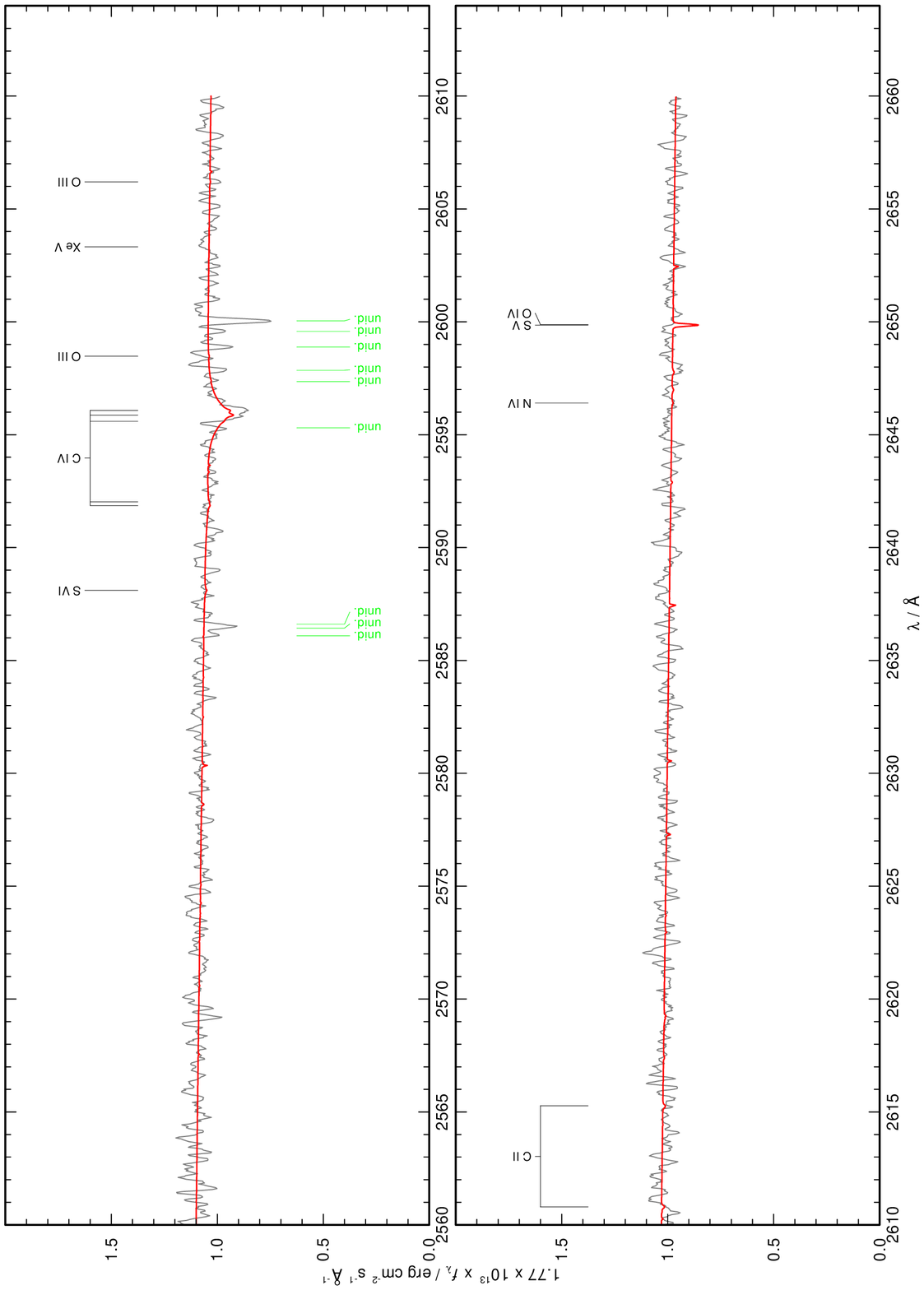}
  \caption{Figure\,\ref{fig:STIS_complete} continued.} 
\end{figure*}

\clearpage

\addtocounter{figure}{-1} 
\begin{figure*}
   \includegraphics[trim=0 -0 0 -8,height=23.0cm,angle=0]{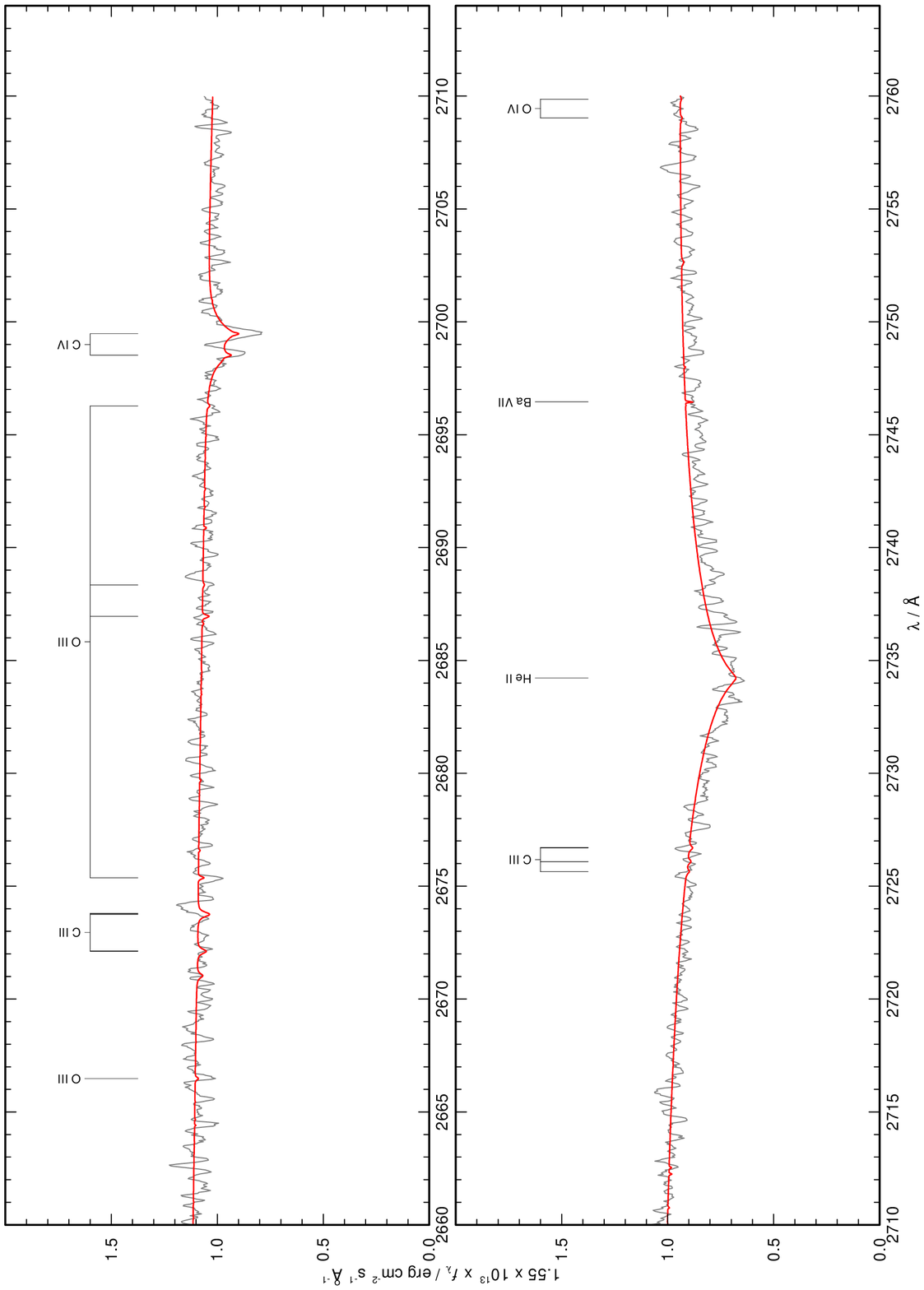}
  \caption{Figure\,\ref{fig:STIS_complete} continued.} 
\end{figure*}

\clearpage

\addtocounter{figure}{-1} 
\begin{figure*}
   \includegraphics[trim=0 -0 0 -8,height=23.0cm,angle=0]{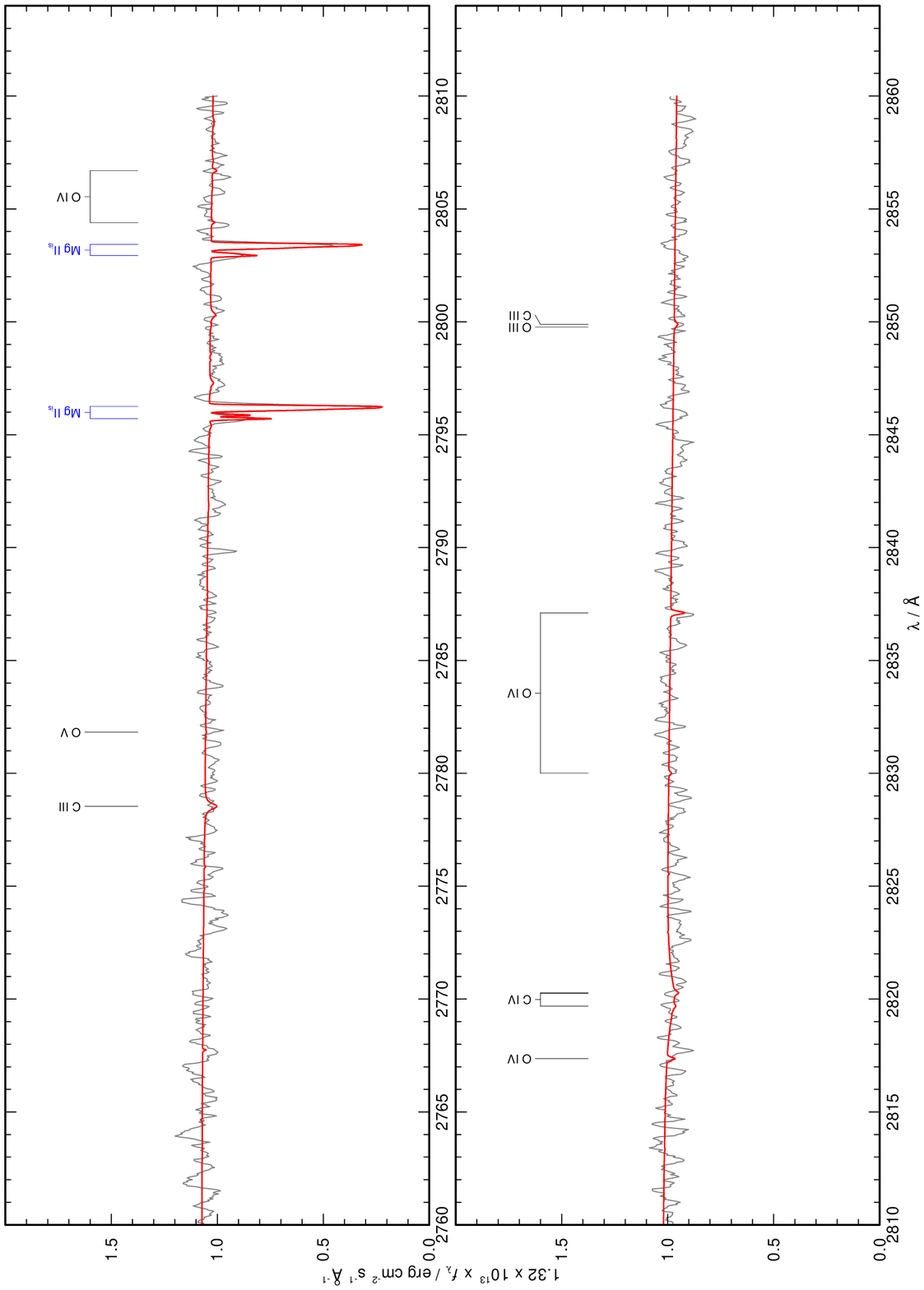}
  \caption{Figure\,\ref{fig:STIS_complete} continued.} 
\end{figure*}

\clearpage

\addtocounter{figure}{-1} 
\begin{figure*}
   \includegraphics[trim=0 -0 0 -8,height=23.0cm,angle=0]{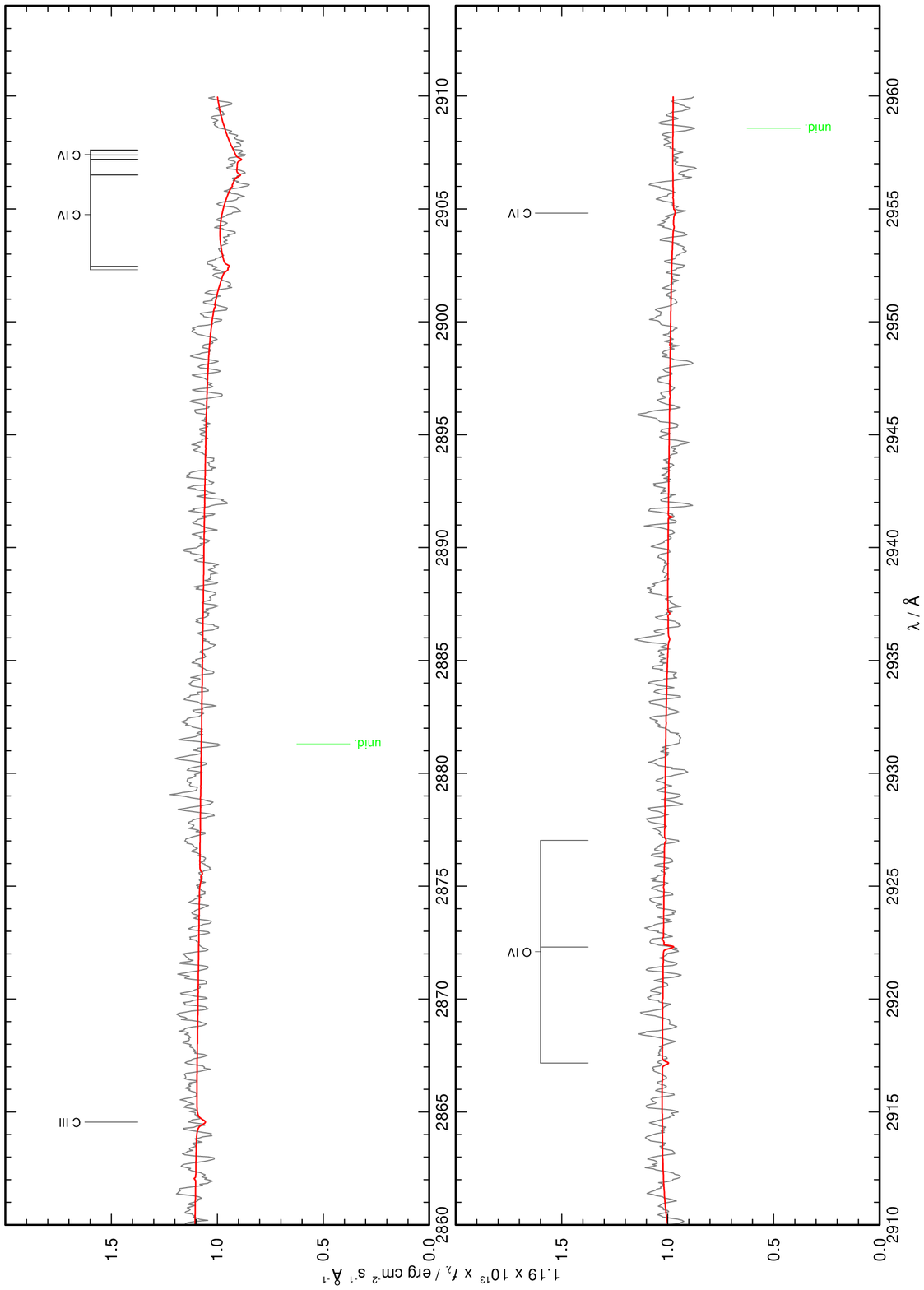}
  \caption{Figure\,\ref{fig:STIS_complete} continued.} 
\end{figure*}

\clearpage

\addtocounter{figure}{-1} 
\begin{figure*}
   \includegraphics[trim=0 -0 0 -8,height=23.0cm,angle=0]{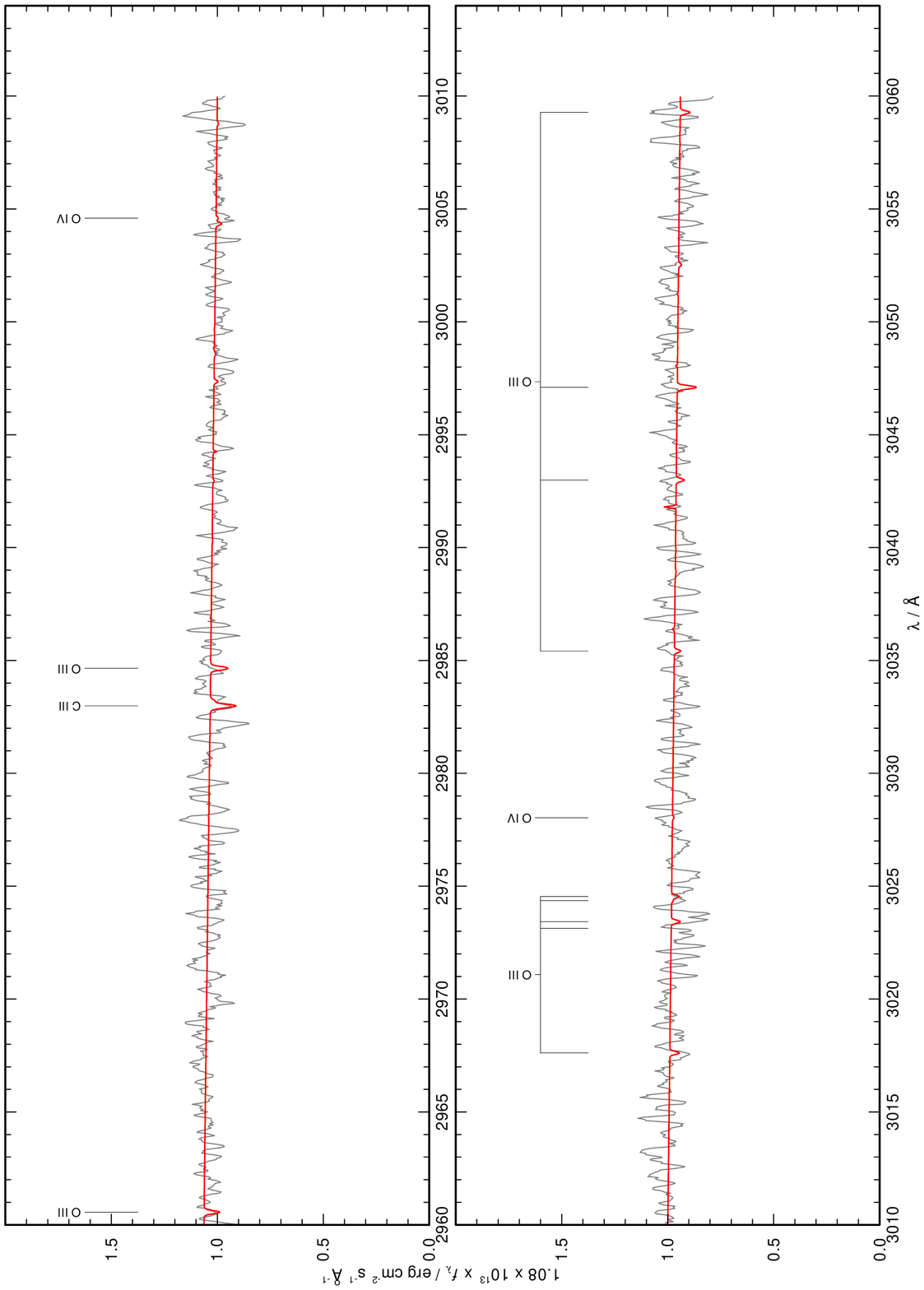}
  \caption{Figure\,\ref{fig:STIS_complete} continued.} 
\end{figure*}

\clearpage

\begin{figure*}
  \includegraphics[trim=0 -0 0 -8,height=23.0cm,angle=0]{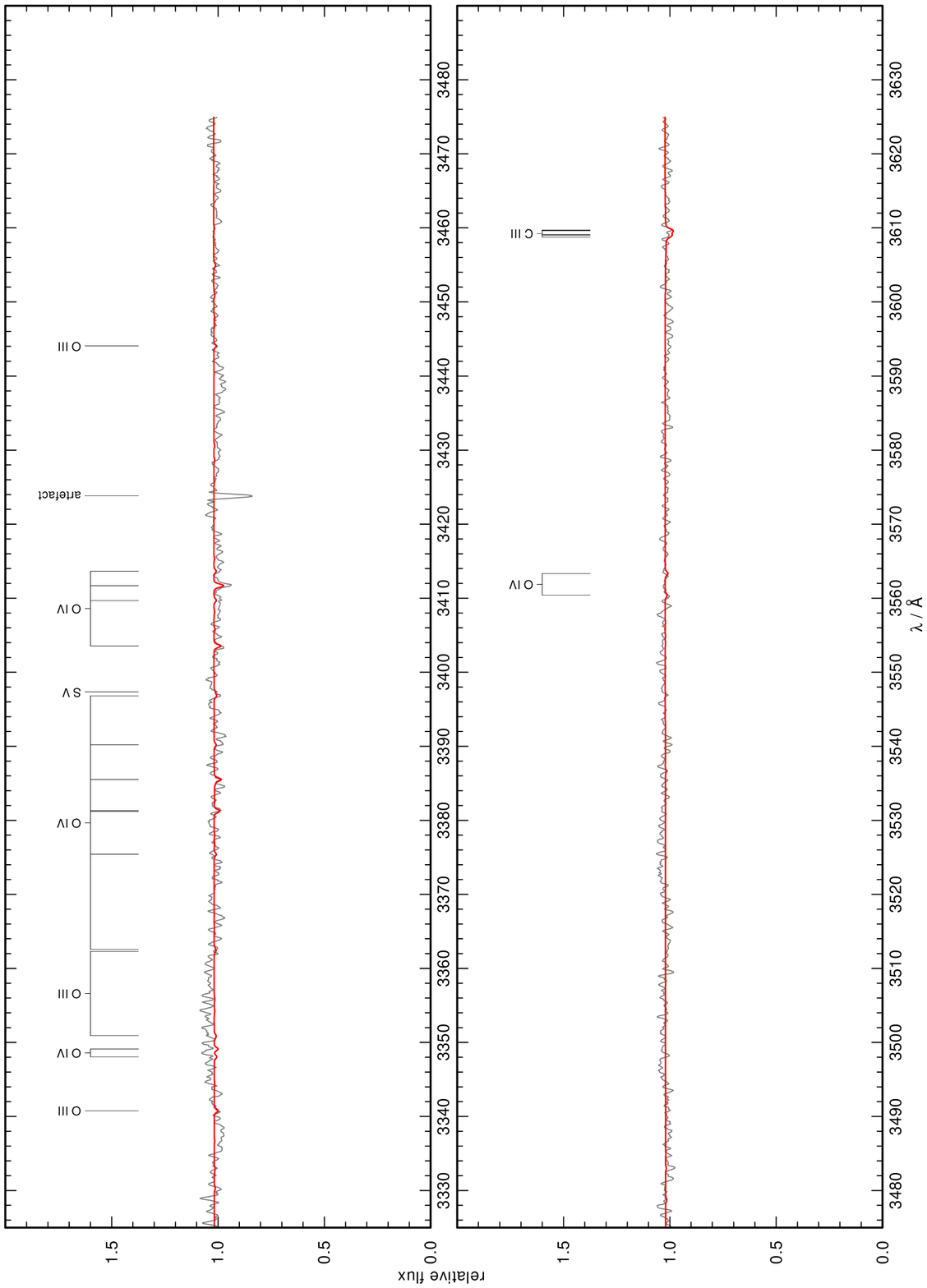}
  \caption{Optical (SPY) observation (gray) compared with the best model (red).
           Stellar lines are identified at top. ``unid.'' denotes unidentified lines.} 
  \label{fig:OPTI_complete}
\end{figure*}

\clearpage

\addtocounter{figure}{-1} 
\begin{figure*}
  \includegraphics[trim=0 -0 0 -8,height=23.0cm,angle=0]{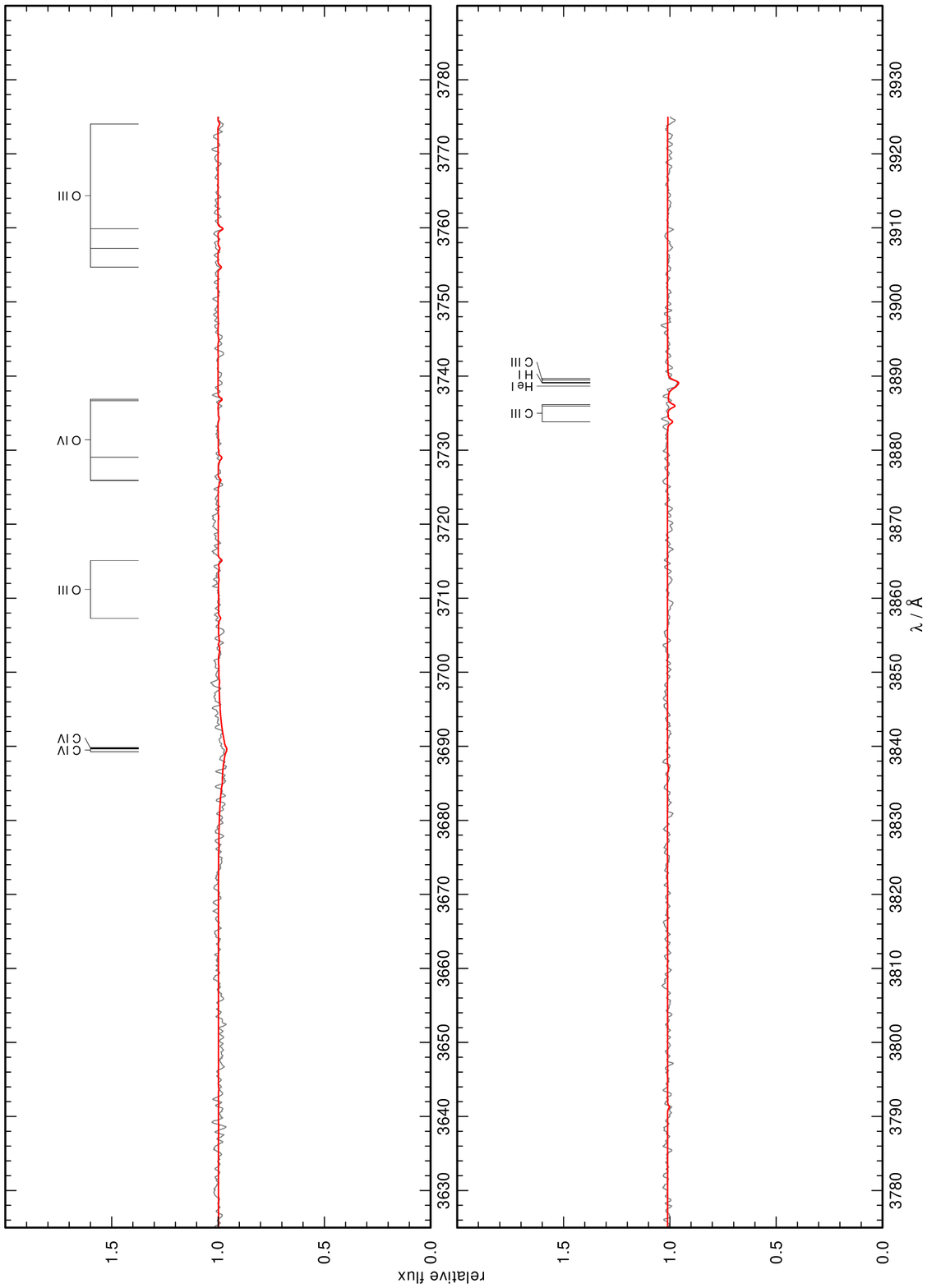}
  \caption{Figure\,\ref{fig:OPTI_complete} continued.} 
  \label{fig:OPTI_complete}
\end{figure*}

\clearpage

\addtocounter{figure}{-1} 
\begin{figure*}
  \includegraphics[trim=0 -0 0 -8,height=23.0cm,angle=0]{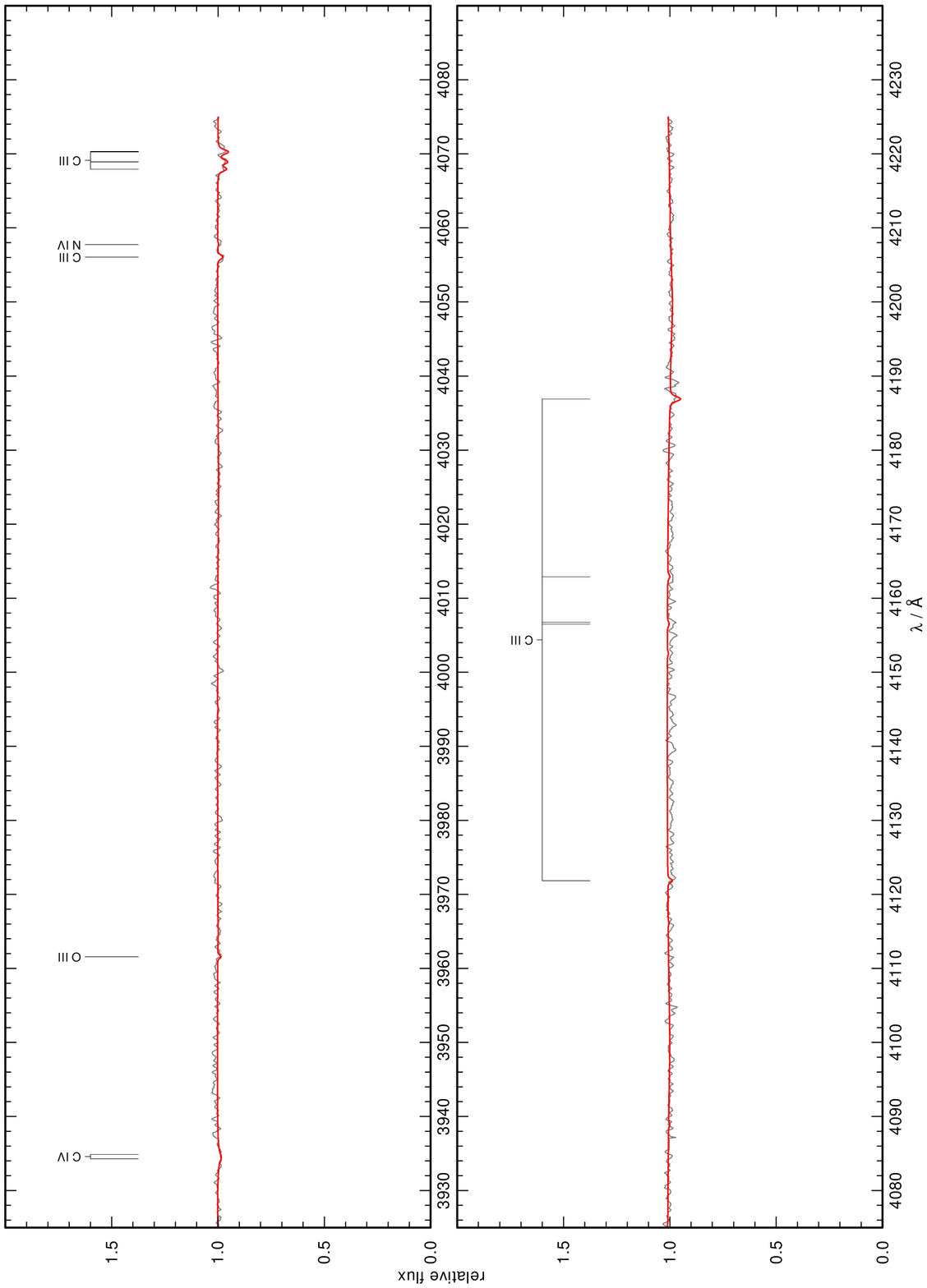}
  \caption{Figure\,\ref{fig:OPTI_complete} continued.} 
  \label{fig:OPTI_complete}
\end{figure*}

\clearpage

\addtocounter{figure}{-1} 
\begin{figure*}
  \includegraphics[trim=0 -0 0 -8,height=23.0cm,angle=0]{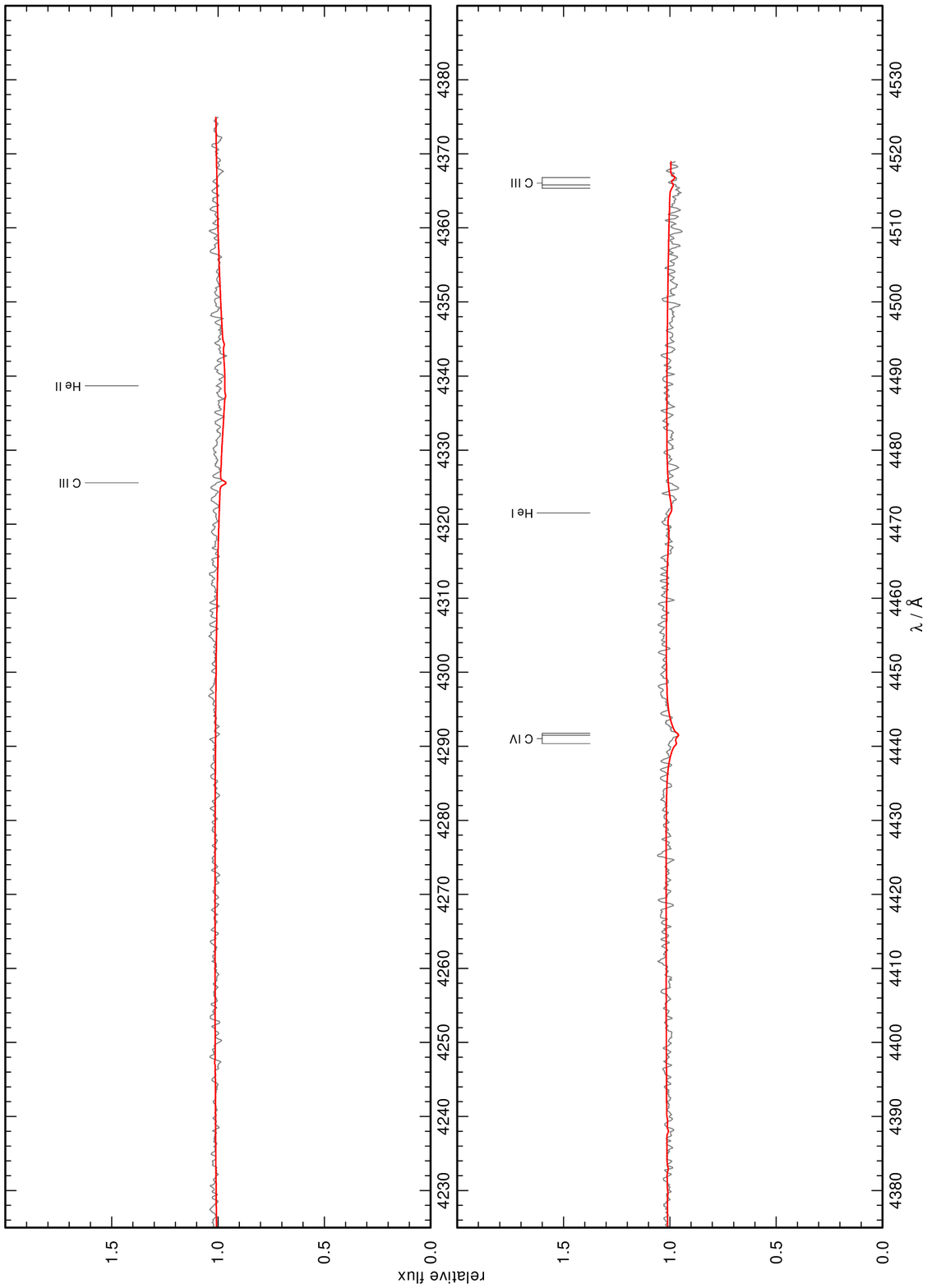}
  \caption{Figure\,\ref{fig:OPTI_complete} continued.} 
  \label{fig:OPTI_complete}
\end{figure*}

\clearpage

\addtocounter{figure}{-1} 
\begin{figure*}
  \includegraphics[trim=0 -0 0 -8,height=23.0cm,angle=0]{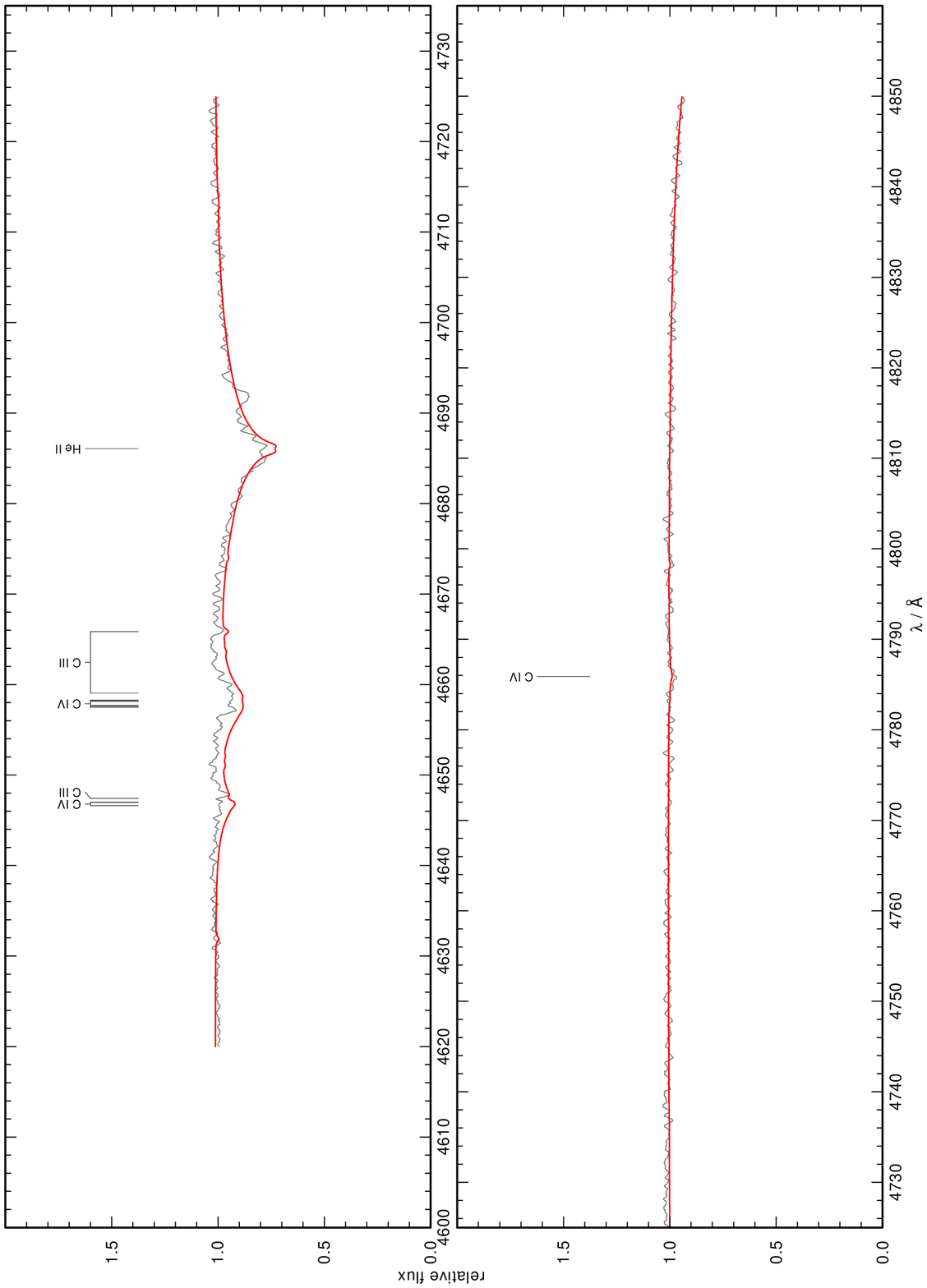}
  \caption{Figure\,\ref{fig:OPTI_complete} continued.} 
  \label{fig:OPTI_complete}
\end{figure*}

\clearpage

\addtocounter{figure}{-1} 
\begin{figure*}
  \includegraphics[trim=0 -0 0 -8,height=23.0cm,angle=0]{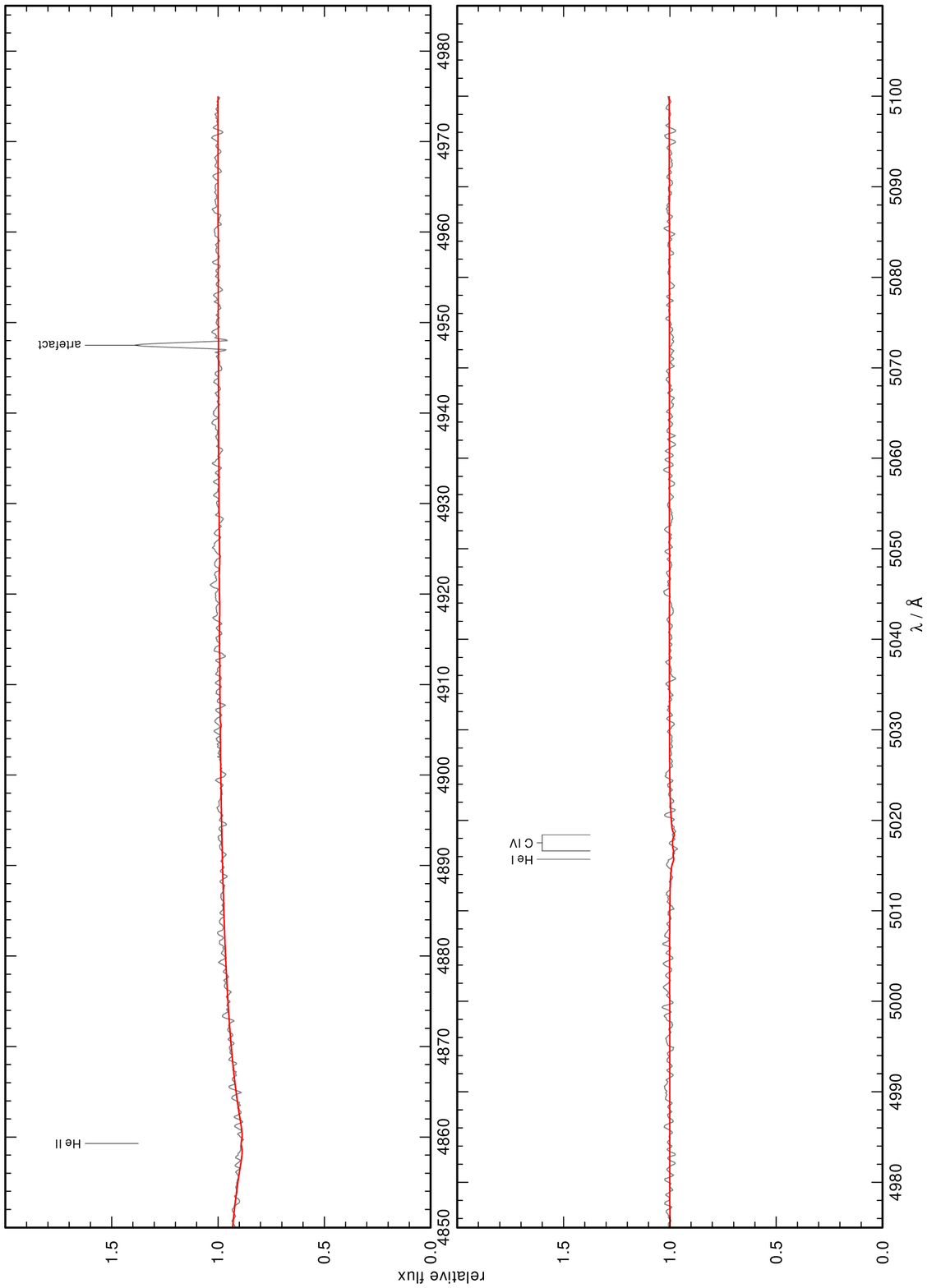}
  \caption{Figure\,\ref{fig:OPTI_complete} continued.} 
  \label{fig:OPTI_complete}
\end{figure*}

\clearpage

\addtocounter{figure}{-1} 
\begin{figure*}
  \includegraphics[trim=0 -0 0 -8,height=23.0cm,angle=0]{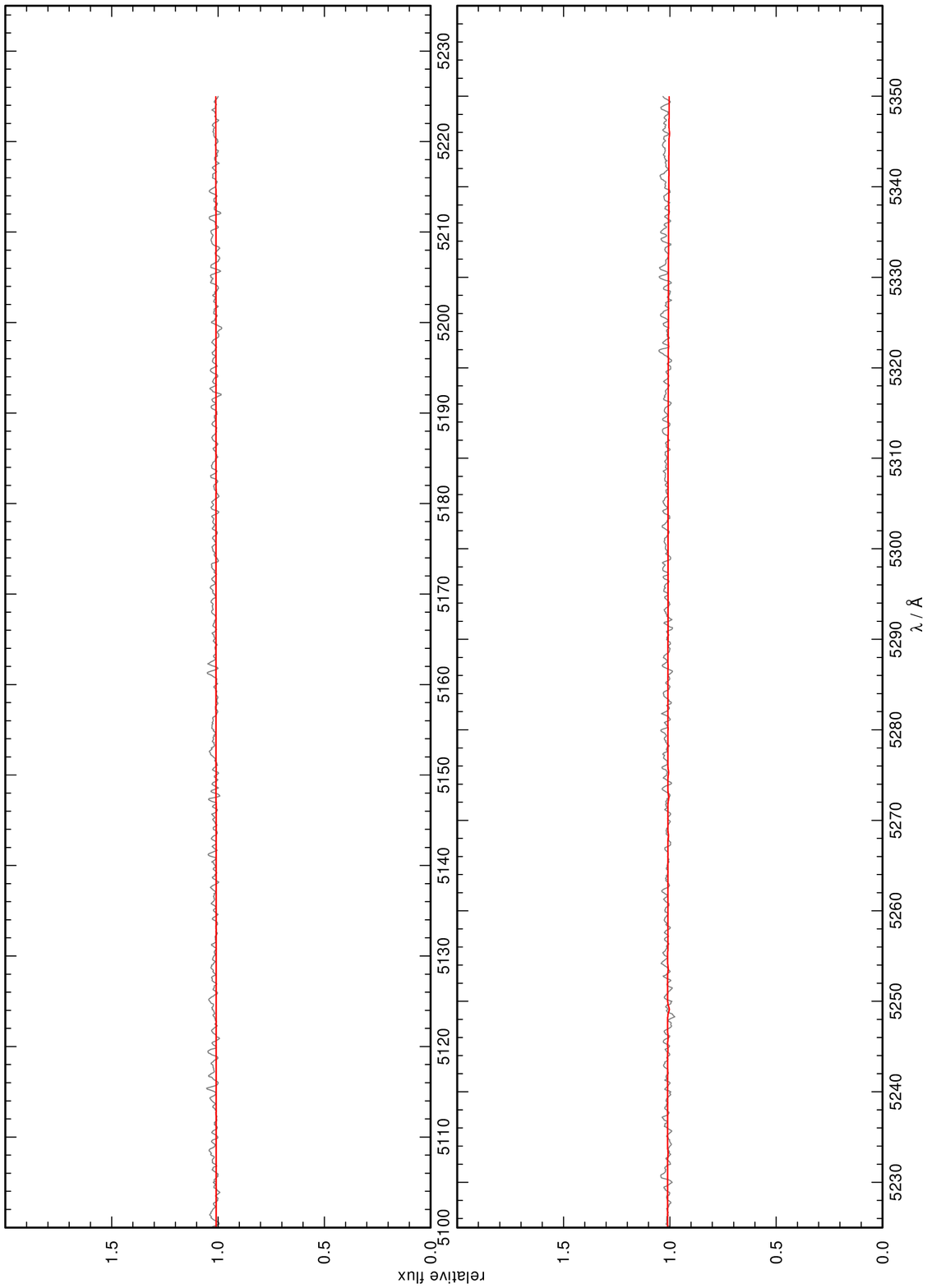}
  \caption{Figure\,\ref{fig:OPTI_complete} continued.} 
  \label{fig:OPTI_complete}
\end{figure*}

\clearpage

\addtocounter{figure}{-1} 
\begin{figure*}
  \includegraphics[trim=0 -0 0 -8,height=23.0cm,angle=0]{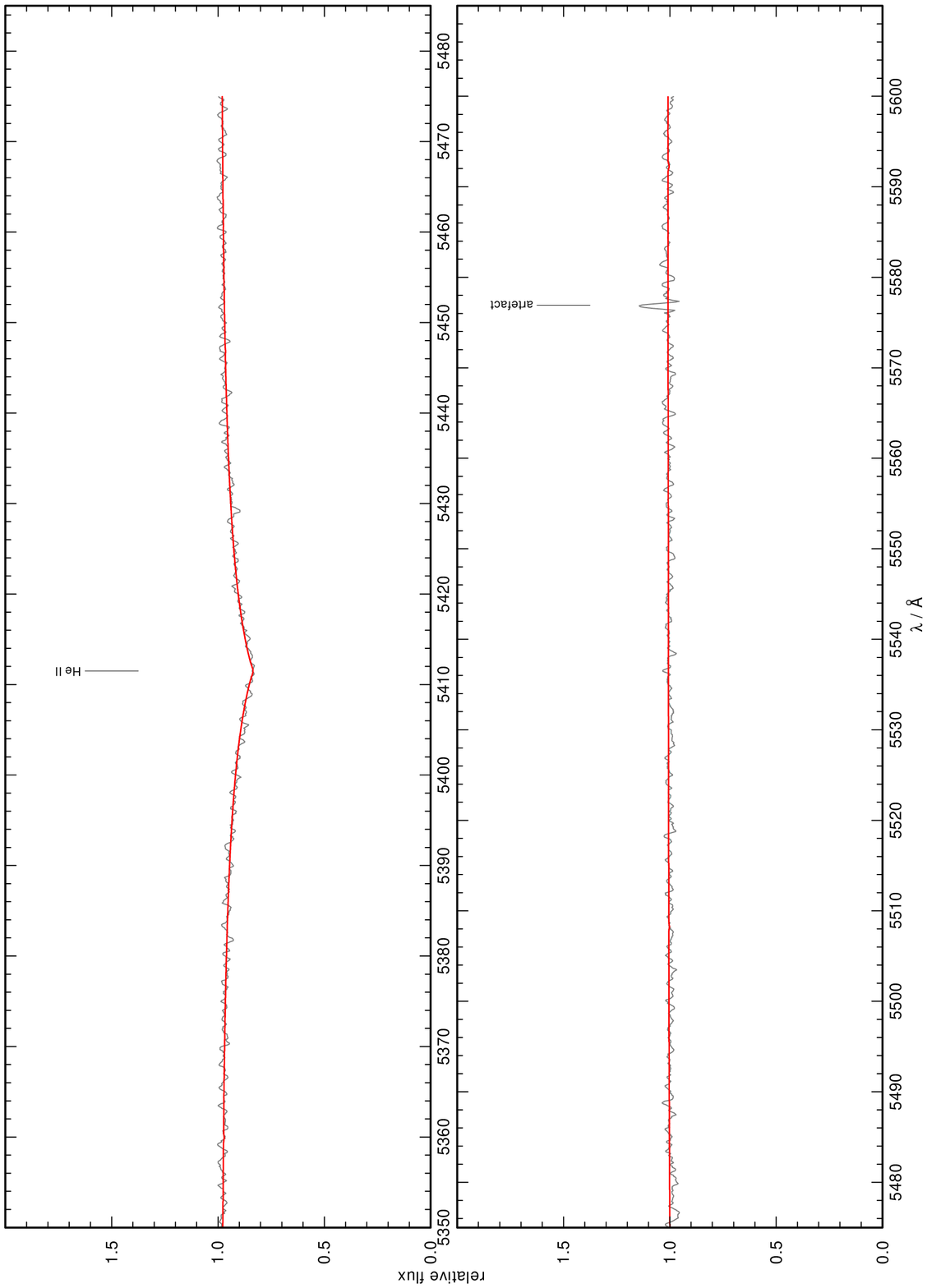}
  \caption{Figure\,\ref{fig:OPTI_complete} continued.} 
  \label{fig:OPTI_complete}
\end{figure*}

\clearpage

\addtocounter{figure}{-1} 
\begin{figure*}
  \includegraphics[trim=0 -0 0 -8,height=23.0cm,angle=0]{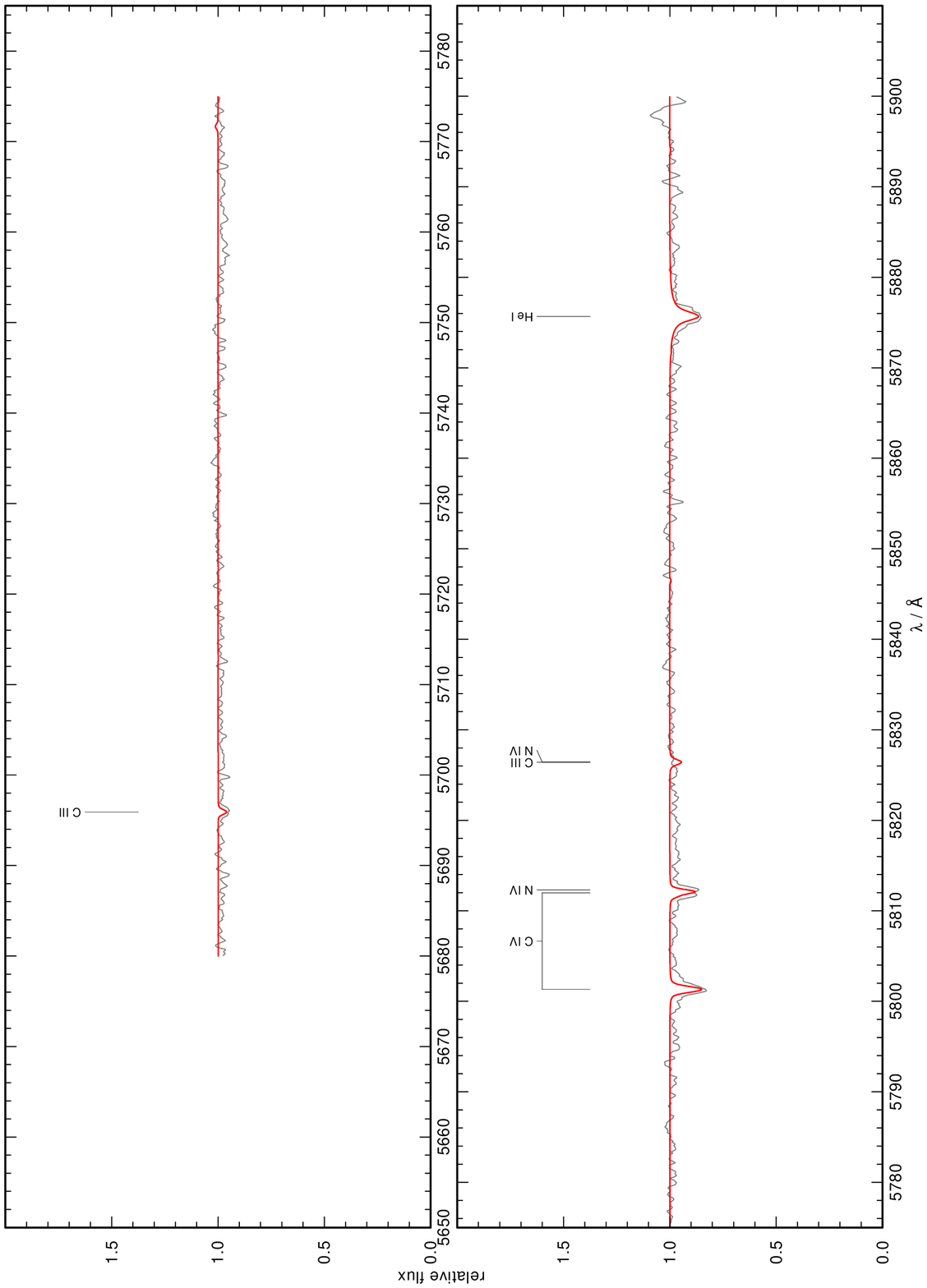}
  \caption{Figure\,\ref{fig:OPTI_complete} continued.} 
  \label{fig:OPTI_complete}
\end{figure*}

\clearpage

\addtocounter{figure}{-1} 
\begin{figure*}
  \includegraphics[trim=0 -0 0 -8,height=23.0cm,angle=0]{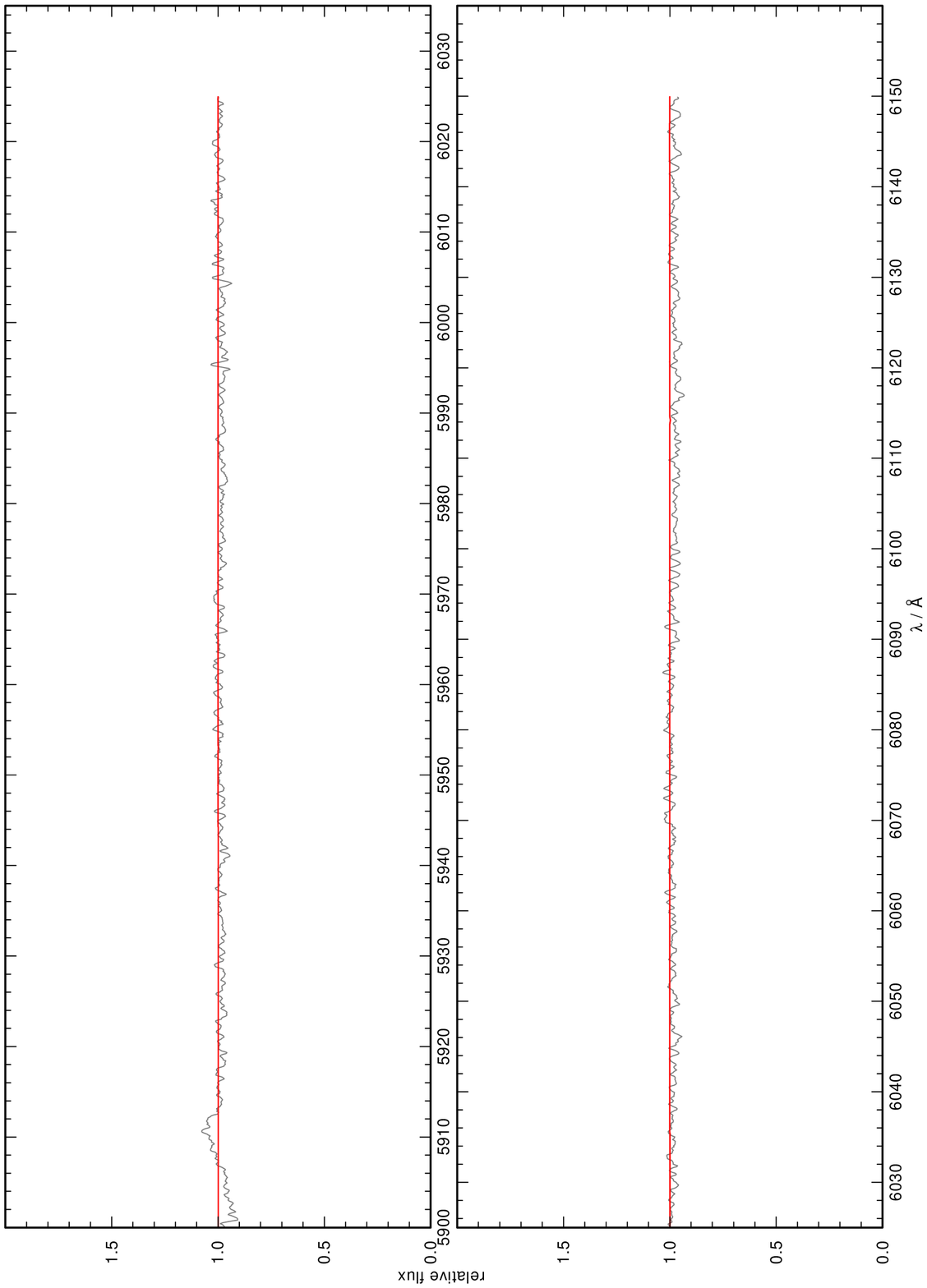}
  \caption{Figure\,\ref{fig:OPTI_complete} continued.} 
  \label{fig:OPTI_complete}
\end{figure*}

\clearpage

\addtocounter{figure}{-1} 
\begin{figure*}
  \includegraphics[trim=0 -0 0 -8,height=23.0cm,angle=0]{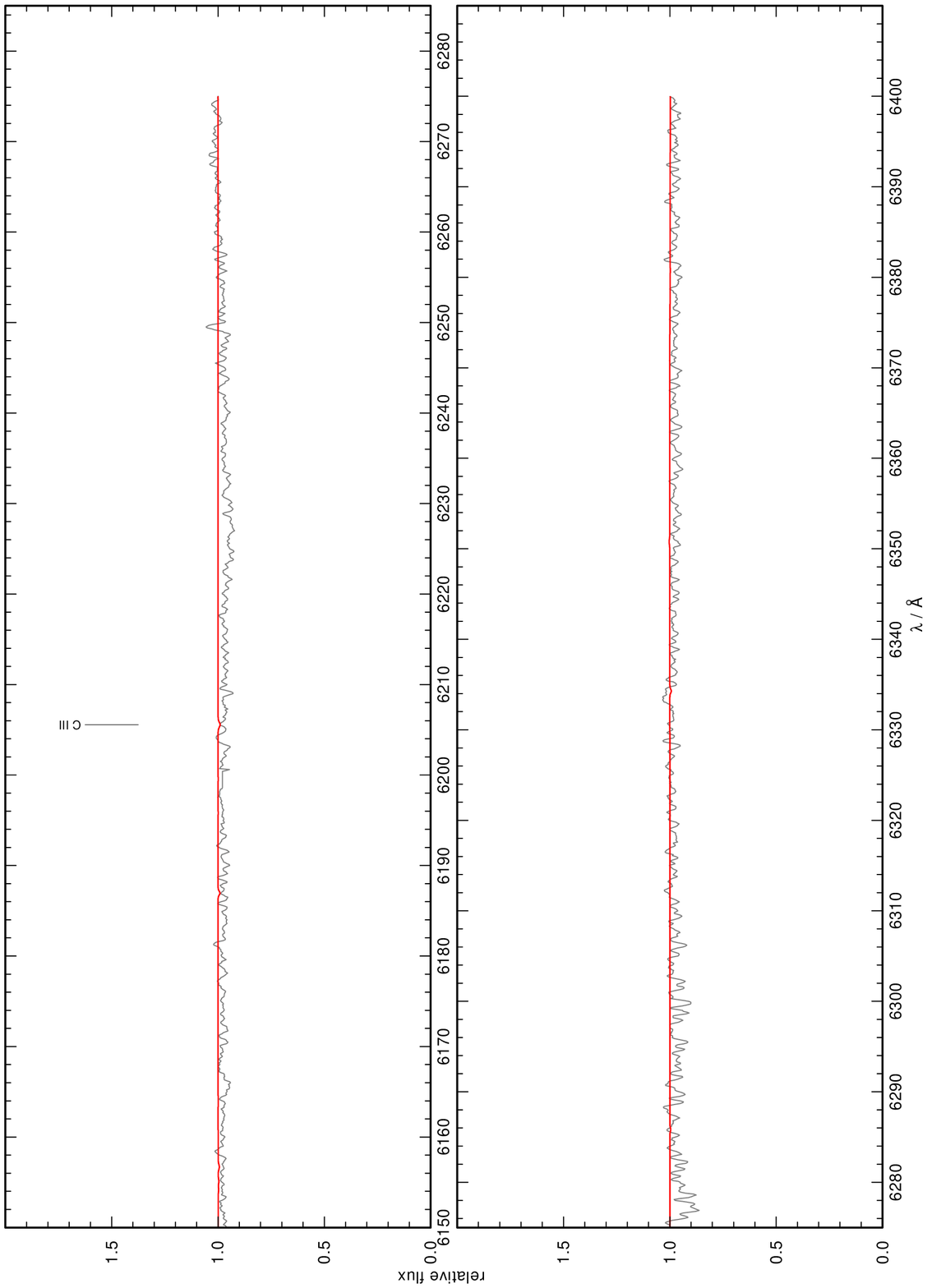}
  \caption{Figure\,\ref{fig:OPTI_complete} continued.} 
  \label{fig:OPTI_complete}
\end{figure*}

\clearpage

\addtocounter{figure}{-1} 
\begin{figure*}
  \includegraphics[trim=0 -0 0 -8,height=23.0cm,angle=0]{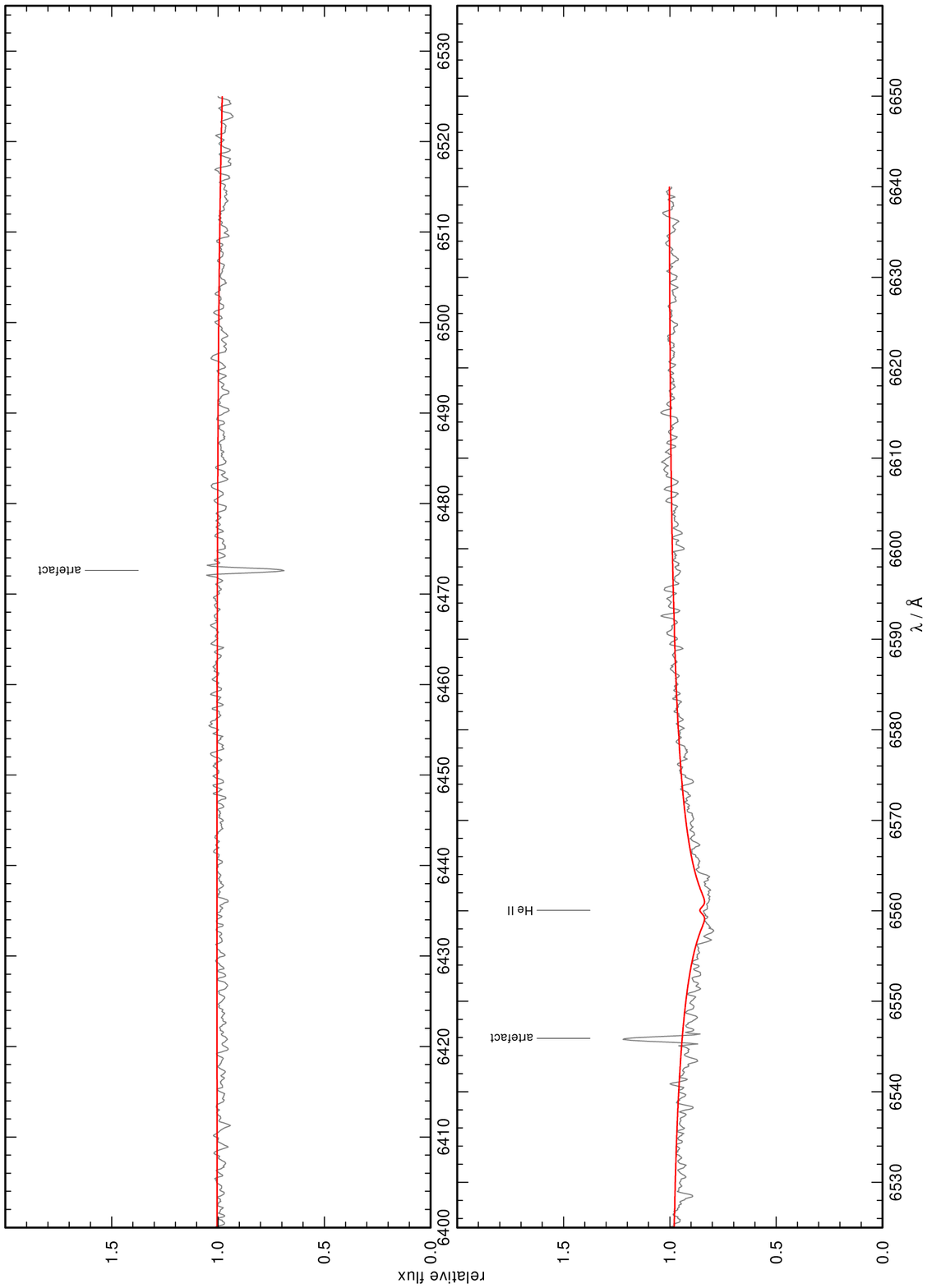}
  \caption{Figure\,\ref{fig:OPTI_complete} continued.} 
  \label{fig:OPTI_complete}
\end{figure*}

\clearpage

\section{WWW interfaces of TEUV, TGRED, and TVIS}
\label{app:gavo}

\begin{landscape}

\addtolength{\textwidth}{6.3cm} 
\addtolength{\evensidemargin}{-1mm}
\addtolength{\oddsidemargin}{-1mm}

\begin{figure*}
  \caption{TEUV WWW interface. Not shown on astro-ph, please visit the WWW page.} 
  \label{fig:TEUV}
\end{figure*}

\begin{figure*}
  \caption{TGRED WWW interface. Not shown on astro-ph, please visit the WWW page.} 
  \label{fig:TGRED}
\end{figure*}

\begin{figure*}
  \caption{TVIS WWW interface. Not shown on astro-ph, please visit the WWW page.} 
  \label{fig:TVIS}
\end{figure*}

\end{landscape}

\end{document}